\documentclass{aa}

\usepackage{graphicx}
\usepackage{txfonts}
\usepackage{natbib}

\bibpunct{(}{)}{;}{a}{}{,} 

\newcommand{\matrixtt}[4]{\left( \begin{array}{cc}#1&#2\\#3&#4\\\end{array} \right)}

\newcommand{\herm}{H}
\newcommand{\jones}[2]{\vec {#1}_{#2}}
\newcommand{\jonesinv}[2]{\vec {#1}^{-1}_{#2}}
\newcommand{\jonesT}[2]{\vec {#1}^{\herm}_{#2}}
\newcommand{\jonesTinv}[2]{\vec {#1}^{{\herm}-1}_{#2}}
\newcommand{\coh}[2]{\mathsf{{#1}}_{{#2}}}

\begin{document}

\title{Revisiting the radio interferometer measurement equation.\\III. Addressing direction-dependent effects in 21 cm WSRT observations of 3C 147}

\author{O.M.\ Smirnov}

\institute{Netherlands Institute for Radio Astronomy (ASTRON)\\
  P.O. Box 2, 7990AA Dwingeloo, The Netherlands \\
  \email{smirnov@astron.nl}}

\date{Received 5 Nov 2010 / Accepted 5 Jan 2011}

\titlerunning{Revisiting the RIME. III. Addressing DDEs in WSRT observations of 3C 147.}
\authorrunning{O.M.\ Smirnov}

\abstract%
{Papers I and II of this series have extended the radio interferometry measurement equation (RIME) formalism to the full-sky case,
and provided a RIME-based description of calibration and the problem of direction-dependent effects (DDEs).}
{This paper aims to provide a practical demonstration of a RIME-based approach to calibration, via
an example of extremely high-dynamic range calibration of WSRT observations of 3C 147 at 21 cm, with full treatment of DDEs.}
{A version of the RIME incorporating differential gains has been implemented in MeqTrees, and applied to the 3C 147 data. This was
used to perform regular selfcal, then solve for interferometer-based errors and for differential gains.}%
{The resulting image of the field around 3C 147 is thermal noise-limited, has a very high dynamic range (1.6 million), 
and none of the off-axis artefacts that plague regular selfcal. The differential gain solutions show a high signal-to-noise 
ratio, and may be used to extract information on DDEs and errors in the sky model.
}%
{The differential gain approach can eliminate DDE-related artefacts, and provide information for iterative improvements of sky models. 
Perhaps most importantly, sources as faint as 2 mJy have been shown to yield meaningful differential gain solutions, 
and thus can be used as potential calibration beacons in other DDE-related schemes.}

\keywords{Methods: numerical - Methods: analytical - Methods: data analysis - Techniques:
interferometric - Galaxies: quasars: individual: 3C 147}

\maketitle

\section*{Introduction}

\begin{figure*}
\sidecaption
\centering
\includegraphics[width=12cm]{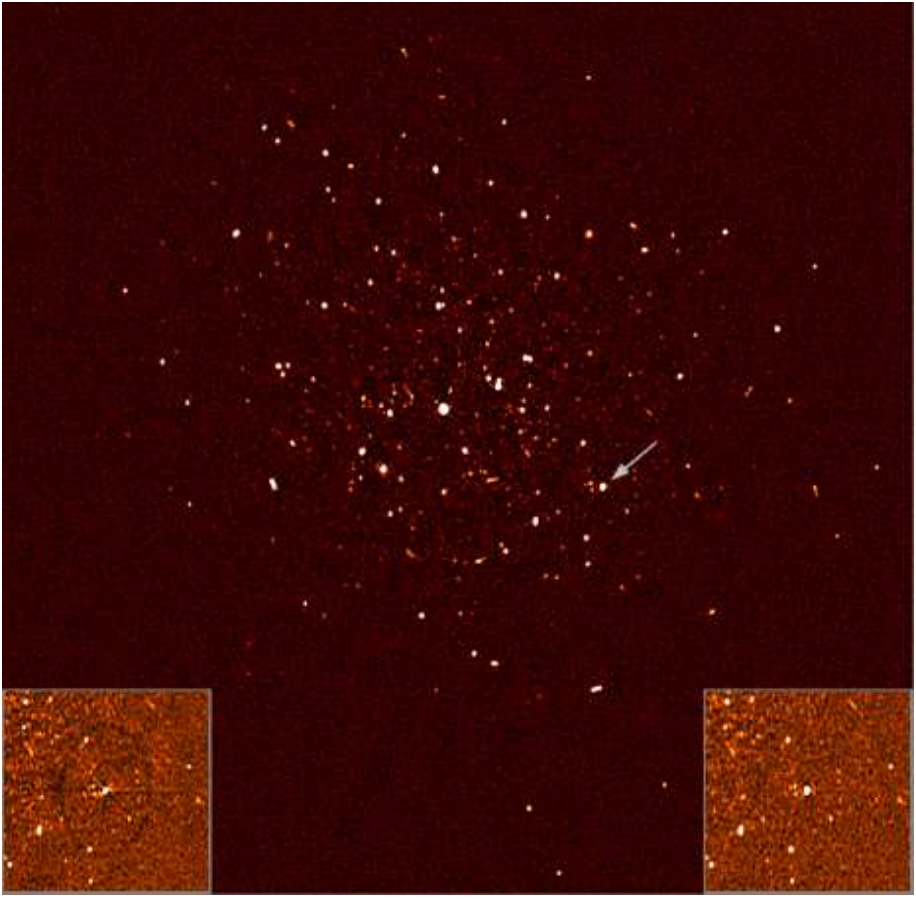}
\caption{\label{fig:3c147} ``Showcase'' image of the field around the bright radio source 3C 147, produced after reduction with MeqTrees. The image is noise-limited, and has a dynamic range of 1.6 million. This DR was already achieved by de Bruyn using regular selfcal in NEWSTAR, but the resulting images contained artefacts around off-axis sources (left inset) due to DDEs. A MeqTrees reduction incorporating differential gains, as described in this paper, has completely eliminated the artefacts (right inset). This image also appears in \citet{meqtrees}.}
\end{figure*}

The field around the bright radio source 3C 147 is a favourite showcase for dynamic range (DR) demonstrations. 3C 147 itself is very bright (22 Jy at 21 cm), which ensures a high SNR for selfcal solutions, while the surrounding field boasts a spectacular collection of mostly point-like fainter sources. The absence of significant extended emission has allowed very accurate sky models to be constructed. It is then not surprising that 3C 147 was the first field to break the $10^6$ dynamic range barrier \citep{deBruyn:million,deBruyn:3c147}, on a single 12-hour synthesis. This spectacular result was achieved with regular selfcal implemented in the NEWSTAR package. A major contributing factor is the relatively low level of beam-related DDEs at the WSRT, as discussed in Paper II \citep[Sect.~2.1]{RRIME2}\footnote{By contrast, even the post-upgrade Expanded VLA (EVLA), with its more significant DDEs, has not yet (at time of writing) exceeded $10^6$.}. An image of the 3C 147 field at 1,600,000:1 dynamic range is shown in Fig.~\ref{fig:3c147}. The making of this image is the subject of Section~\ref{sec:calibration}.

The very same benign properties that allow the WSRT to achieve record DR also make it a perfect instrument for studying DDEs. The latter are prominent enough to clearly show up in high-DR maps (see e.g. left inset of Fig.~\ref{fig:3c147}), but not severe enough to hinder the building up of very deep sky models during normal selfcal\footnote{Strictly speaking, this is only true in continuum mode. In spectral line mode, the 17 MHz ``ripple'' discussed in Paper II \citep[Sect.~2.1.1]{RRIME2} becomes a very troublesome DDE. Further work is required on this subject.}. The 3C 147 observations by \citet{deBruyn:3c147} are a perfect example of this. I have therefore decided to reprocess these data using MeqTrees, to see if DDEs can be eliminated through the use of a suitable form of the RIME.

In the presence of a dominant source (in this case 3C 147 itself), selfcal solutions will tend to subsume all effects in the direction of that source. DDEs will then manifest themselves as artefacts around other sources, which need to be at a certain distance from the dominant source for the effect to become apparent. Since the dominant source is usually placed at or near the pointing centre (i.e. on-axis), DDE-related artefacts are also called {\em off-axis artefacts.} The artefacts themselves are quite clear (Fig.~\ref{fig:3c147}, left inset): the nature of the effects responsible for them, far less so. At least four possibilities have been postulated:

\begin{itemize}
\item Pointing errors \citep[Paper II,][Sect.~2.1.4]{RRIME2}.
\item Differential tropospheric refraction ({\em ibid.}, Sect.~2.2.4).
\item Errors in antenna positions, including non-coplanarity.
\item Other systematic coordinate errors.
\end{itemize}

One of the objects of this study was to narrow down these possibilities. Section~\ref{sec:de-analysis} therefore analyses the differential gain solutions obtained during this calibration, with a view to characterizing the DDEs.

\section{Calibration approaches and results}
\label{sec:calibration}

\subsection{Observations and NEWSTAR reduction}

The observational data in question were obtained by de Bruyn in 2003. A single 12-hour synthesis was taken, using $8\times20$ MHz bands (of 64 channels each) from 1300 to 1460 MHz, with 30 s integration time. Due to a back-end problem, one of the cross-correlations was corrupted, so only the $XX$ and $YY$ correlations were usable. De Bruyn then successfully reduced the observation using the NEWSTAR package, achieving a world record 1,600,000:1 dynamic range \citep{deBruyn:3c147}. Only regular selfcal was done and no peeling was attempted, so the resulting image showed DDE-related artefacts around off-axis sources (Fig.~\ref{fig:3c147}, left inset). One result of the reduction was a very deep NEWSTAR-format sky model for the field, containing about 300 point sources. This provided a fantastic platform from which to begin my DDE study with MeqTrees. I had a ready-made sky model that was known to be good enough to reach the thermal noise with this particular dataset, and I had intermediate images from de Bruyn's reduction that could be used as checkpoints.

The same observation was repeated in 2006 with somewhat different correlator settings \citep[for details, see]{deBruyn:3c147}, and reduced in a similar manner.

For his NEWSTAR reduction, de Bruyn self-calibrated each channel independently, rather than explicitly calibrating for a bandpass (see below). This procedure is described in Paper II \citep[Sect.~1.1]{RRIME2}. It uses the following implicit form of the RIME:

\begin{eqnarray}\label{eq:newstar-rime}
\coh{V}{pq} & = & \jones{G}{p} \left ( \coh{M}{pq} \ast \coh{X}{pq} \right ) \jonesT{G}{q}, \\
\nonumber \coh{X}{pq} & = & \sum_{s} E^2_s \coh{X}{spq},
\end{eqnarray}

and consists of the following steps:

\begin{enumerate}
\item Find $\jones{\tilde{G}}{p}$ that minimizes $|\coh{X}{pq} - \coh{D}{pq}|$ in a least-squares sense. Compute ``corrected data'' as $\jones{D}{pq}' = \jonesinv{\tilde{G}}{p} \coh{D}{pq} \jonesinv{\tilde{G}}{q}.$ (The solution interval here was one timeslot, one frequency channel. Only 56 channels in each band were usable; these were further averaged down using Hanning tapering, so in the end only 28 frequency points per band were used.)

\item Find $\coh{\tilde{M}}{pq}$ that minimizes $|\coh{M}{pq} \ast \coh{X}{pq} - \coh{D}{pq}'|$.
Compute ``corrected data'' as $\jones{D}{pq}'' = \jones{D}{pq}' \div \coh{\tilde{M}}{pq}$, where ``$\div$'' is element-by-element division -- the inverse of ``$\ast$''. (The solution interval here was the full 12 hours, one band.)

\item Compute ``residual data'' as $\coh{R}{pq} = \coh{D}{pq}'' - \coh{X}{pq}$. 
\end{enumerate}

This procedure was repeated for each band. Residual visibilities were imaged and summed across all 8 bands, then  deconvolved using H\"ogbom CLEAN. The sky model was then added back in using a Gaussian restoring beam.

\subsection{Calibration in MeqTrees: an overview}

In broad terms, selfcal in MeqTrees also consists of a least-squares fit of an equation such as (\ref{eq:newstar-rime}) to the data. However, the following features are different:

\begin{itemize}
\item The structure of the equation is not fixed: arbitrary forms of the RIME may be constructed. Crucially for my purposes, these may include DDE terms.
\item The elements of the Jones matrices are not necessarily simple solvable parameters (though they may be), but can be represented by arbitrary functions. For example, rather than solve for $E$-Jones (or $Z$-Jones) elements directly, MeqTrees can derive them from some model of the primary beam (or ionosphere) and solve for the parameters of the model. An example of this is given in Sect.~\ref{sec:3C 147:pointing}.
\item Different solvables may have different solution intervals, even in a simultaneous solution.
\end{itemize}

Because of the inherent flexibility of MeqTrees, calibration can be pursued in a great variety of ways. During this study, a specific methodology was narrowed down, implemented and tested. This became the basis of the ``Calico'' calibration framework that is now included with MeqTrees. The steps (and terminology) of calibration with Calico are as follows:

\begin{enumerate}
\item A desired form of the RIME is constructed, by selecting a sky model, and picking a series of Jones terms (plus, optionally, interferometer-based errors). For example, a form similar to NEWSTAR's implicit RIME (Eq.~\ref{eq:newstar-rime}), would be:
\[
\coh{V}{pq} =  \coh{M}{pq} \ast ( \jones{G}{p} \coh{X}{pq} \jonesT{G}{q} ) 
\]
$\coh{V}{pq}$ is usually called the ``corrupted predict''.
\item ``Corrupted predict'' is fitted to ``data''. That is, the MeqTrees solver is instructed to find values of solvable parameters that minimize $|\coh{D}{pq}-\coh{V}{pq}|$ in a least-squares sense. This can be (and usually is) done in multiple stages, e.g. $\jones{\tilde{G}}{p}$ first, followed by $\coh{\tilde{M}}{pq}$, etc.
\item Output visibilities are computed as either ``corrected data''
\[
\coh{D}{pq}' = \jonesinv{\tilde{G}}{p} (\coh{D}{pq} \div \coh{\tilde{M}}{pq}) \jonesTinv{\tilde{G}}{q},
\]
or as ``corrupted residuals'' $\coh{R}{pq} = \coh{D}{pq}-\coh{V}{pq}$, or as ``corrected residuals''
\[
\coh{R}{pq}' = \jonesinv{\tilde{G}}{p} (\coh{R}{pq} \div \coh{\tilde{M}}{pq}) \jonesTinv{\tilde{G}}{q}.
\]
\end{enumerate}

Subsequent imaging steps are not considered part of Calico or MeqTrees per se, since they apply equally to data calibrated using any other means, and are accomplished via external tools such as the {\tt lwimager} program (part of the {\tt casarest} package). These steps may also differ from project to project. I routinely image the per-band corrected residuals $\coh{R}{pq}'$, since these provide the best visual indicator of the quality of a calibration. For this particular study, I subsequently made 8-band residual images in MFS mode, using all data. The images were further deconvolved using Cotton-Schwab CLEAN \citep{Schwab:csclean}, and the sky model was added back in using a Gaussian restoring beam.

While Calico can produce ``corrected data'' (for purposes of imaging, etc.) at any step of the reduction, it does not use it as input for subsequent stages like NEWSTAR and other 2GC packages do. The fitting at step 2 is always done using the original\footnote{Or at most pre-averaged.} observed data $\coh{D}{pq}$, and calibration consists of building up the RIME until the ``corrupted predict'' fits the observations. Once DDEs enter the picture, ``corrected data'' (in the conventional understanding of data corrected for instrumental errors) do not really exist any more, since visibilities can only be corrected for the value of a DDE in a particular direction. Hence the Calico philosophy is to work with the original data at all times.

\subsection{Bandpass selfcal}

A per-channel, per-timeslot solution for $\jones{G}{p}$ in Eq. (\ref{eq:newstar-rime}) has $2N$ complex unknowns, and $N(N-1)$ complex measurements, where $N$ is the number of stations (14 for the WSRT). While an acceptable ratio (this is, after all, why selfcal works in the first place!), it does not leave a lot of room for introducing DDE-related parameters. In general, we want to allow our solutions as few DoF's as possible, and making the solution intervals (in time and/or frequency) larger is one way of ensuring this.

The WSRT has a reasonably stable bandpass, so an obvious way to reduce the parameter count is to separate $\jones{G}{p}$ into a bandpass component to capture the frequency structure (with little to none variation in time), and a frequency-independent, rapidly varying complex gain. Per each station/receiver, this replaces $N_\mathrm{chan}\times N_\mathrm{time}$ parameters with only $N_\mathrm{chan}+N_\mathrm{time}$ of them.

For the MeqTrees reduction, I therefore started with the following RIME:

\begin{equation}\label{eq:3C 147:bandpass}
\coh{V}{pq} = \coh{M}{pq} \ast \left ( \jones{G}{p} \jones{B}{p} \left( \sum_s E_s \coh{X}{spq} E^{\herm}_s \right)  \jonesT{B}{q} \jonesT{G}{q} \right )
\end{equation}

Here, $\coh{X}{spq}$, $E_s$, $\jones{G}{p}$ and $\coh{M}{pq}$ have the same meaning as in the NEWSTAR equation above (including a similar smearing correction term in $\coh{X}{spq}$). The $\jones{B}{p}$ term is a second diagonal Jones matrix representing the bandpass. Note that MeqTrees itself makes no special distinction between $\jones{G}{}$ and $\jones{B}{}$. Both are generic diagonal complex Jones matrices, with solvable real and imaginary parts. It is only when we specify the solution intervals that these Jones terms acquire their intended meanings:

\begin{itemize}
\item The $\jones{G}{p}$ solution interval is 30 s, all channels (thus, one independent solution every timeslot, per the entire band).
\item The $\jones{B}{p}$ solution interval was initially set to 30 minutes, and one channel (thus, one independent solution every channel, per 60 timeslots). Note that as the bandpass has significant structure, I did not attempt to fit it with any smooth function, but rather allowed each channel to be fitted as an independent parameter, with a timescale of 30 minutes. The latter interval was intended to accommodate slow drift in the bandpass.
\end{itemize}

\begin{figure}
\begin{centering}
\includegraphics[width=.5\columnwidth]{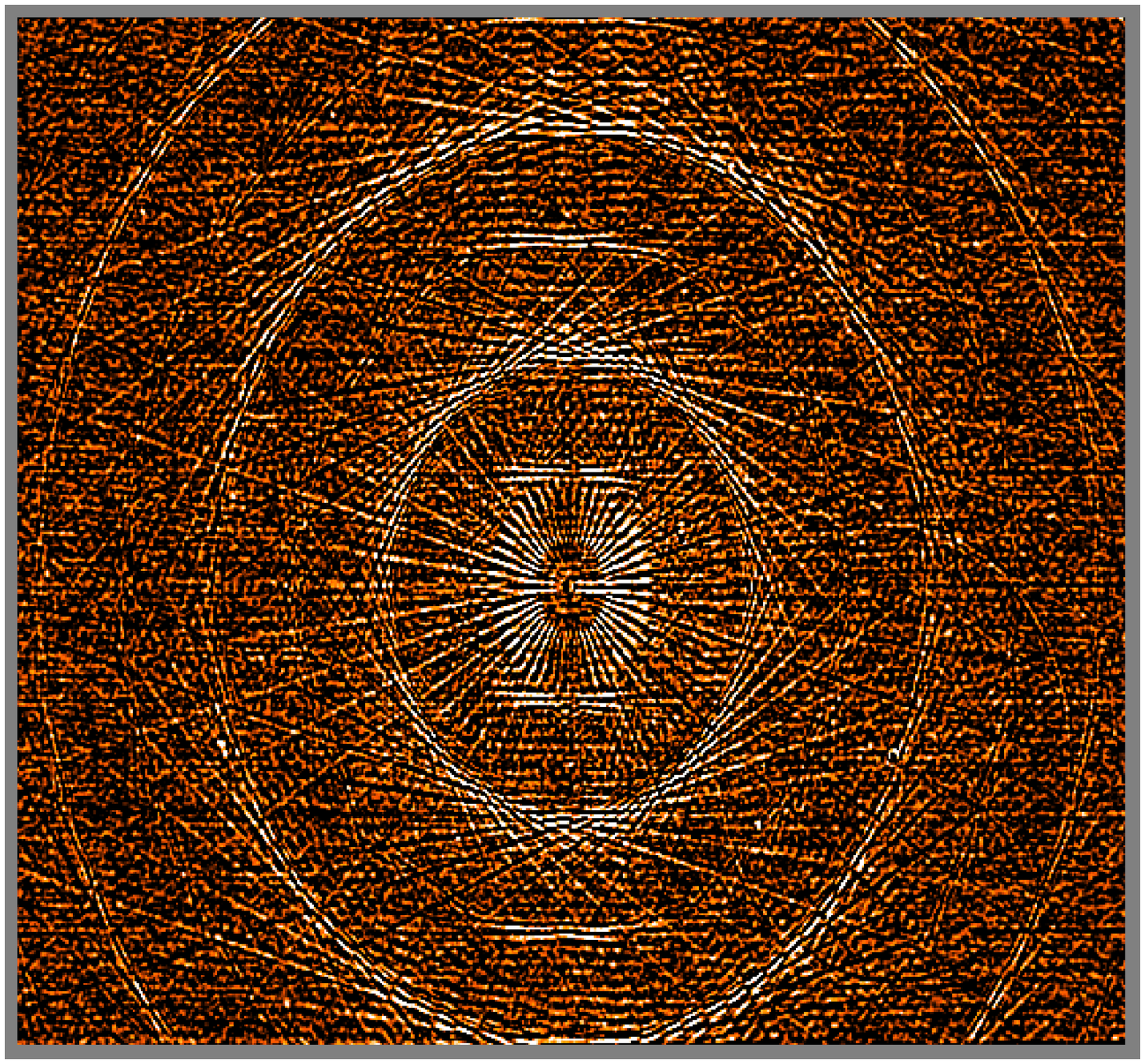}%
\includegraphics[width=.5\columnwidth]{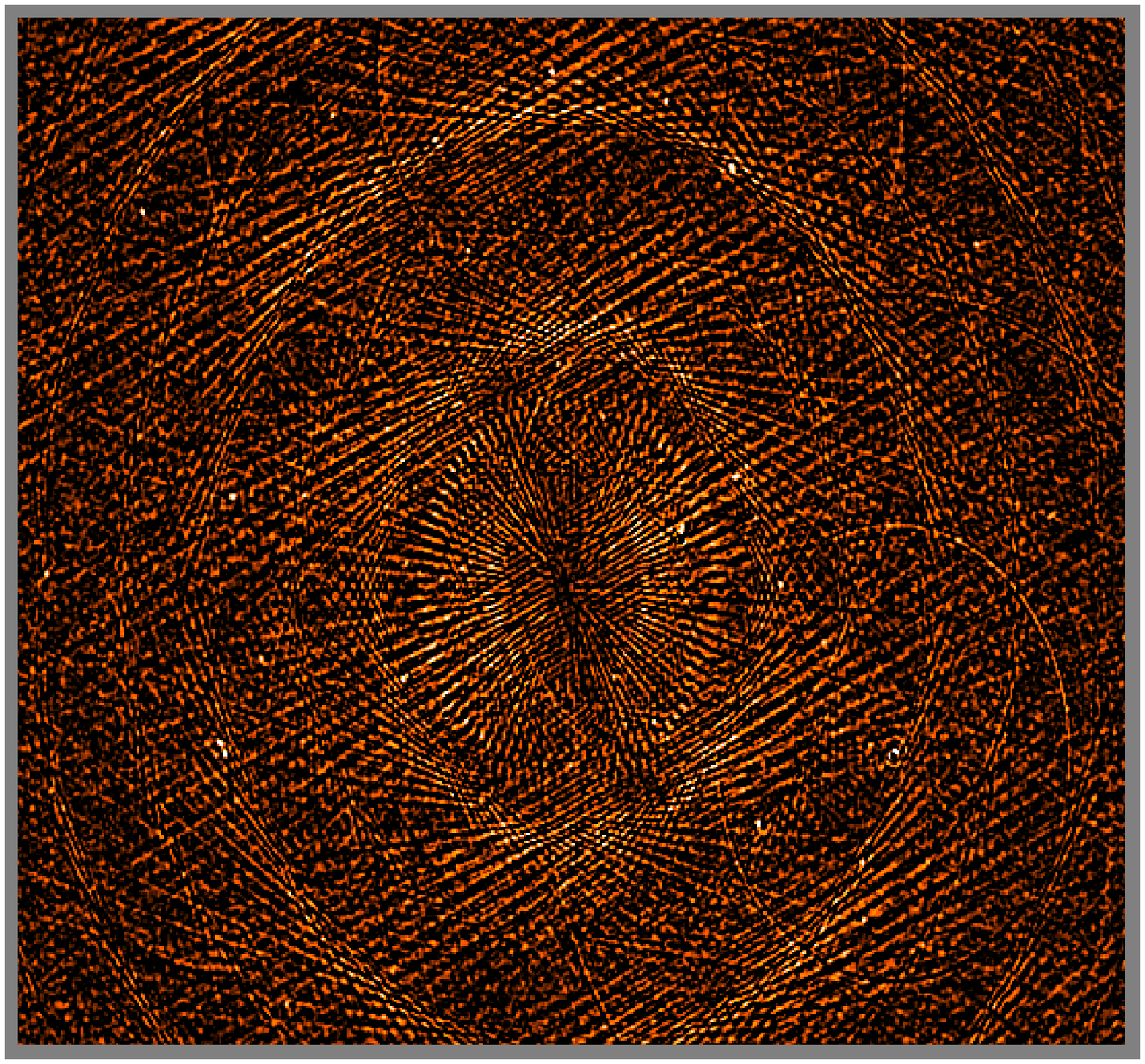}\par
\includegraphics[width=.5\columnwidth]{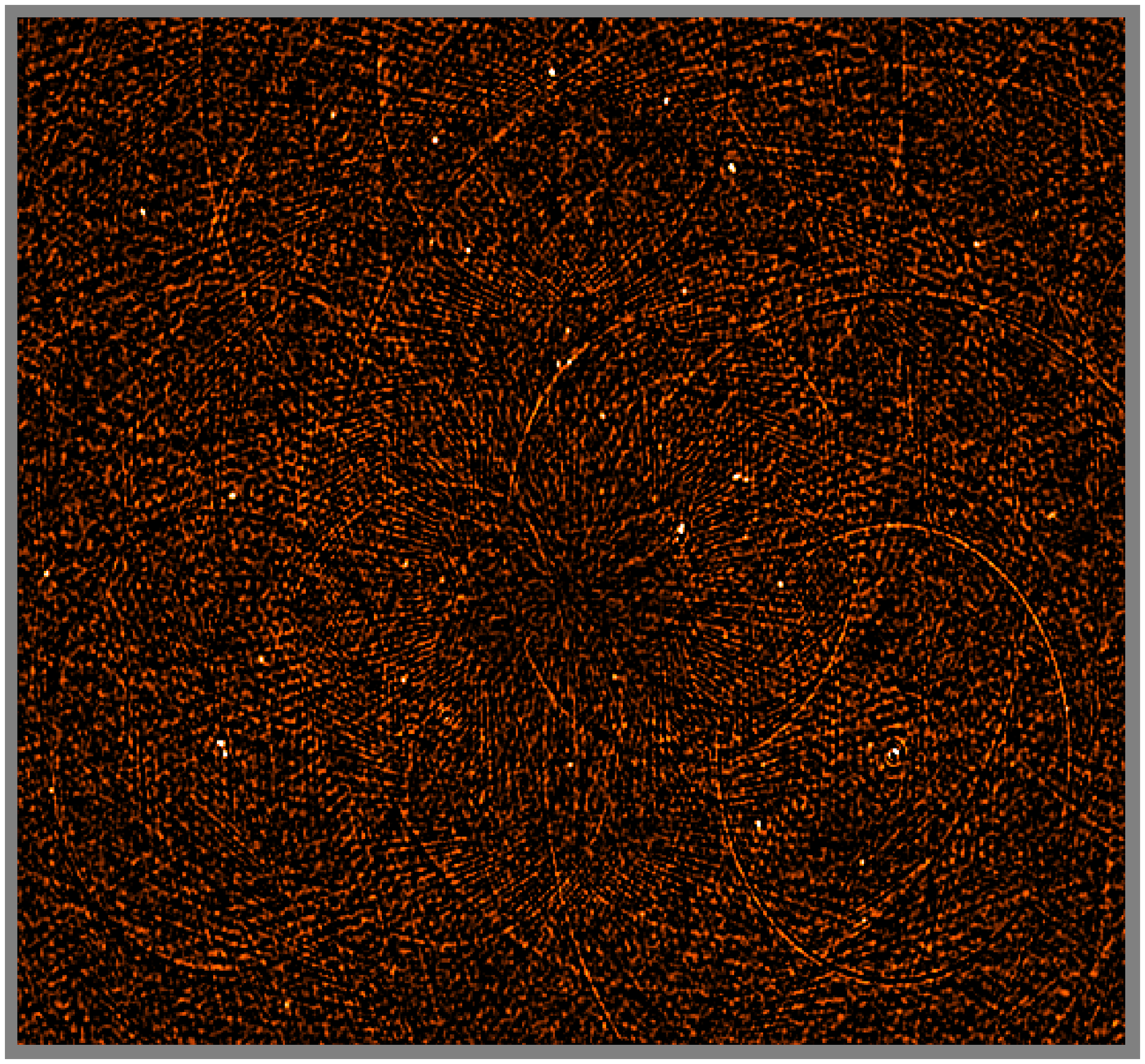}%
\includegraphics[width=.5\columnwidth]{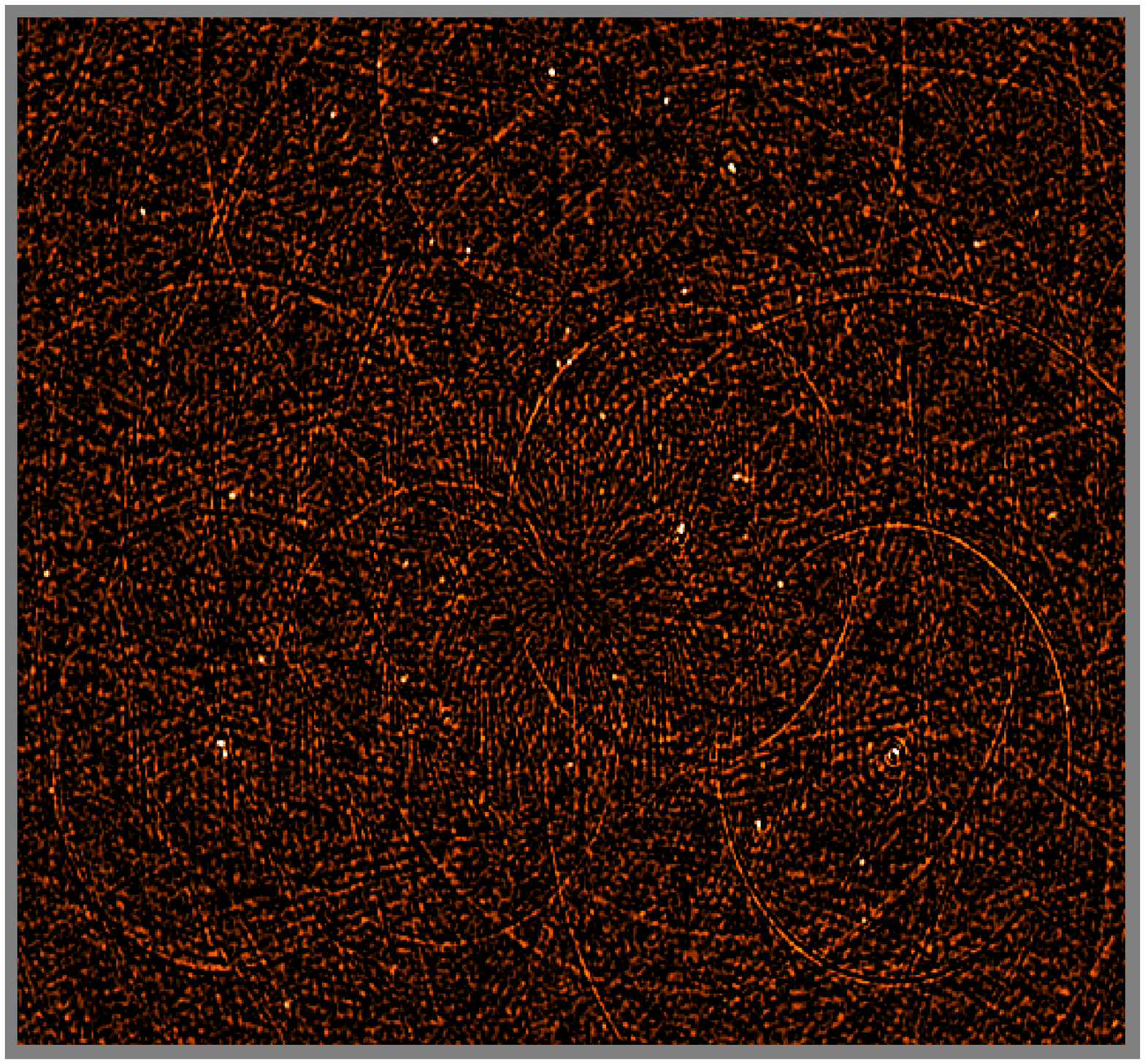}\par
\end{centering}
\caption{\label{fig:Bsol}Single-band residual images produced via bandpass selfcal with different solution intervals for $\jones{B}{p}$: 30 minutes (upper left), 15 minutes (upper right), 7.5 minutes (lower left), and with the 7.5 minute solution smoothed using splines (lower right). Even in the best-case image, the dominant source 3C 147 was not subtracted out perfectly, leaving behind DR-limiting artefacts.}
\end{figure}

The initial result of this calibration was profoundly unsatisfactory (Fig.~\ref{fig:Bsol}, upper left). The residual image was dominated by spoke-like artefacts centred on 3C 147, at about 10,000:1 level (relative the flux of 3C 147 itself). These spokes correspond to edges of the 30-minute solution intervals (being an E-W array, WSRT has a one-dimensional instantaneous PSF). The obvious explanation for the error is short-term bandpass instability. Decreasing the solution interval of $\jones{B}{p}$ to 15 and 7.5 minutes reduced the artefacts to levels of 50,000:1 to 100,000:1, but did not eliminate them entirely. Smoothing the 7.5 minute solution with a spline (along the time axis, per each channel independently) produced a marginal improvement (Fig.~\ref{fig:Bsol}, lower right). Subtraction artefacts are still plainly visible in the map, although at a level not significantly above thermal noise.

At this stage I had to conclude that the WSRT bandpass exhibits some very low-level, but extremely short-term jitter, precluding a separate bandpass selfcal at extreme dynamic ranges. On the other hand, this result also shows that where a single-band dynamic range of no more than 100,000 is expected (as is the case for many other observations), bandpass selfcal provides perfectly adequate results, and can cut down on the number of solvable parameters significantly. In the meantime, for the 3C 147 study I had to revert to the per-channel selfcal approach of de Bruyn.

\subsection{Per-channel selfcal}

The RIME for per-channel selfcal is just Eq.~(\ref{eq:3C 147:bandpass}) without the bandpass term. It is, in fact, very similar to NEWSTAR's implicit equation (\ref{eq:newstar-rime}):\footnote{With the exception of the position of the $\coh{M}{pq}$ term, which is on the inside of Eq.~(\ref{eq:newstar-rime}), and on the outside here. For this particular dataset it makes no difference: since only the $XX$ and $YY$ correlations are used, all matrices in Eq.~(\ref{eq:newstar-rime}) are diagonal, and for diagonal matrices the ``$\ast$'' operator is equivalent to matrix multiplication, and commutes. For the full-polarization case, the two approaches are \emph{not} equivalent.}
 
\begin{equation}\label{eq:3C 147:perchan}
\coh{V}{pq} = \coh{M}{pq} \ast \left ( \jones{G}{p} \left( \sum_s E_s \coh{X}{spq} E^{\herm}_s \right)  \jonesT{G}{q} \right )
\end{equation}

\begin{figure}
\begin{centering}
\includegraphics[width=.5\columnwidth]{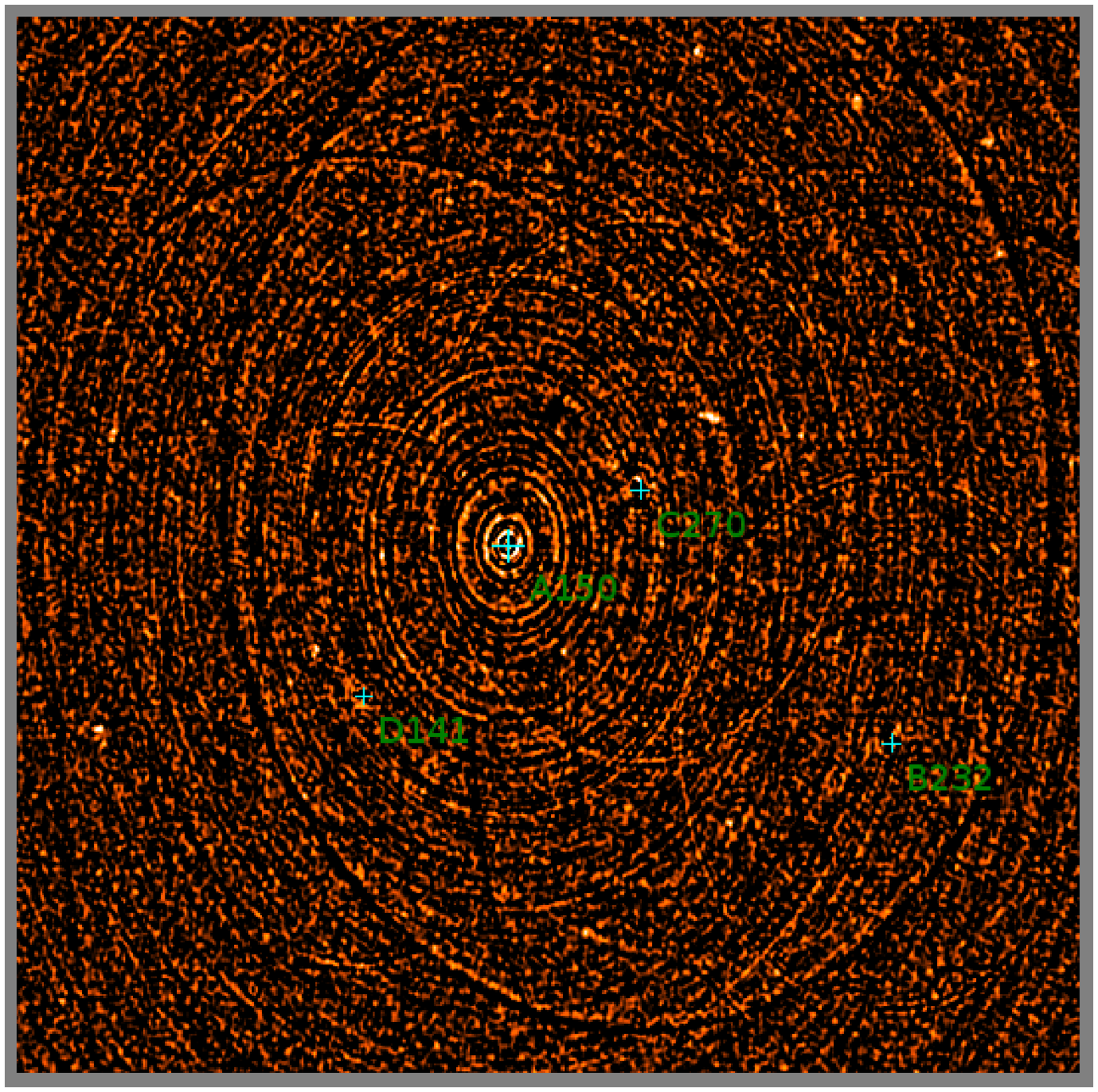}%
\includegraphics[width=.5\columnwidth]{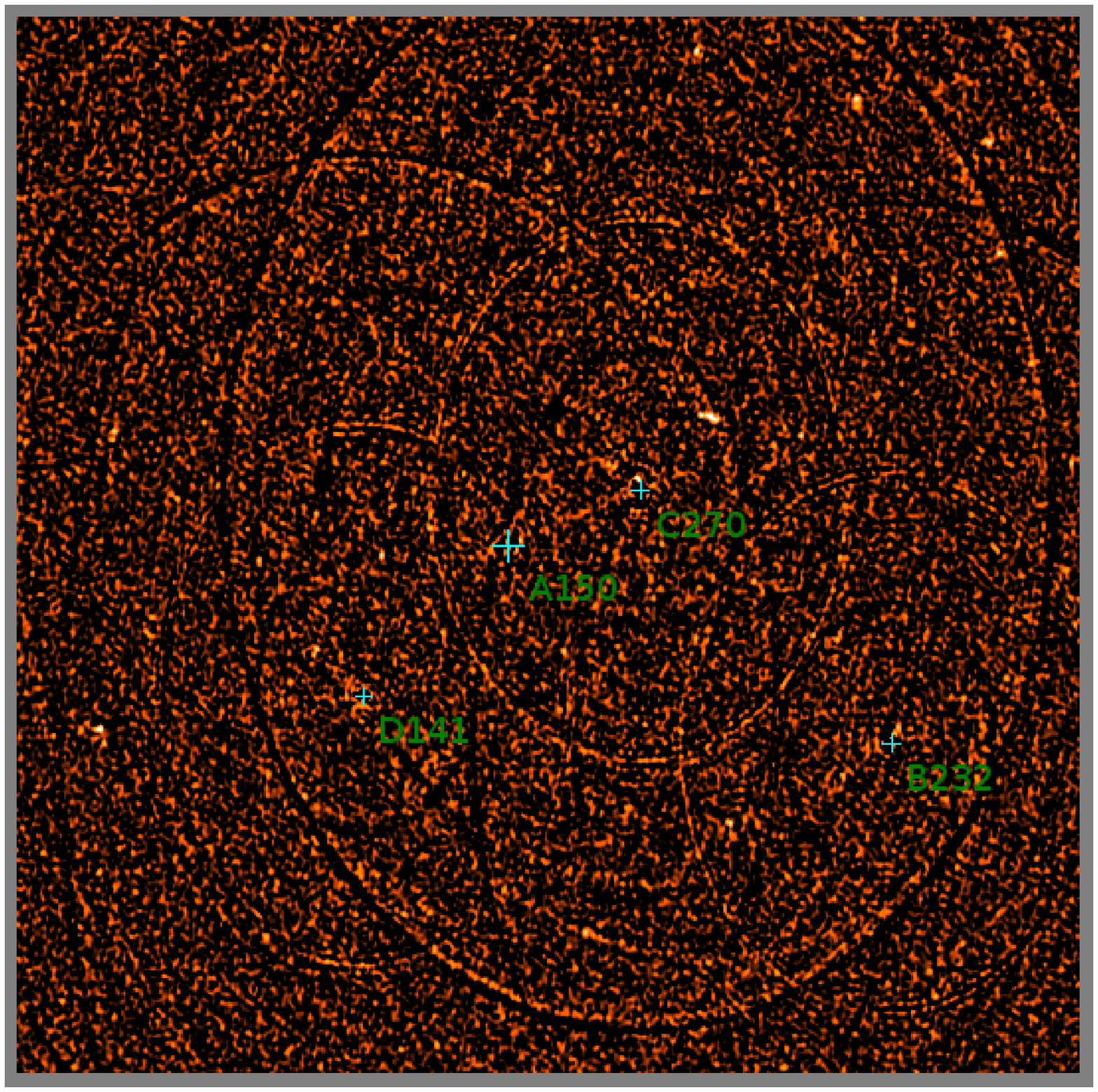}\par
\end{centering}
\caption{\label{fig:Gsol}Single-band residual images produced via per-channel selfcal. The left image is the result of solving for $\jones{G}{p}$. It is dominated by concentric rings centred on 3C 147 (designated as ``A150'' here). These are caused by closure errors, and go away once a solution for interferometer-based errors $\coh{M}{pq}$ is done (right image). The remaining artefacts are associated with off-axis sources B, C and D, and are due to DDEs. 
}
\end{figure}

Per-channel selfcal is achieved by setting the solution interval of $\jones{G}{p}$ to one channel and one timeslot. The resulting single-band residual images are dominated by closure errors at a level of $\sim 100 \mu$Jy (or 1:200,000 relative to 3C 147 itself); these go away after an $\coh{M}{pq}$ solution (Fig.~\ref{fig:Gsol}). These images are qualitatively very similar to those obtained by de Bruyn during his NEWSTAR calibration (which is to be expected, given the similarity of the equations).

The remaining artefacts in Fig.~\ref{fig:Gsol} are associated with the three next-brightest sources\footnote{The source IDs used here follow the ``COPART'' (Clustering, Order, Position Angle, Radius, Type) convention, as implemented by the Tigger sky model management tool (available with MeqTrees). A COPART source ID starts with an alphabetic designator (A, B, ... Z, aa, ab, ...) assigned to sources in order of decreasing brightness. This is already a unique identifier, and is sometimes used by itself for brevity, as in the paragraph above. A full ID also encodes approximate position relative to field centre: two digits for the position angle (in units of $10\degr$), and one digit for radial distance (in units of $10\arcmin$). Optional suffixes indicate source type and clustering. Thus the full ID of source B is B232; being slightly extended, in this particular sky model it is actually represented by a cluster of six delta functions: B232, B232a, ... B232e.} in 
the field: B (42~mJy), C (52~mJy), and D (22~mJy). The furthest of these (B) is about 20$\arcmin$ away from centre. The artefacts themselves are under $50 \mu$Jy (thermal noise in one band being $\sim 30 \mu$Jy), or at a level of 1:1000 relative to the associated sources. This is why \citet{deBruyn:million} talks about an ``off-axis dynamic range'' of a 1000: while 3C 147 itself (22~Jy) is subtracted without a trace (down to the thermal noise), the off-axis sources are only subtracted to a precision of about 1000. Some of the artefact structure is doubtlessly due to slightly under- or overestimated sky model fluxes (this produces regular rings matching the WSRT PSF), which can be taken care of during subsequent deconvolution. Most of it, however, is due to DDEs and does not deconvolve, producing artefacts in the final 8-band images such as the one shown in the left inset of Fig.~\ref{fig:3c147} (the inset is, in fact, a close-up of source B).

\subsection{Interferometer-based errors\label{sec:3C 147:closure-errors}}

It is not clear what causes closure errors at the WSRT. Common sense suggests the analogue part of the signal chain is to blame, but there is no hard evidence either way. What is evident is that high-DR images exhibit concentric ring-like artefacts such as those in the left image of Fig.~\ref{fig:Gsol}, and that these go away once a solution for an interferometer-based  multiplicative error -- the $\coh{M}{pq}$ term of Eq.~(\ref{eq:3C 147:perchan}) -- is applied. A single solution per band, per the entire 12 hours (per correlation and interferometer) is sufficient. The $\coh{M}{pq}$ solutions are usually within $10^{-3}-10^{-4}$ of unity (as is the case here), but can be much higher in some observations, for reasons that remain mysterious. 

\begin{figure}
\begin{centering}
\includegraphics[width=.5\columnwidth]{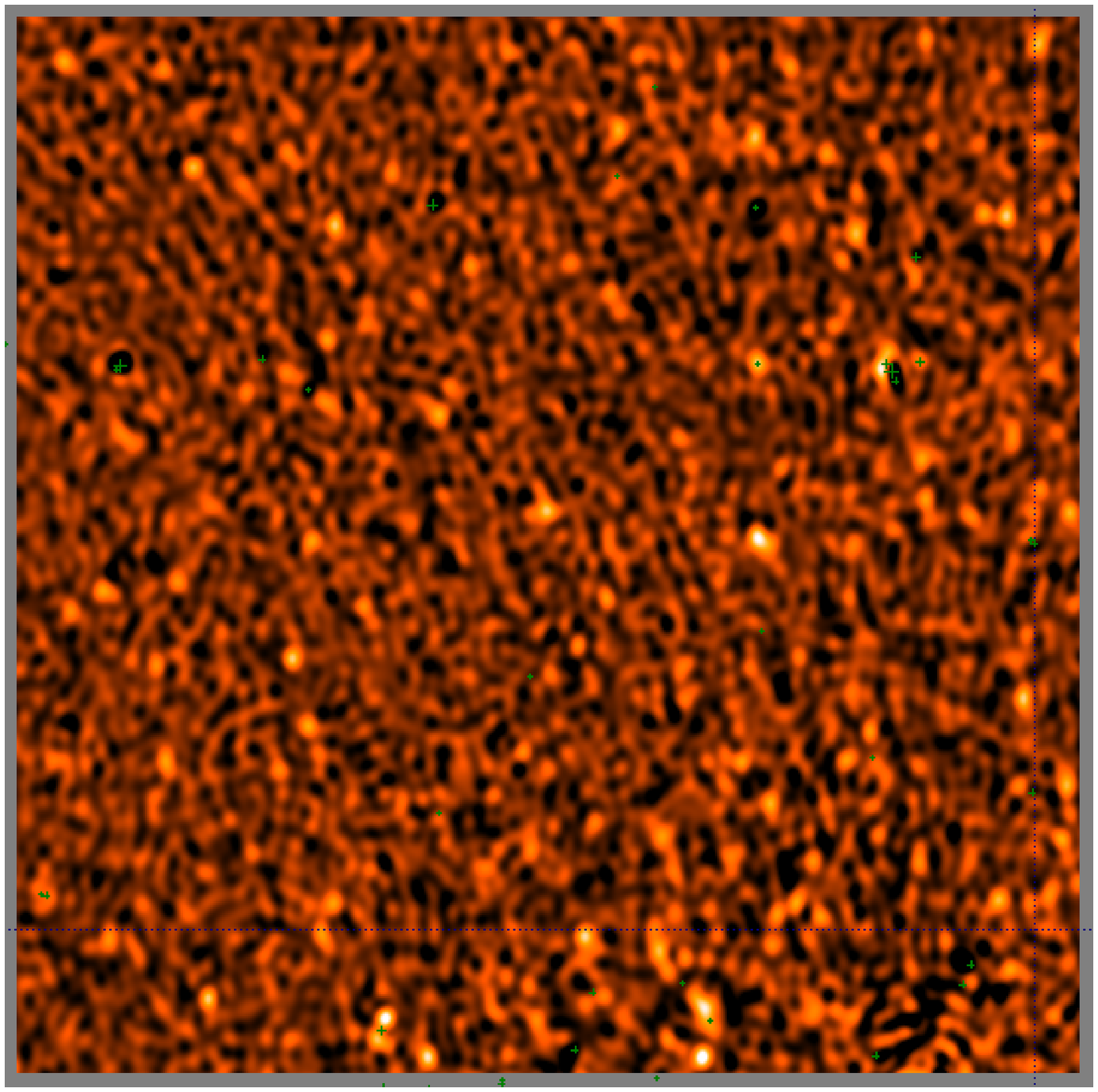}%
\includegraphics[width=.5\columnwidth]{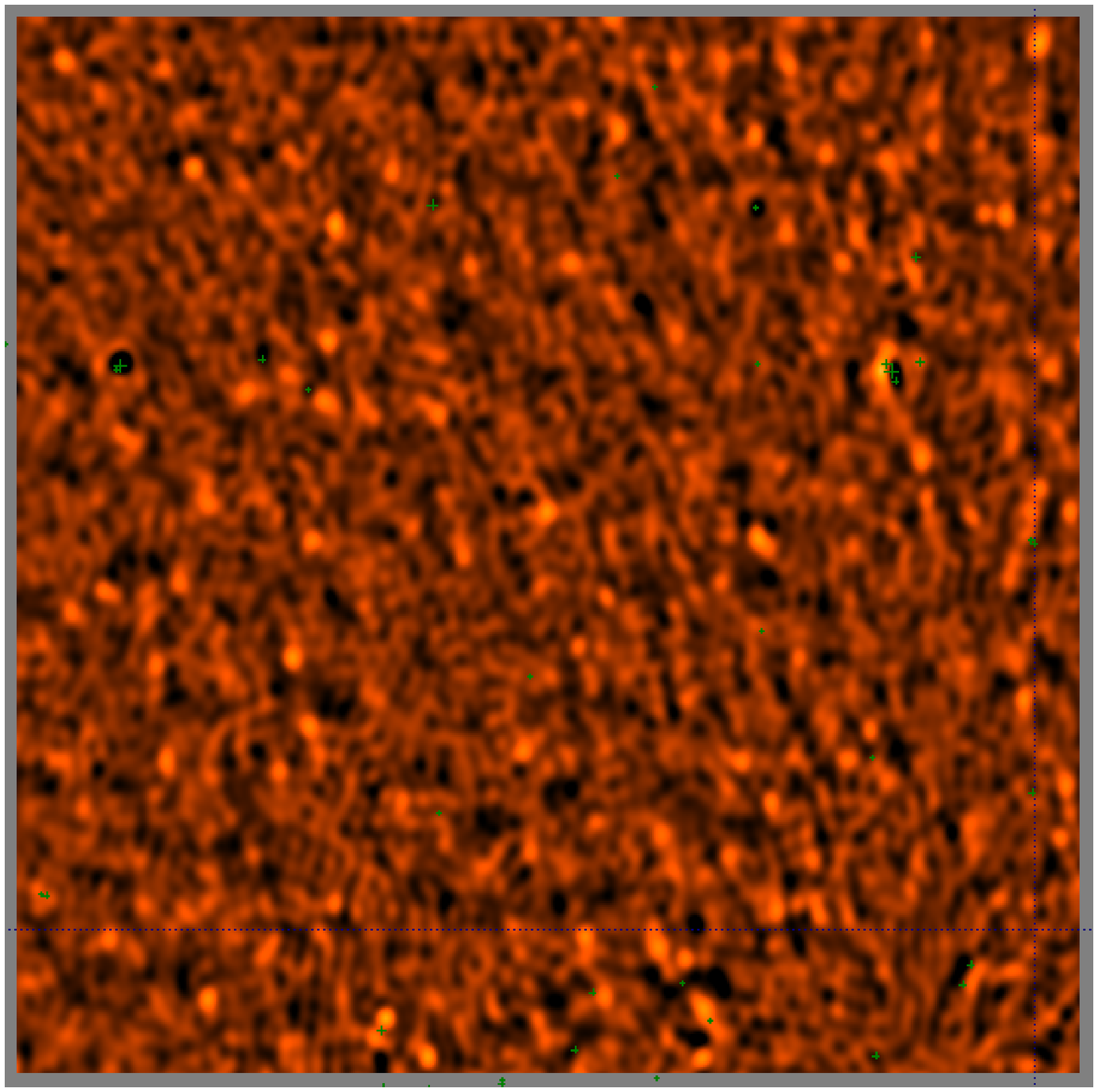}\par
\end{centering}
\caption{\label{fig:source-suppression}Source suppression through interferometer-based error solutions. On the left is a deconvolved 8-band residual image of the centre of the field, with 12 hour solutions for $\coh{M}{pq}$. On the right is the same image with 30 minute solutions. The positions of (subtracted) model sources are indicated by crosses. Suppression of unmodelled sources is evident in the right image.
}
\end{figure}

The latter fact suggests an intrinsic time variability, but solving for $\coh{M}{pq}$ on short time intervals is very dangerous. Any solution for $\coh{M}{pq}$ will also try to compensate for observed flux that is not present in the sky model. Unless the solution interval is sufficiently long, there will be unmodelled sources with a fringe rate slow enough that their vector average visibility over the solution interval will be significantly non-zero. These sources will then tend to be attenuated by the $\coh{M}{pq}$ solutions. The 3C 147 observations provide a perfect example (Fig.~\ref{fig:source-suppression}). On the left is an 8-band residual image with a 12 hour $\coh{M}{pq}$ solution; on the right is one with a 30 minute solution. Model sources are indicated by crosses. Attenuation of unmodelled sources towards phase centre is clearly visible in the right image. 

This implies that closure errors cannot be reliably solved for on observations shorter than 12 hours, unless a ``complete'' (i.e. down to the noise) sky model of the center of the field is available.

\subsection{Pointing selfcal\label{sec:3C 147:pointing}}

Since the main cause of artefacts in WSRT images is commonly considered to be pointing error \citep[see Paper II,][Sect.~2.1.4]{RRIME2}, I decided to implement a form of the pointing selfcal algorithm suggested by \citet{SB:pointing}. This proved to be a straightforward exercise in MeqTrees, since only a small modification of the RIME of Eq.~(\ref{eq:3C 147:perchan}) was required: 

\begin{equation}\label{eq:3C 147:pointing}
\coh{V}{pq} = \coh{M}{pq} \ast \left ( \jones{G}{p} \left( \sum_s E_{sp} \coh{X}{spq} E^{\herm}_{sq} \right) \jonesT{G}{q} \right )
\end{equation}

Instead of a per-source beam gain $E_s=E(l_s,m_s)$ \citep[where $E(l,m)$ is the $\cos^3$ approximation given by Paper II,][Sect.~2.1.1]{RRIME2}, which I had been using in the previous equations, here I used a per-antenna, per-source beam gain $E_{sp}$, defined as follows:

\begin{equation}\label{eq:3C 147:offset-beam}
E_{sp} = E(l_s+\delta l_p,m_s+\delta m_p)
\end{equation}

Per-antenna pointing offsets $\delta l_p,\delta m_p$ were then treated as solvable parameters. 

The results of this proved inconclusive. Even though the solution converged to some physically-sensible offsets (on the order of arcmin), no tangible reduction of artefacts was observed in residual images. This could be due to a number of reasons:

\begin{enumerate}
\item The $\cos^3$ approximation is not good enough -- unlikely, as it has been independently verified at least for the core of the main lobe, which the sources in question sources are well within.
\item With only a few sources sufficiently bright to exhibit DDEs, we simply don't have enough constraints for a pointing solution on this field.
\item The model fluxes/positions for the sources in question are not sufficiently accurate.
\item The dominant DDE affecting this observation is not pointing error. This will be elaborated on further in Sect.~\ref{sec:de-analysis}. 
\item There is something wrong with my implementation of pointing selfcal, especially since the figures in Sect.~\ref{sec:de-analysis} suggest that mispointings \emph{are} detectable.
\end{enumerate}

Trying to get a better grip on the problem, I eventually settled on a ``controlled experiment'':
locating a field containing multiple bright off-axis sources, and observing it with deliberately exaggerated mispointings, to see if these can be more readily recovered from the data. This experiment became known as the ``QMC Project'' (in honour of the long-defunct Quality Monitoring Committee of WSRT), and was successfully carried out. The results of this are still being processed, and will be presented in a follow-up paper. 
In the meantime, I had to look for alternative approaches to DDEs in the 3C 147 field.

\subsection{Differential gains: the ``flyswatter''\label{sec:diffgains}}

In the spirit of ``phenomenological'' equations discussed in Paper II \citep[Sect.~1.3]{RRIME2}, I decided to introduce a {\em differential gain} Jones ($\Delta E$-Jones) into my form of the RIME:

\begin{equation}\label{eq:3C 147:de}
\coh{V}{pq} = \coh{M}{pq} \ast \left ( \jones{G}{p} \left( \sum_s \Delta\jones{E}{sp} E_s \coh{X}{spq} E^{\herm}_s \Delta\jonesT{E}{sq} \right)  \jonesT{G}{q} \right )
\end{equation}

The $\Delta\jones{E}{sp}$ term is meant to subsume {\bf all} DDEs associated with source $s$ and antenna $p$ (with the exception of the nominal primary beam gain, which is already represented by $E_s$). Solving for this term requires some caution, lest too many DoF's be introduced into the equation. I approached this as follows:

\begin{itemize}
\item $\Delta\jones{E}{}$ was fixed at unity for all sources except B, C, and D;
\item For B, C and D, $\Delta\jones{E}{}$ was set to a diagonal matrix with solvable complex elements.
\item The solution intervals for the $\Delta\jones{E}{}$ elements were set to one solution per 30 minutes, per entire band (and per source, antenna, receptor).
\end{itemize}

\begin{figure}
\begin{centering}
\includegraphics[width=.5\columnwidth]{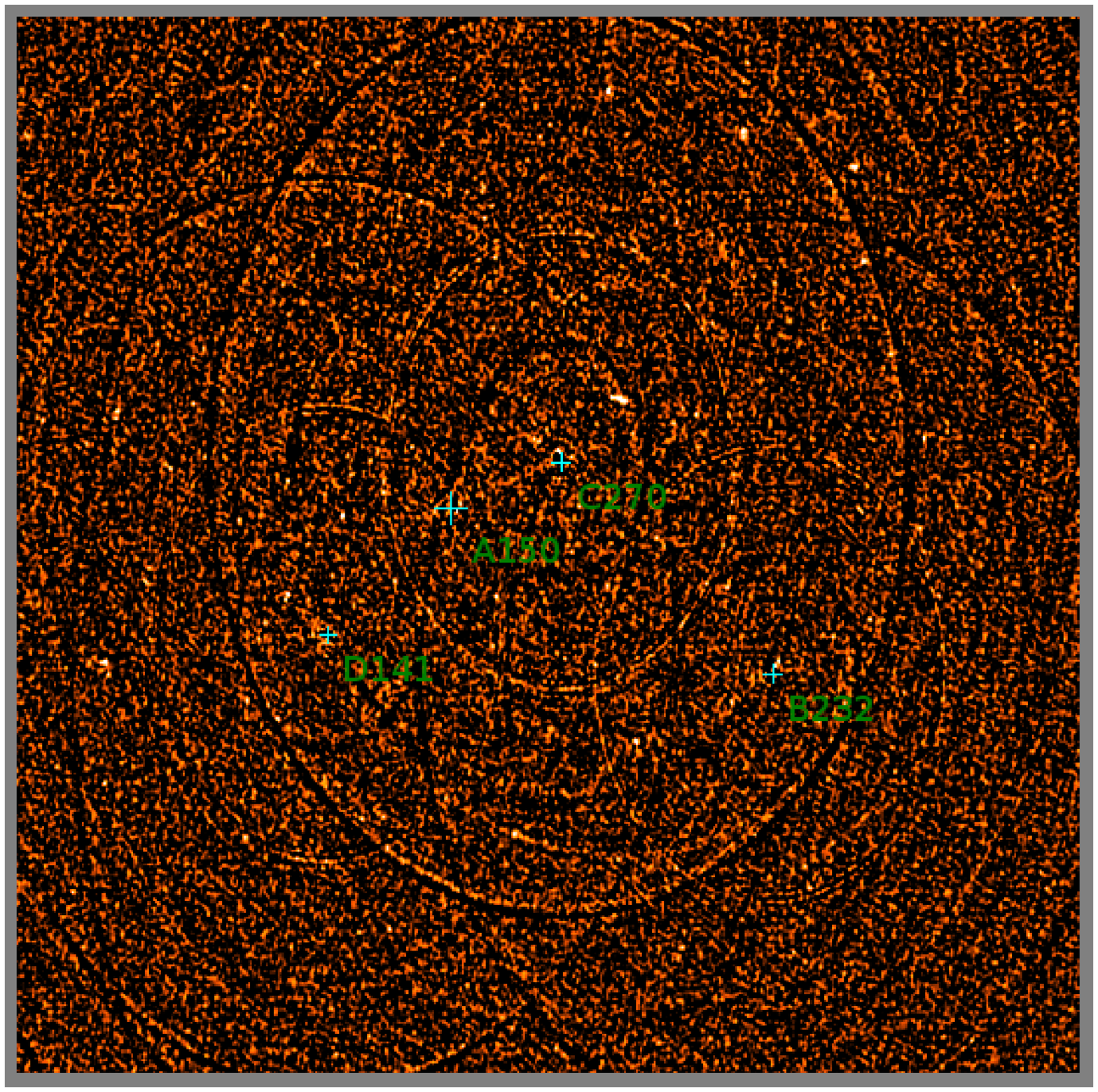}%
\includegraphics[width=.5\columnwidth]{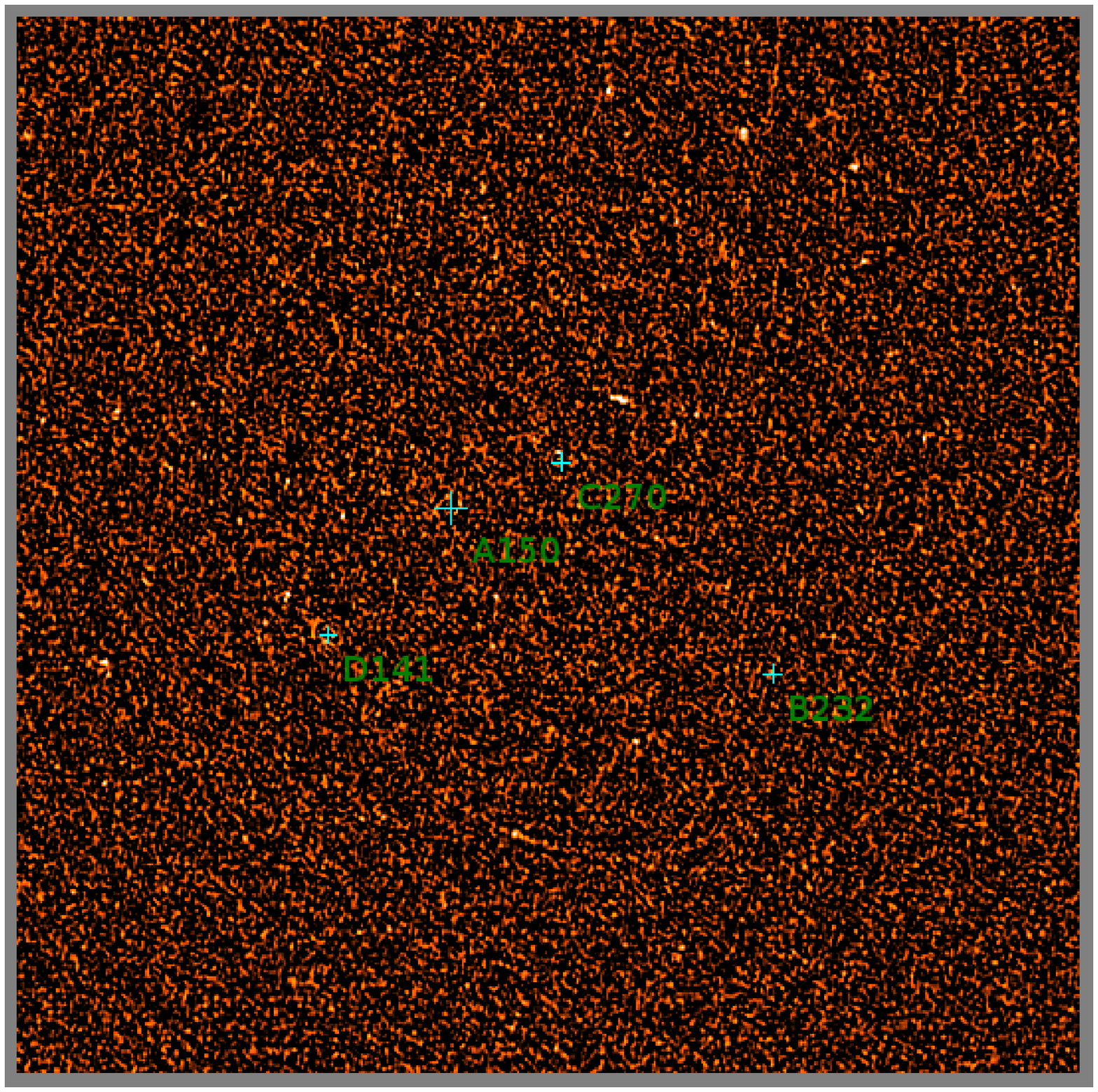}\par
\end{centering}
\caption{\label{fig:dEsol}Results if the flyswatter. On the left is a single-band residual image after  $\jones{G}{p}$ and $\coh{M}{pq}$ solutions only. On the right is the same image with differential gain solutions for sources B, C, and D.}
\end{figure}

Figure~\ref{fig:dEsol} shows the effect of $\Delta\jones{E}{}$ solutions. All artefacts associated with sources B, C and D disappeared completely, to the point where artefacts around fainter sources became just about visible. These were later eliminated once $\Delta\jones{E}{}$ solutions were enabled for those sources. In fact, any source to which a solvable $\Delta\jones{E}{}$ term was assigned promptly vanished from the residual maps, hence my name for the differential gains approach: the ``flyswatter'' algorithm. 

Once 8-band residual images were created, the increased sensitivity made it apparent that four more sources were exhibiting a small amount of DDEs. A more in-depth look at the sky model also showed that all seven sources were in fact slightly extended; the NEWSTAR model represented each by a tight cluster of point sources. Since all ``sources'' in such a cluster are subject to the same DDE, it seemed sensible to make sure the same $\Delta\jones{E}{sp}$ term was applied to all of them. This was easily done using Tigger, a sky model manager/viewer tool included with MeqTrees. Tigger automatically detects source clusters and assigns source identifiers appropriately (e.g. B232, B232a, ...  B232e) It was then a simple matter to tell the Calico framework to use the \emph{same} $\Delta\jones{E}{s_0 p}$ term for all sources $s$ associated with cluster $s_0$. 

\begin{figure}
\begin{minipage}[c]{.45\columnwidth}
\includegraphics[width=\textwidth]{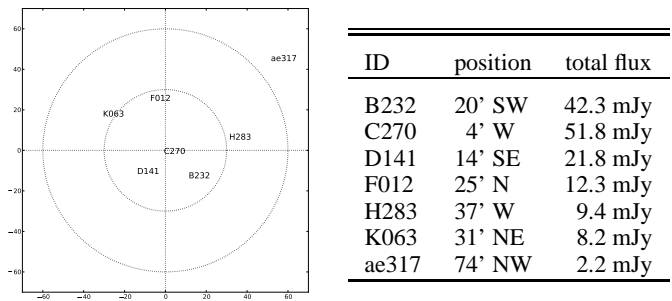}
\end{minipage}\hfill\begin{minipage}[c]{.5\columnwidth}
\begin{tabular}{lr@{ }lr@{.}l}
\hline
\hline
& & \\ [-1ex]
ID & \multicolumn{2}{c}{position} & \multicolumn{2}{c}{total flux} \\
\hline
& & \\ [-1ex]
B232  & 20' & SW & 42 & 3 mJy\\
C270  &  4' & W  & 51 & 8 mJy\\
D141  & 14' & SE & 21 & 8 mJy\\
F012  & 25' & N  & 12 & 3 mJy \\
H283  & 37' & W  &  9 & 4 mJy \\
K063  & 31' & NE &  8 & 2 mJy \\
ae317 & 74' & NW &  2 & 2 mJy \\
\hline
\end{tabular}
\end{minipage}

\caption{\label{fig:source-plot}Positions (relative to nominal pointing centre) and aggregate fluxes (apparent) of the seven off-axis source clusters for which $\Delta\jones{E}{}$ solutions were obtained. Circles are at a radius of $30\arcmin$ and $1\degr$. For reference, the FWHM of the WSRT voltage beam is $\sim50\arcmin$ at 1.4~GHz.}
\end{figure}

The seven source clusters for which differential gain solutions were eventually obtained are summarized in Fig.~\ref{fig:source-plot}. Two of them are somewhat noteworthy. Source {\bf C270} is very close to centre, and therefore shouldn't be affected by DDEs as much as the other sources. It is, however, a complicated and highly polarized source, so perhaps the artefacts it exhibits after regular selfcal are primarily due to sky model inaccuracies, which the $\Delta\jones{E}{}$ solutions absorb (see discussion in Sect.~\ref{sec:de-analysis-model}). Source {\bf ae317} is almost the opposite: it is very faint, but far enough off-axis to be in a sidelobe of the primary beam, and so subject to especially severe DDEs.

\subsection{The showcase result} 

The ultimate result of my calibration of the 2003 observation is shown in Fig.~\ref{fig:3c147}. This image is a true showcase for the differential gains approach. The precise steps leading to this image were as follows:

\begin{enumerate}

\item Each of the 8 bands was independently calibrated using per-channel selfcal, interferometer-based errors, and $\Delta\jones{E}{}$ solutions on seven source clusters, as described above. Corrected residuals were generated.

\item The residuals for all 8 bands were imaged together (in MFS mode) to produce a single residual image. This revealed a large number of fainter sources not visible in the per-band maps.

\item The 8-band image was deconvolved using Cotton-Schwab CLEAN \citep{Schwab:csclean}.

\item The sky model was added back into the deconvolved image, using a Gaussian restoring beam.

\end{enumerate}

The resulting image is completely artefact-free. Presumably, all other sources in the field are too faint to exhibit any DDE-related artefacts. With a dynamic range of 1,600,000:1, this image is the deepest and cleanest single-synthesis radio map in the world to date. 

\subsection{Flyswatter limitations\label{sec:dE-limitations}}

While it can help produce spectacular images, the flyswatter has some serious caveats and drawbacks that need to be explored. First of all, it is a brute-force approach, in the sense that it squashes all effects into a single $\Delta\jones{E}{}$ term. This includes inaccuracies in the sky model! Indeed, any missing source flux or error in source position can be accommodated with a suitable $\Delta\jones{E}{}$. Even unmodelled source structure can ``leak'' into differential gain solutions (Sect.~\ref{sec:de-analysis-model}). Thus, differential gains are good for subtracting sources, but at the cost of mashing up information on the source per se. (One does not use a flyswatter to probe a fly's anatomy!)

Secondly, solvable differential gains can lead to a proliferation of DoF's. Per-channel selfcal has $N_\mathrm{ant}$ unknowns per $\frac{N_\mathrm{ant}(N_\mathrm{ant}-1)}{2}$ measurements, or $\frac{2}{N_\mathrm{ant}-1}$ unknowns per measurement; differential gains add $\frac{2N_\mathrm{src}}{(N_\mathrm{ant}-1)N_\mathrm{time}N_\mathrm{freq}}$ unknowns per measurement, where $N_\mathrm{time}$ and $N_\mathrm{freq}$ are the sizes of the solution interval for $\Delta\jones{E}{}$. This ratio remains favourable for small $N_\mathrm{src}$ and large $N_\mathrm{time}$ and/or $N_\mathrm{freq}$ (as is the case for my 3C 147 reduction), but one must be careful.

The third caveat is processing cost. While usually not as expensive in terms of I/O or CPU as peeling (which, in addition to the solutions themselves, requires repeated subtraction and phase shifting steps), the flyswatter is not free. Every source with a differential gain solution adds $4N_\mathrm{ant}$ unknowns (assuming a diagonal complex $\Delta\jones{E}{sp}$ term, hence 4 real values per matrix) to the equations. As the number of unknowns ($N_\mathrm{unk}=4N_\mathrm{ant}N_\mathrm{src}$) grows, inversion of the normal matrix within the least-squares solver becomes a CPU bottleneck, since it scales as $O(N_\mathrm{unk}^3)$. This makes it impractical to solve for $\Delta\jones{E}{}$'s for more than a handful\footnote{The precise meaning of a ``handful'' here depends on additional factors such as $N_\mathrm{ant}$, size of solution intervals, etc. In effect, these factors influence the constant of the overall cubic scaling law.} of sources at a time. 

One way to mitigate the solver bottleneck is to decompose the $\Delta\jones{E}{sp}$'s into nearly-orthogonal sets of unknowns. For example, we can treat the set of $\Delta\jones{E}{sp}$'s associated with one source $s$ as independent from all other sources. The $(4N_\mathrm{ant}N_\mathrm{src})^2$ normal matrix inside the solver then becomes block-diagonal, composed of $N_\mathrm{src}$ blocks of size $(4N_\mathrm{ant})^2$. Inversion of this matrix then scales as $O(N_\mathrm{src})O(N_\mathrm{ant}^3)$. This scheme was tested in MeqTrees, and it was found that the trade-off is slower convergence, requiring more iterations. For large numbers of sources, however, this becomes very favourable.

\section{Analysis of differential gain solutions\label{sec:de-analysis}}

It is time to see whether any useful information can be gleaned from the differential gains solutions themselves. As a result of the reduction, I had obtained: per each source direction (7 of these: see Fig.~\ref{fig:source-plot}), per each antenna (14), per each band (8), per 30-minute interval (24 of these in a 12-hour synthesis), two complex numbers representing the apparent differential gain of the X and Y dipole (``differential'' being relative to the gain in the direction of 3C 147 -- almost at the centre of the field -- which had been taken care of by regular selfcal).

I then adopted the following approach. Given the relatively low fractional bandwidth, I didn't expect much variation with frequency in $\Delta\jones{E}{}$. I therefore treated each set of 8 per-band solutions as independent samples of the same variable. The mean of the 8 samples was used as an estimator of variable, and the standard deviation of the 8 samples as an estimator of the error (i.e. the error bar).

Since it quickly became apparent that the $\Delta\jones{E}{}$ solutions were exhibiting some very interesting behaviour, I applied exactly the same procedure to the 2006 observations, so that comparisons could be made. The plots below show the results from both observations. 

\begin{figure*}
\sidecaption
\parbox[b]{12cm}{
\includegraphics[width=12cm]{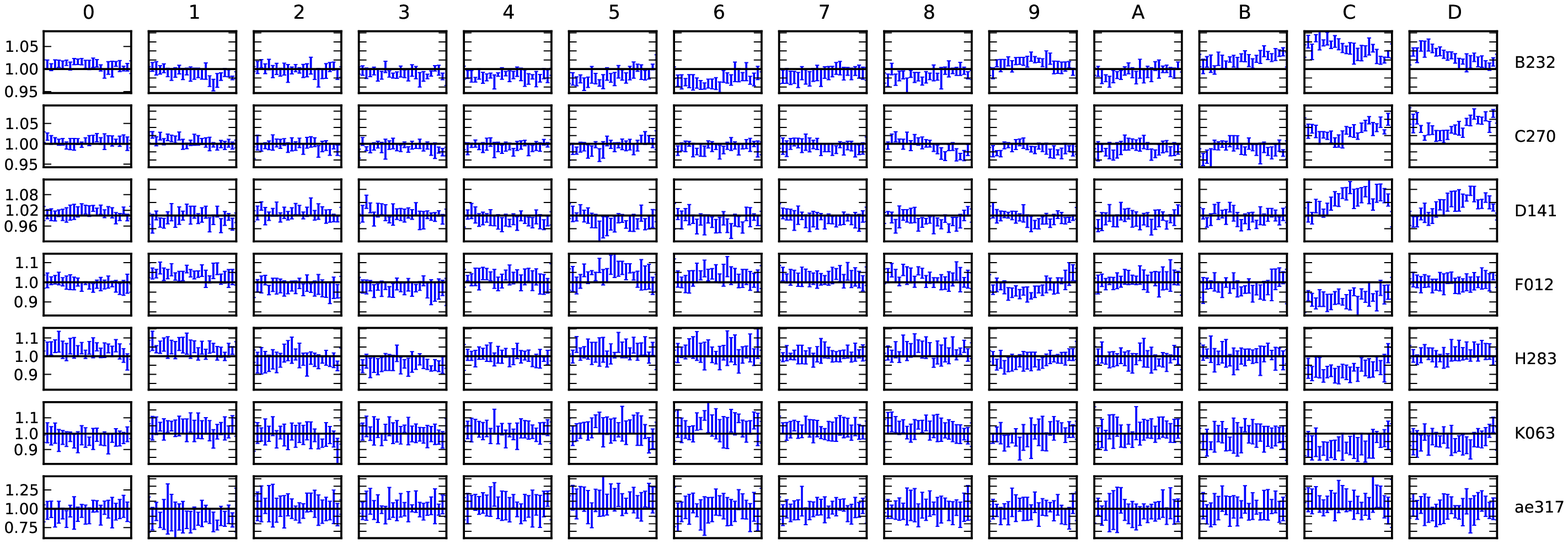} \\
\includegraphics[width=12cm]{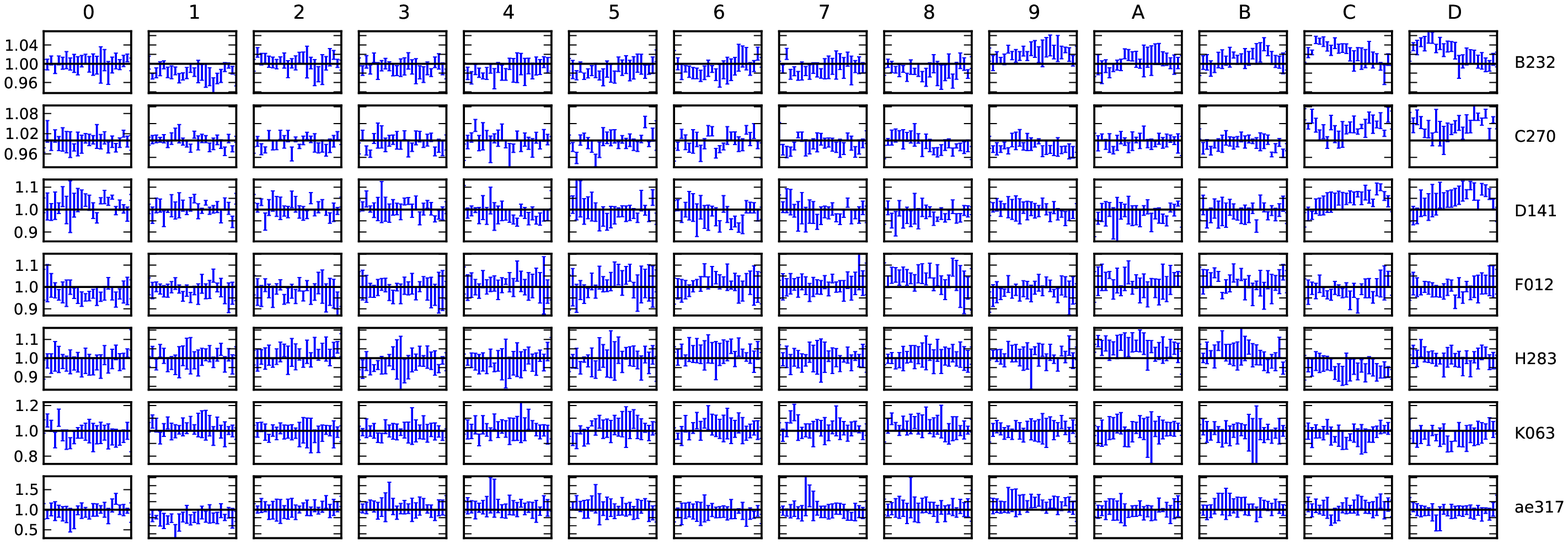}
}
\caption{\label{fig:dEampl}Differential gain-amplitudes ($||\Delta\jones{E}{}||$) as a function of time for the 2003 (top) and 2006 (bottom) observations. Rows correspond to sources, columns to antennas. The vertical plot scale is fixed within each row, but differs from row to row. Horizontal lines indicate the $||\Delta\jones{E}{}||=1$ level.}
\end{figure*}

\begin{figure*}
\sidecaption
\parbox[b]{12cm}{
\includegraphics[width=12cm]{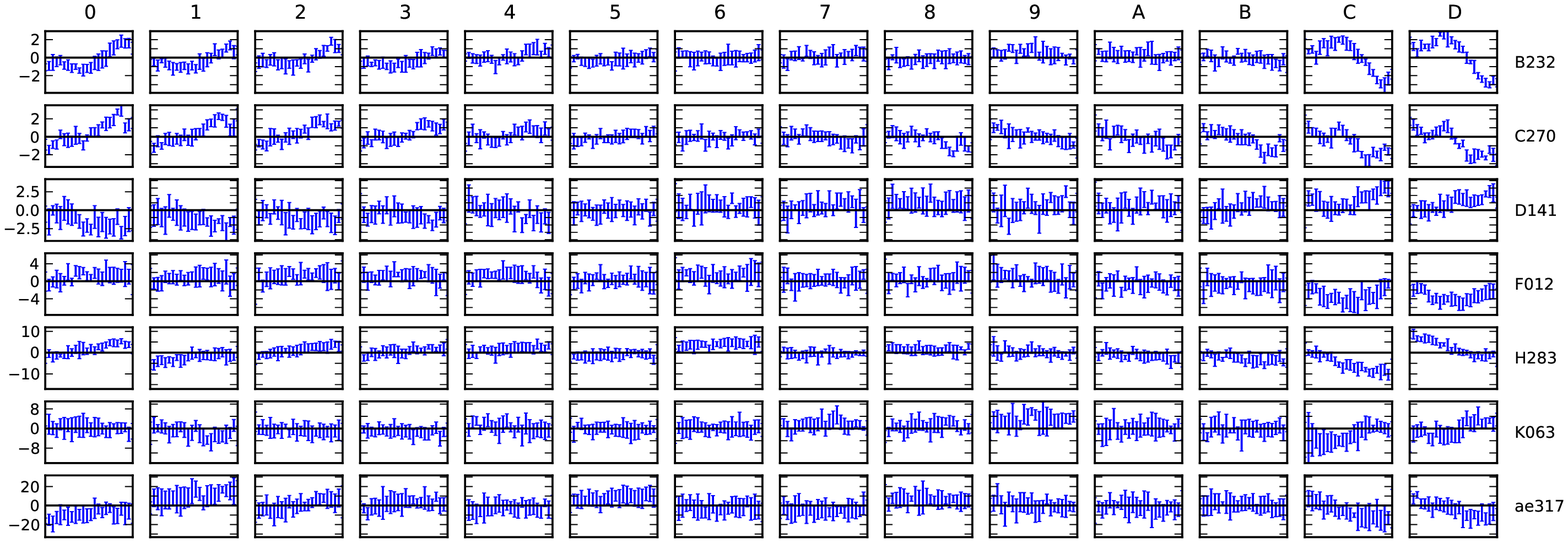}
\includegraphics[width=12cm]{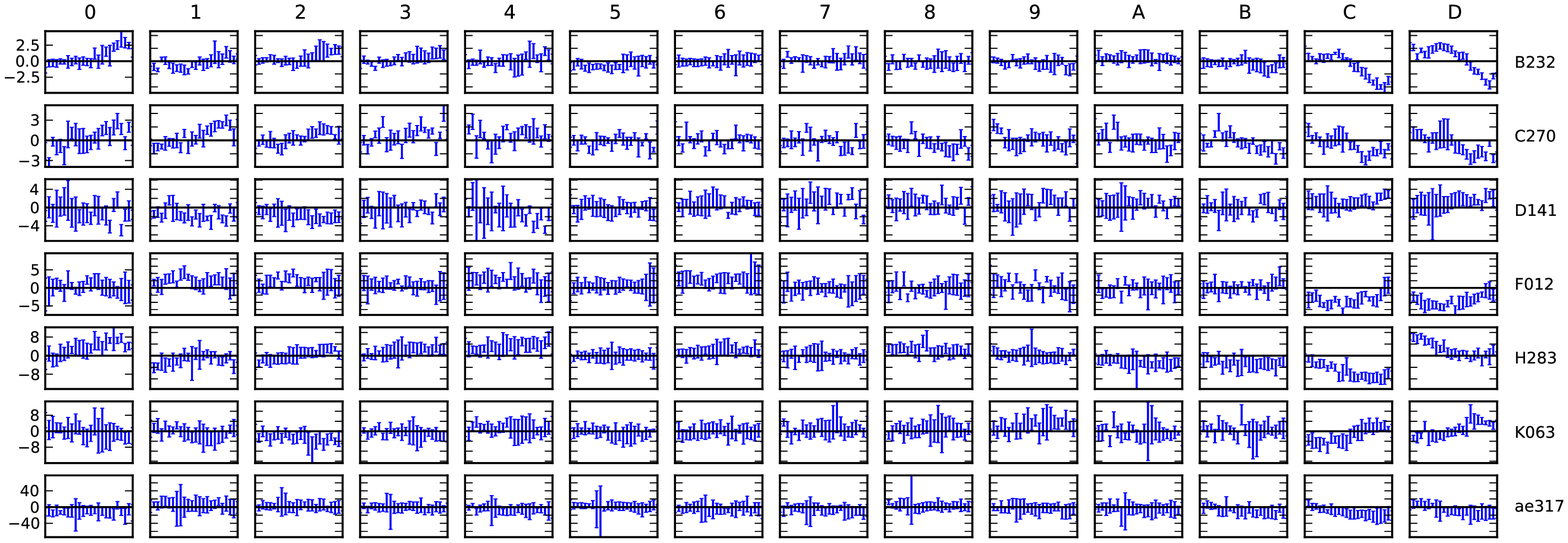}
}
\caption{\label{fig:dEphase}Differential gain-phases ($\arg\Delta\jones{E}{}$, in degrees) as a function of time for the 2003 (top) and 2006 (bottom) observations. Rows correspond to sources, columns to antennas. The vertical plot scale is fixed within each row, but differs from row to row. Horizontal lines indicate the $\arg\Delta\jones{E}{}=0$ level.}
\end{figure*}

Figure~\ref{fig:dEampl} is a summary of the differential gain-amplitudes per source, per antenna. The precise quantity plotted here was computed as follows. First, I computed the norm of the $\Delta\jones{E}{}$ matrix (diagonal by construction) as 

\[
\left|\left|\matrixtt{a}{0}{0}{b}\right|\right| \equiv \sqrt{|a|^2+|b|^2},
\]

I then normalized (divided) this value by the mean value per source (that is, the mean across all time intervals, bands, and antennas), which was meant to take out the effect of incorrect model fluxes (see below). The resulting ``normalized norm'' was then plotted as a function of time, per source, per antenna.

Figure~\ref{fig:dEphase} is a similar plot of the differential gain-phases, computed as

\[
\arg\matrixtt{a}{0}{0}{b} \equiv \frac{\arg a + \arg b}{2}.
\]

The most striking feature of Figs.~\ref{fig:dEampl} and \ref{fig:dEphase} is the high SNR. They show a high degree of temporal continuity in the solutions, and statistically significant structure. This strongly suggests that the solutions represent real physical or numerical effects. As to the nature of these effects, I still do not have satisfactory answers, though it is hoped that the ``QMC Project'' mentioned earlier will shed some more light on the issues. The rest of this section discusses some of the more prominent questions, and proposes some rather speculative explanations.
 
\subsection{Absorbing errors in the sky model\label{sec:de-analysis-model}}

As already mentioned, a major caveat of the flyswatter approach is that $\Delta\jones{E}{}$ solutions will tend to absorb inaccuracies in the source model. By analogy (and for exactly the same reasons), classic selfcal 
alone cannot solve for absolute positions or fluxes. Indeed, if the true position of a source $\vec l = (l,m)$ is offset from the model position $\vec l^\mathrm{(mod)} = \vec l + \delta\vec l = (l+\delta l,m + \delta m)$, while the true brightness $\coh{B}{}$ differs from the model brightness by a multiplicative matrix factor: $\coh{B}{}^\mathrm{(mod)} = \jones{A}{} \coh{B}{} \jonesT{A}{}$ (the latter being a straightforward generalization of a scalar factor $a^2$), then the coherency term of the RIME for the model source may be written out as

\begin{eqnarray*}
\coh{X}{pq}^\mathrm{(mod)} & = & K_{p}(\vec l + \delta\vec l) \coh{B}{}^\mathrm{(mod)} K^\herm_{q}(\vec l + \delta\vec l) \\
 & = & K_{p}(\delta\vec l) K_{p}(\vec l) \jones{A}{}\coh{B}{}\jonesT{A}{} K^\herm_{q}(\vec l) K^\herm_{q}(\delta\vec l) \\
 & = & [ K_{p}(\delta\vec l)\jones{A}{}]\cdot[K_{p}(\vec l)\coh{B}{}K^\herm_{q}(\vec l)]\cdot[K_{q}(\delta \vec l)\jones{A}{}]^\herm \\
 & = & [K_{p}(\delta\vec l)\jones{A}{}] 
       \cdot \coh{X}{pq} \cdot 
       [K_{q}(\delta\vec l)\jones{A}{}]^\herm.
\end{eqnarray*}

If a solvable differential gain is then assigned to the source, the model can be made to fit the data by absorbing the $K_{p}(\delta\vec l)\jones{A}{}$ factor into the $\Delta\jones{E}{ps}$ solutions.

Even more insidiously, differential gains can absorb some source structure. Consider a source that is slightly extended in one direction, enough to be resolved on the longest baselines. An E-W array like the WSRT has a one-dimensional instantaneous {\em fan beam}. It will ``see'' the source as a point source when the fan beam is aligned with the source orientation, and start resolving it when the fan beam becomes perpendicular to the source. In other words, the apparent flux of the source will remain constant in time on short baselines, and vary in time on the long baselines as the source resolves. If such a source is represented by a point source in the sky model, the model flux will be constant on all baselines. Now, if some antennas are predominantly involved in long baselines (RTC and RTD, in the case of WSRT), $\Delta\jones{E}{}$ solutions can compensate for some of the flux discrepancy by changing the gain-amplitudes of these antennas. I would expect to see a variation of $||\Delta\jones{E}{}||$ with a 12-hour period. Since most of the baselines to RTC and RTD are mutually redundant (0-C equals 1-D, etc.), their variation in $||\Delta\jones{E}{}||$ should be very similar. 

This is exactly what we're seeing in Fig.~\ref{fig:dEampl}! The plots very strongly suggest that the top three sources (B, C and D) are indeed slightly extended (more so than in the model, that is). If this is the case, then the dominant contribution to $||\Delta\jones{E}{}||$ on antennas RTC and RTD is due to source structure rather than any actual DDE. 

\subsection{Amplitude behaviour}

If the behaviour of differential amplitude on RTC and RTD is due to source structure, this still leaves effects on the other antennas unexplained. First let us consider what a pointing error would look like. With the exception of source ae317 (for which the solutions are much too noisy anyway), the other six sources are well within the main lobe of the primary beam, where the beam gain can be expected to decrease smoothly with distance from pointing centre. We should therefore expect to see $||\Delta\jones{E}{sp}||>1$ if antenna $p$ mispoints \emph{towards} source $s$ (the source appears brighter on antenna $p$), and $||\Delta\jones{E}{sp}||<1$ if it mispoints \emph{away}. WSRT dishes are equatorially mounted, so mispointings due to mechanical or electronic errors in either axis drive would be stationary with respect to the sky, and thus cause constant 
$||\Delta\jones{E}{}||$ offsets. Mispointings due to wind pressure, thermal or gravitational deformation, on the other hand, would be intrinsically time-variable (but would perhaps correlate between adjacent antennas).

Quite a few plots in Fig.~\ref{fig:dEampl} do show (mostly) static offsets. It can be illuminating to present $||\Delta\jones{E}{}||$ in a format I call the ``rogues gallery'' (Figs.~\ref{fig:rogues-2003} and \ref{fig:rogues-2006}). This shows, for each of the 14 antennas, a 12-hour average $||\Delta\jones{E}{}||$ per source, using circles of varying size placed at the position of the source. The magnitude of $(||\Delta\jones{E}{}||-1)$ is indicated by circle size, and the sign by colour. A static mispointing in, e.g., a Northern direction would show up as blue circles in the top half of the plot (i.e. sources appearing brighter), and red circles in the bottom half.

\newlength{\roguewidth}
\setlength{\roguewidth}{.2\columnwidth} 


\begin{figure}
\centering
\begin{tabular}{@{}c@{}c@{}c@{}c@{}c@{}}
\includegraphics[width=\roguewidth]{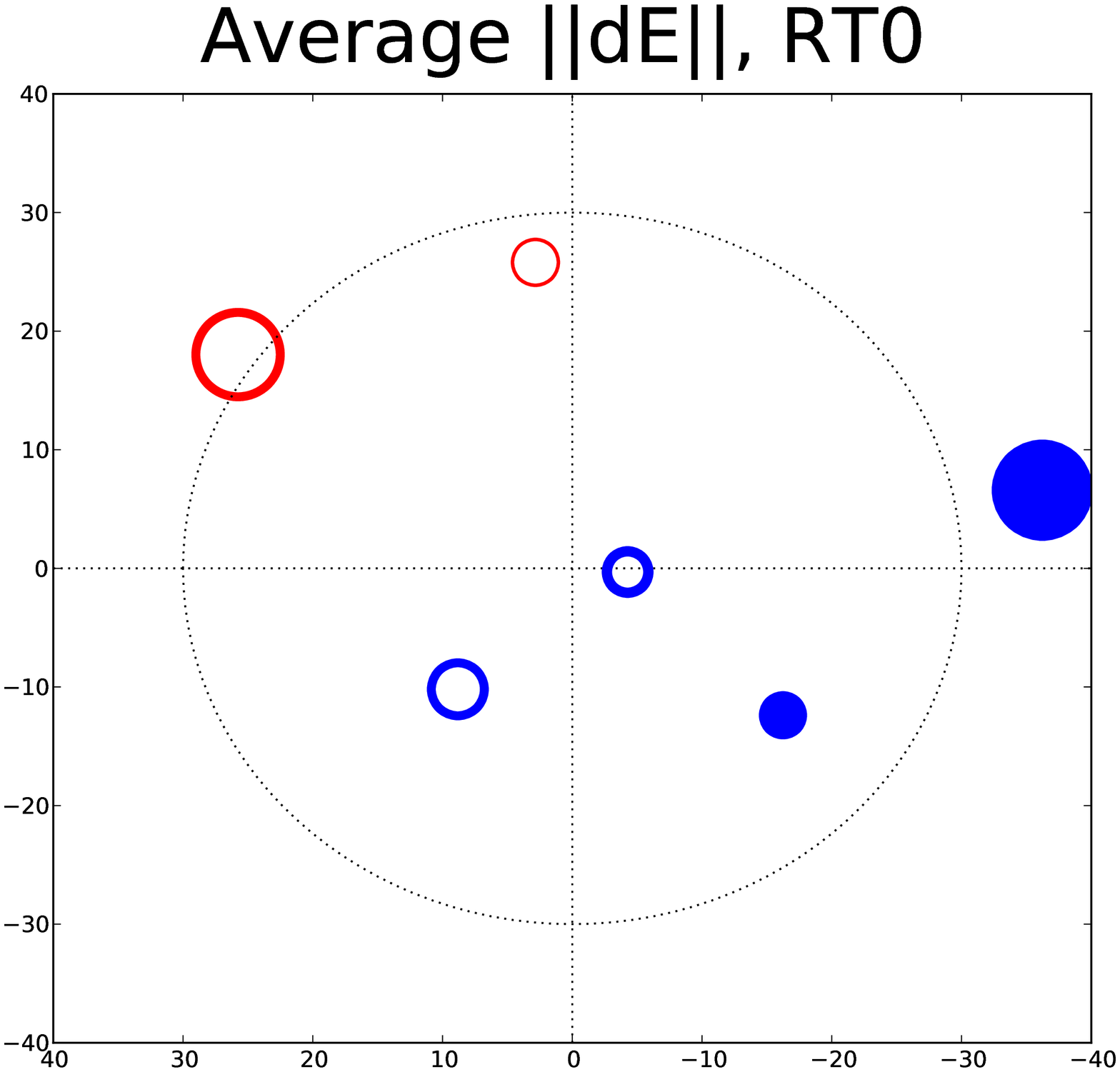} &
\includegraphics[width=\roguewidth]{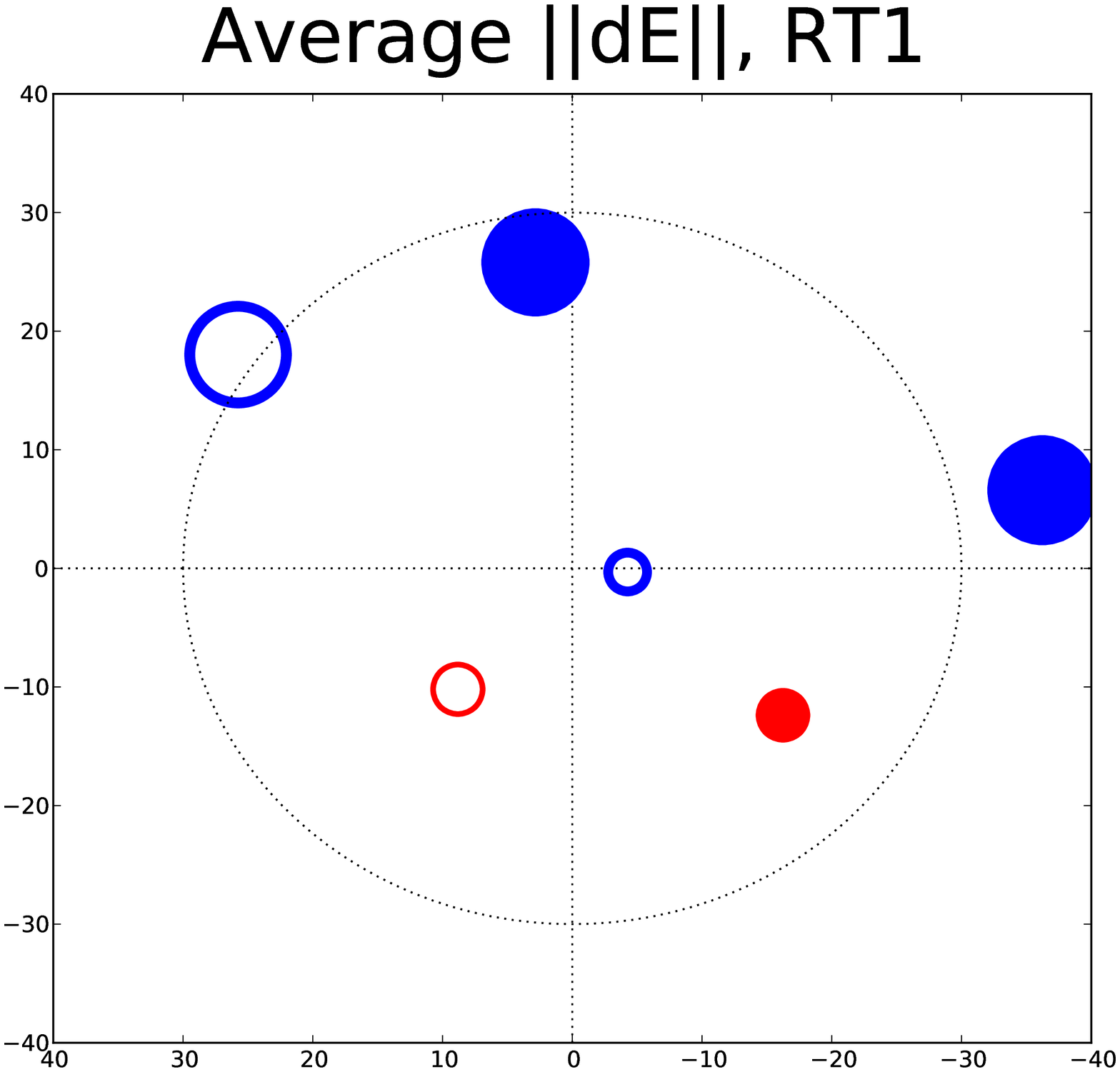} &
\includegraphics[width=\roguewidth]{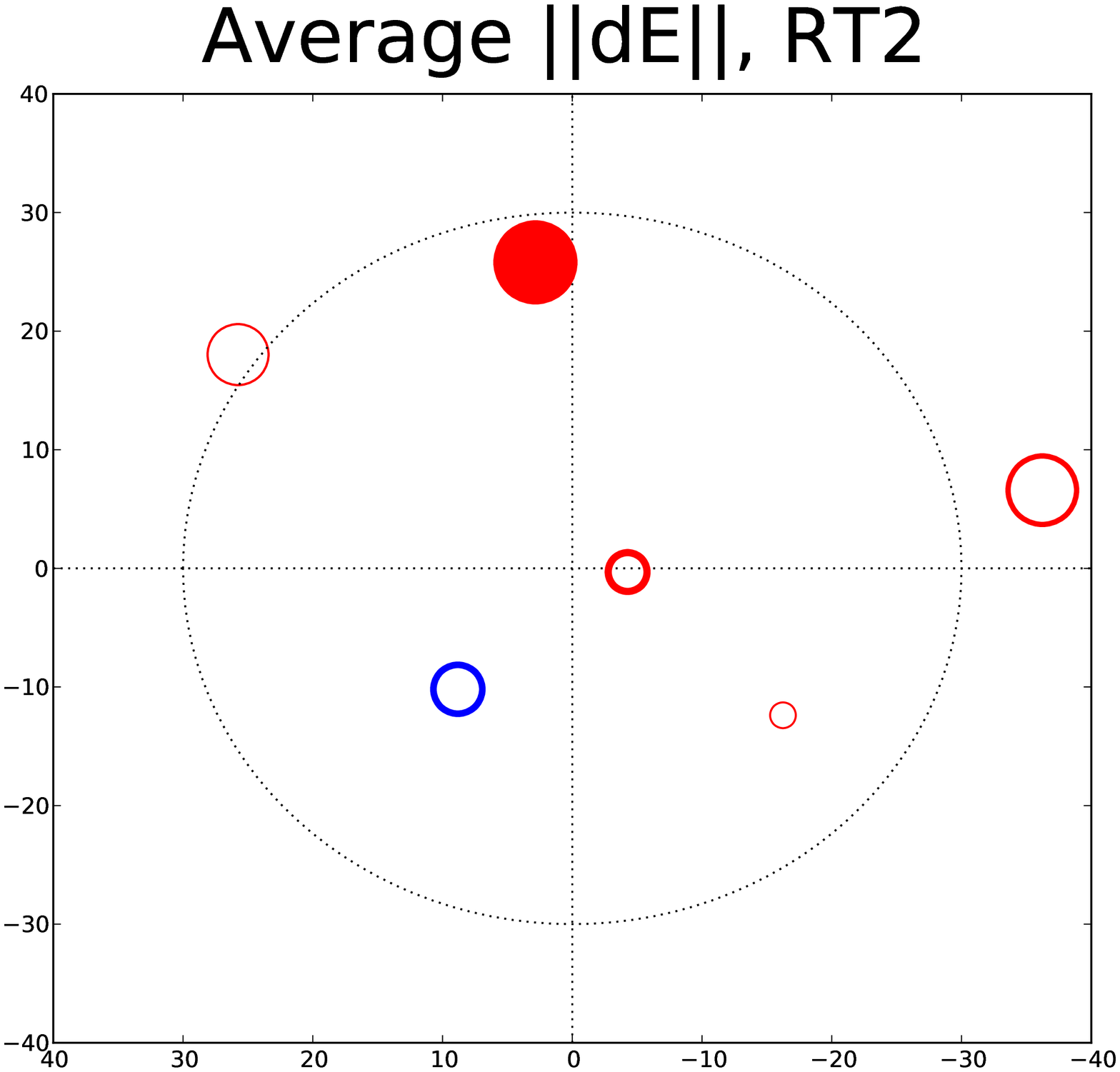} &
\includegraphics[width=\roguewidth]{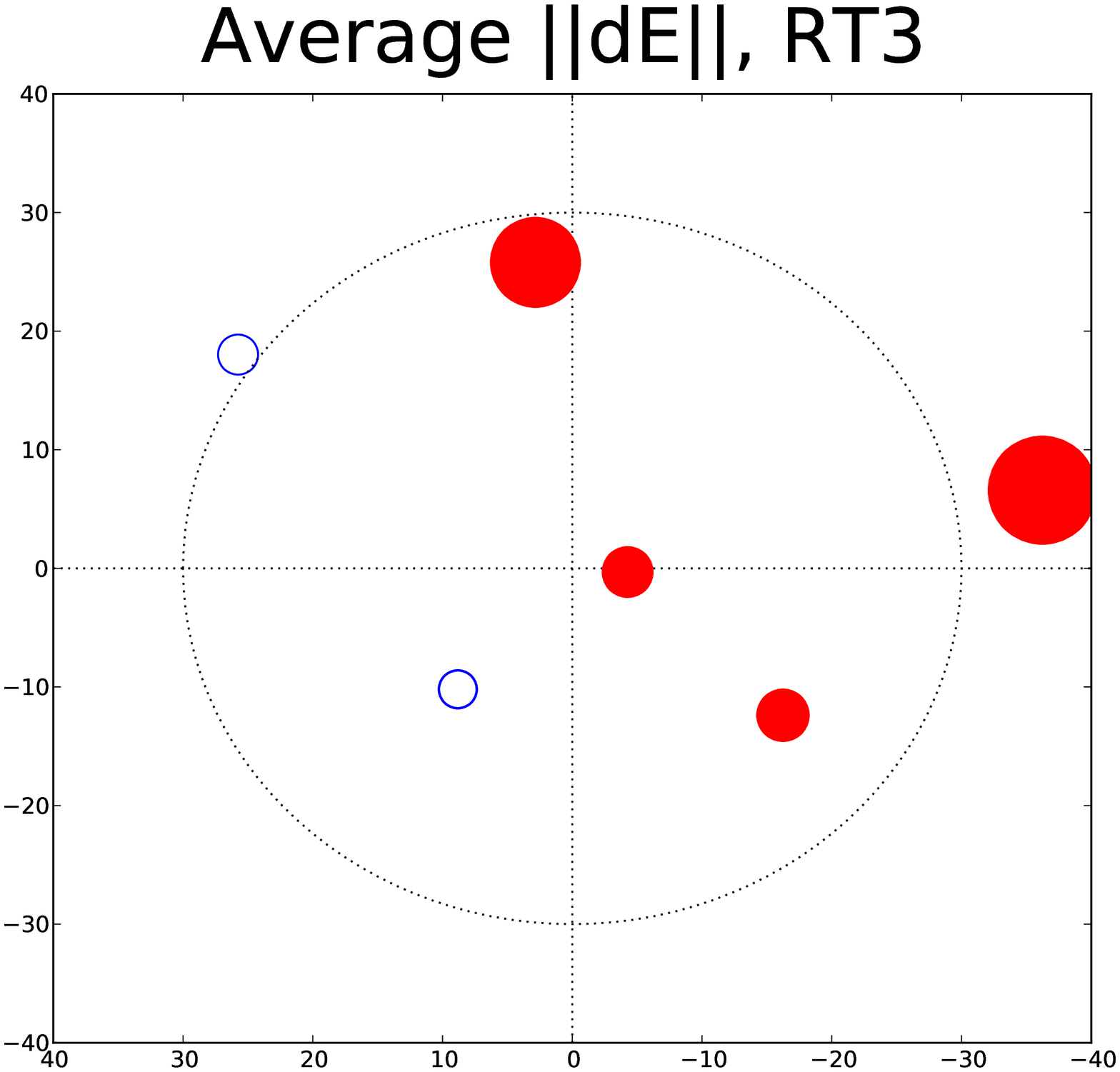} &
\includegraphics[width=\roguewidth]{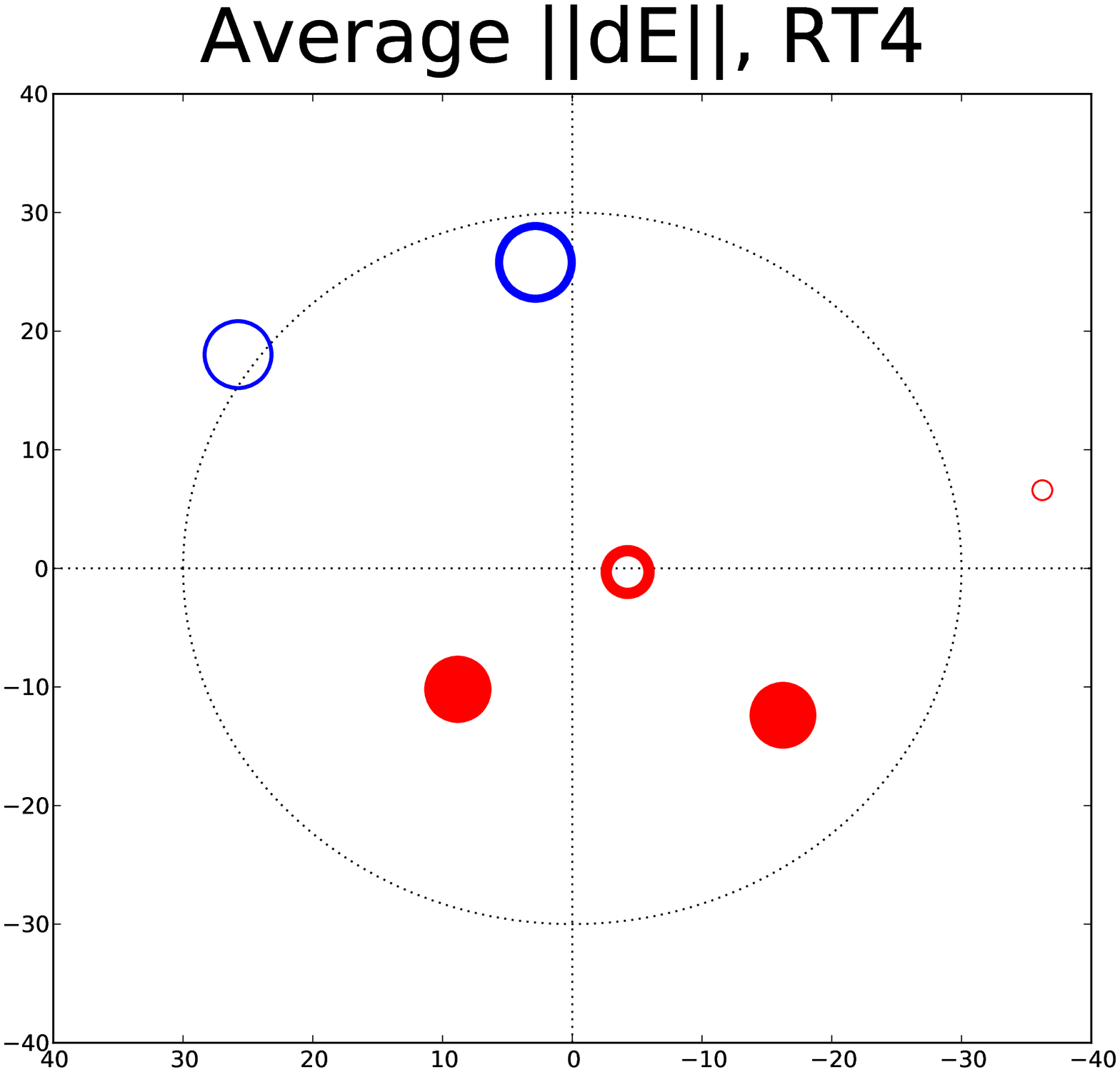} \\
\includegraphics[width=\roguewidth]{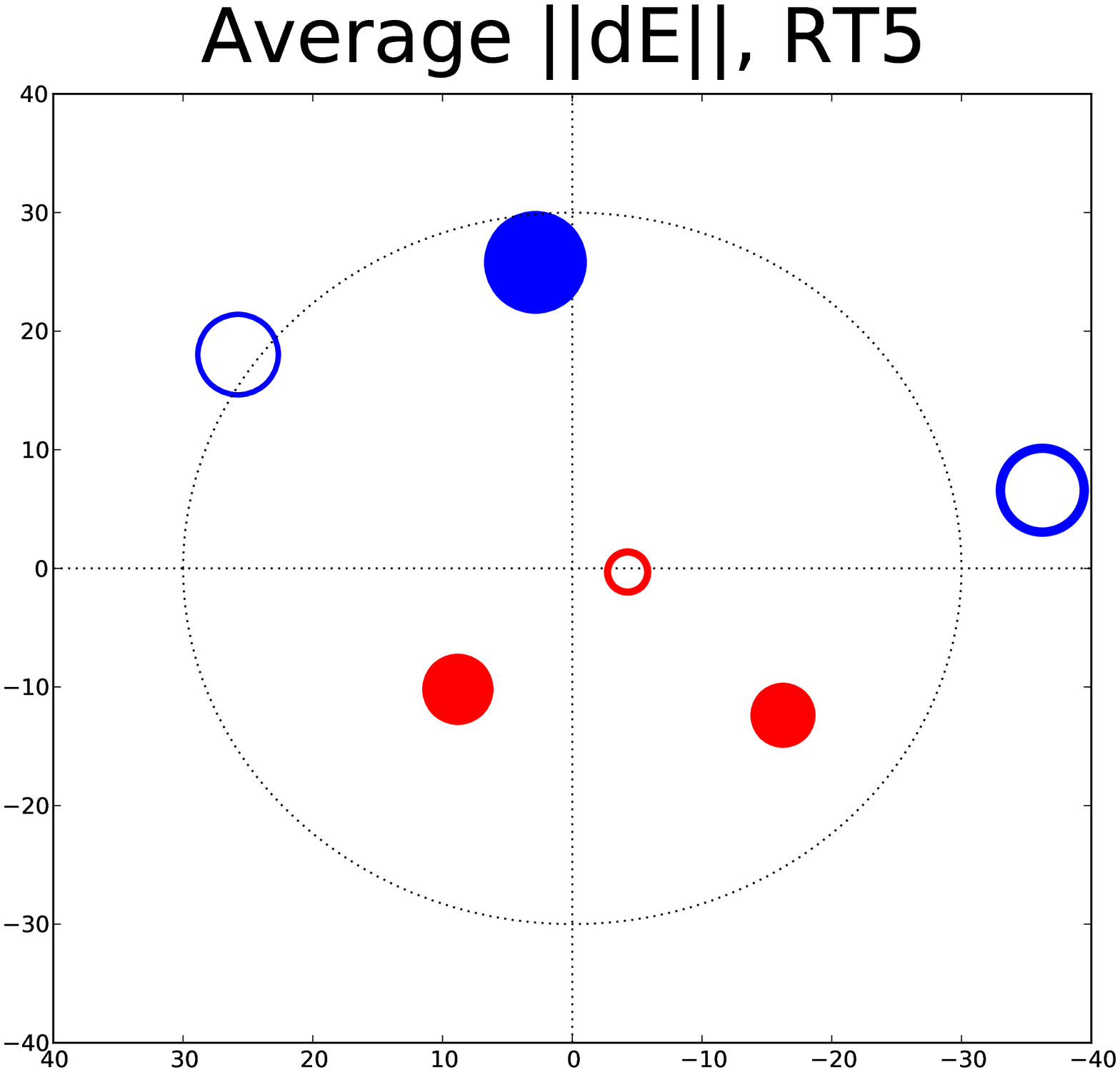} &
\includegraphics[width=\roguewidth]{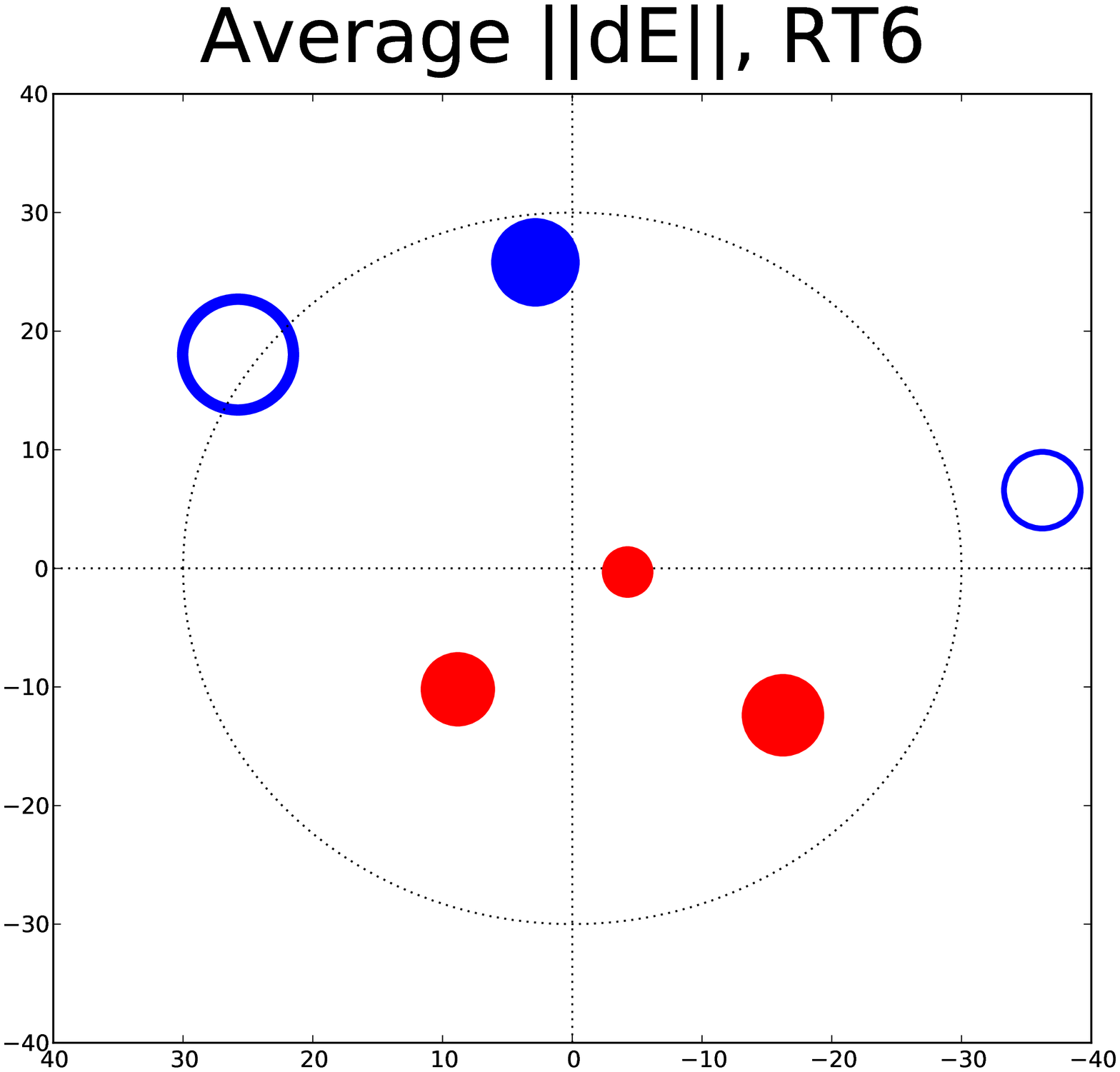} &
\includegraphics[width=\roguewidth]{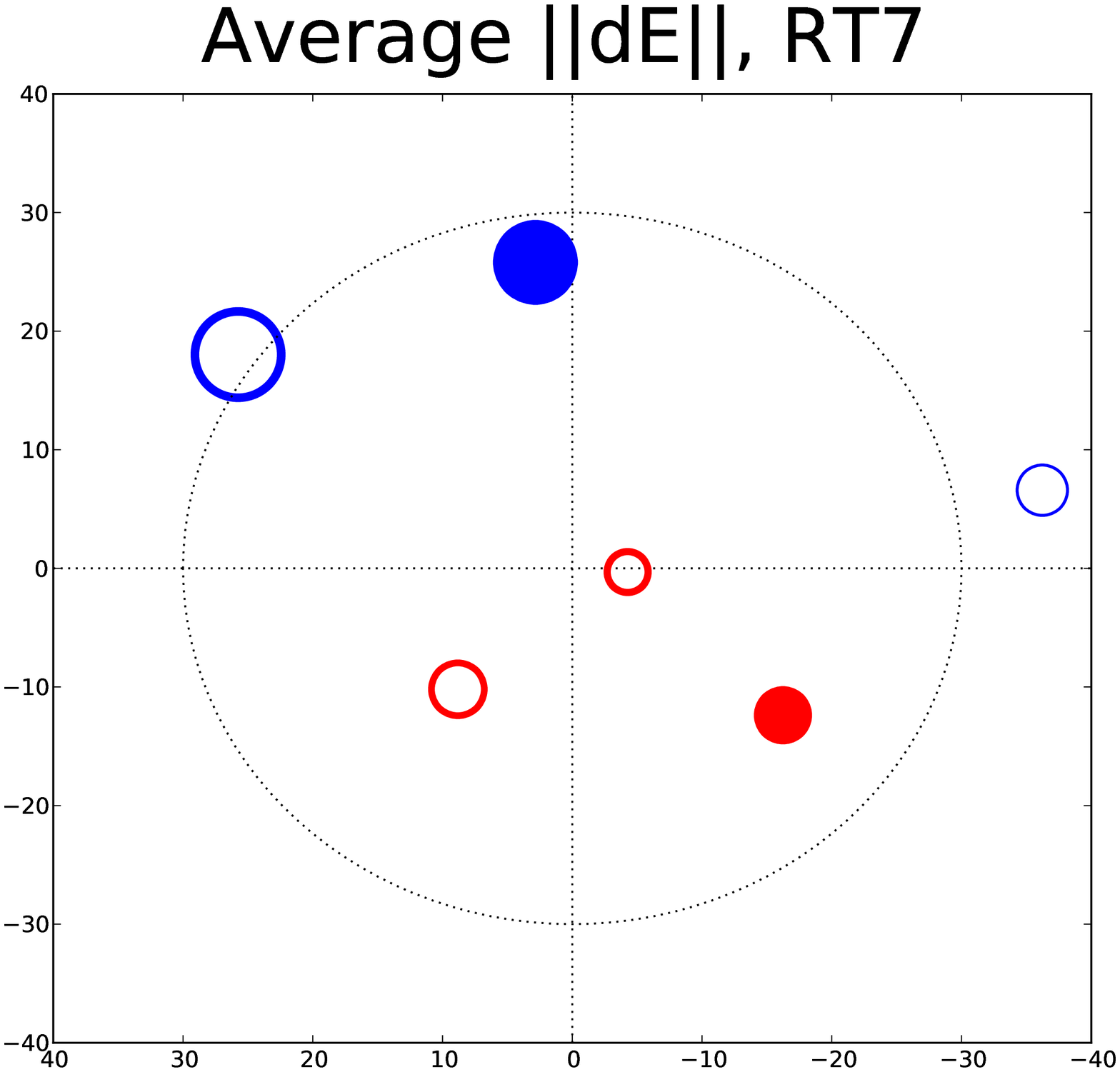} &
\includegraphics[width=\roguewidth]{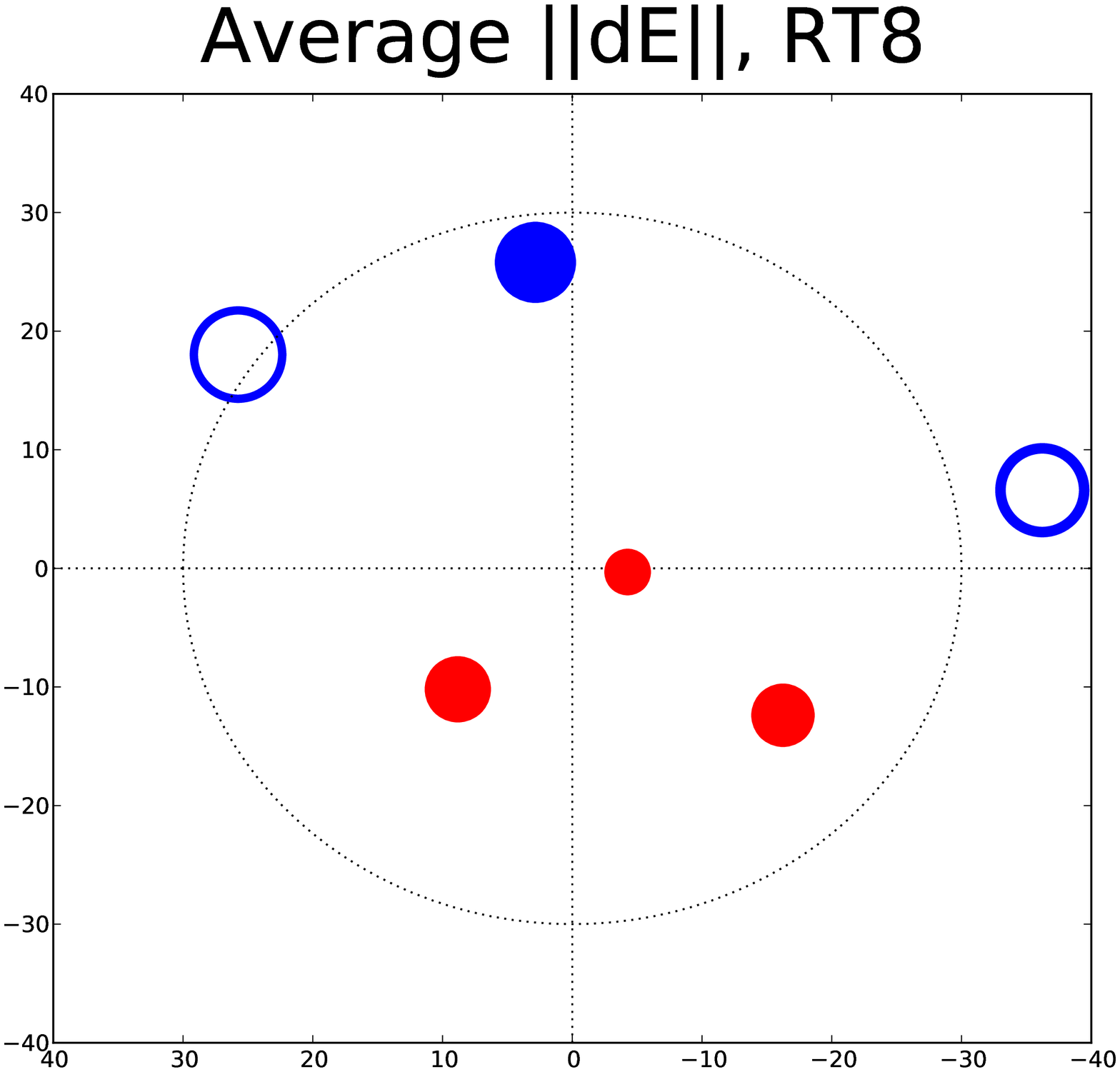} &
\includegraphics[width=\roguewidth]{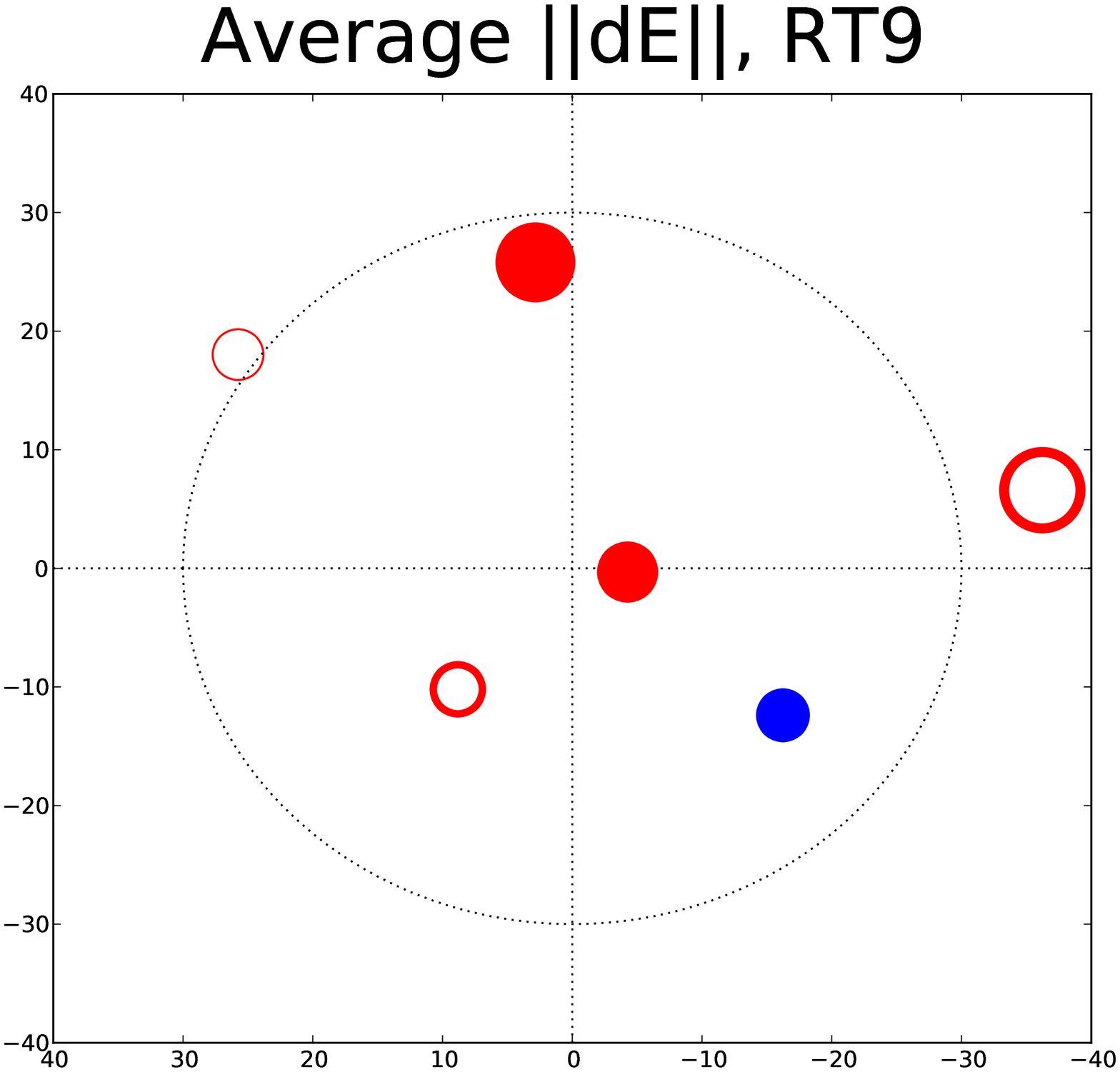} \\
\includegraphics[width=\roguewidth]{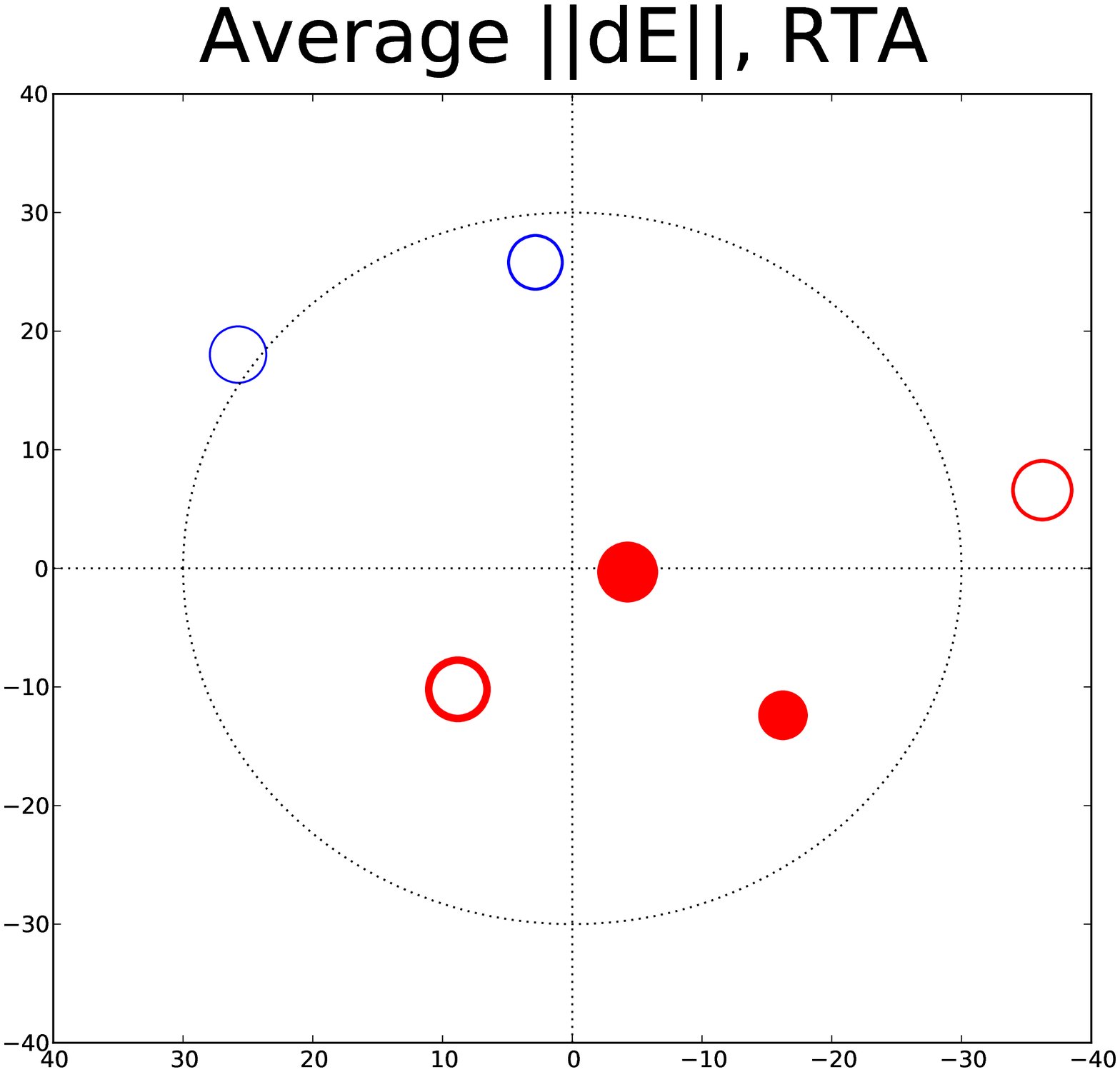} &
\includegraphics[width=\roguewidth]{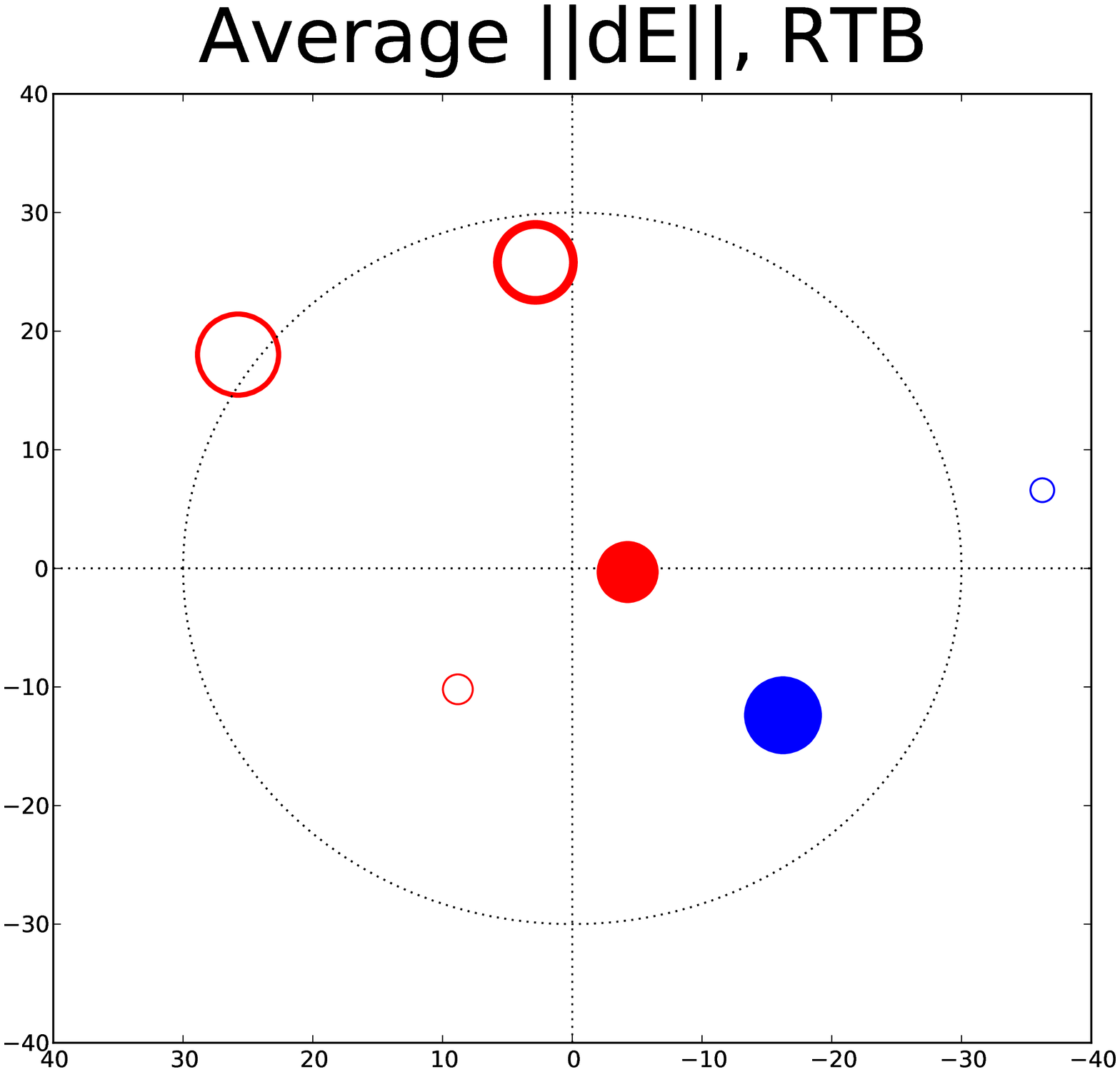} &
\includegraphics[width=\roguewidth]{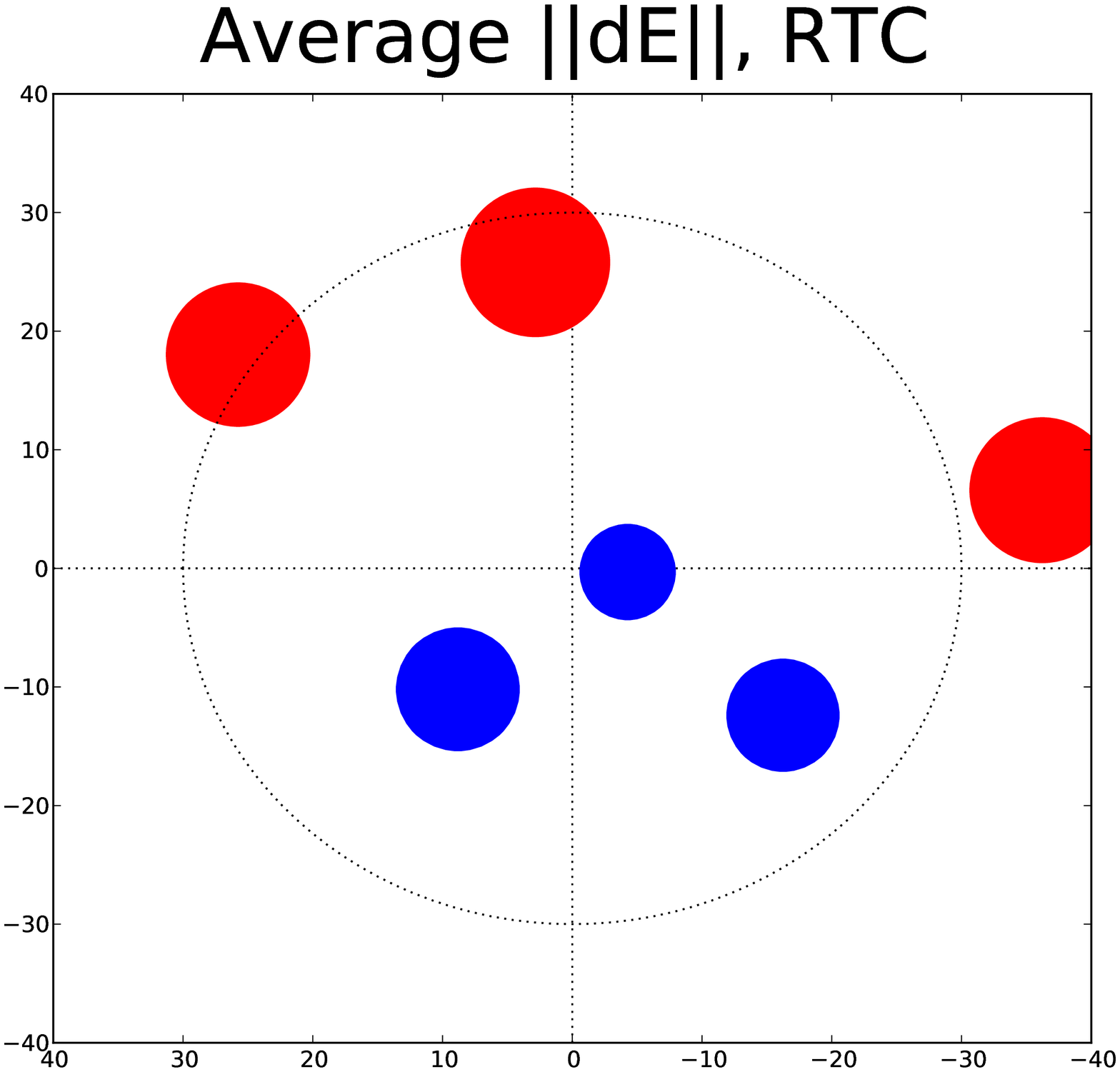} &
\includegraphics[width=\roguewidth]{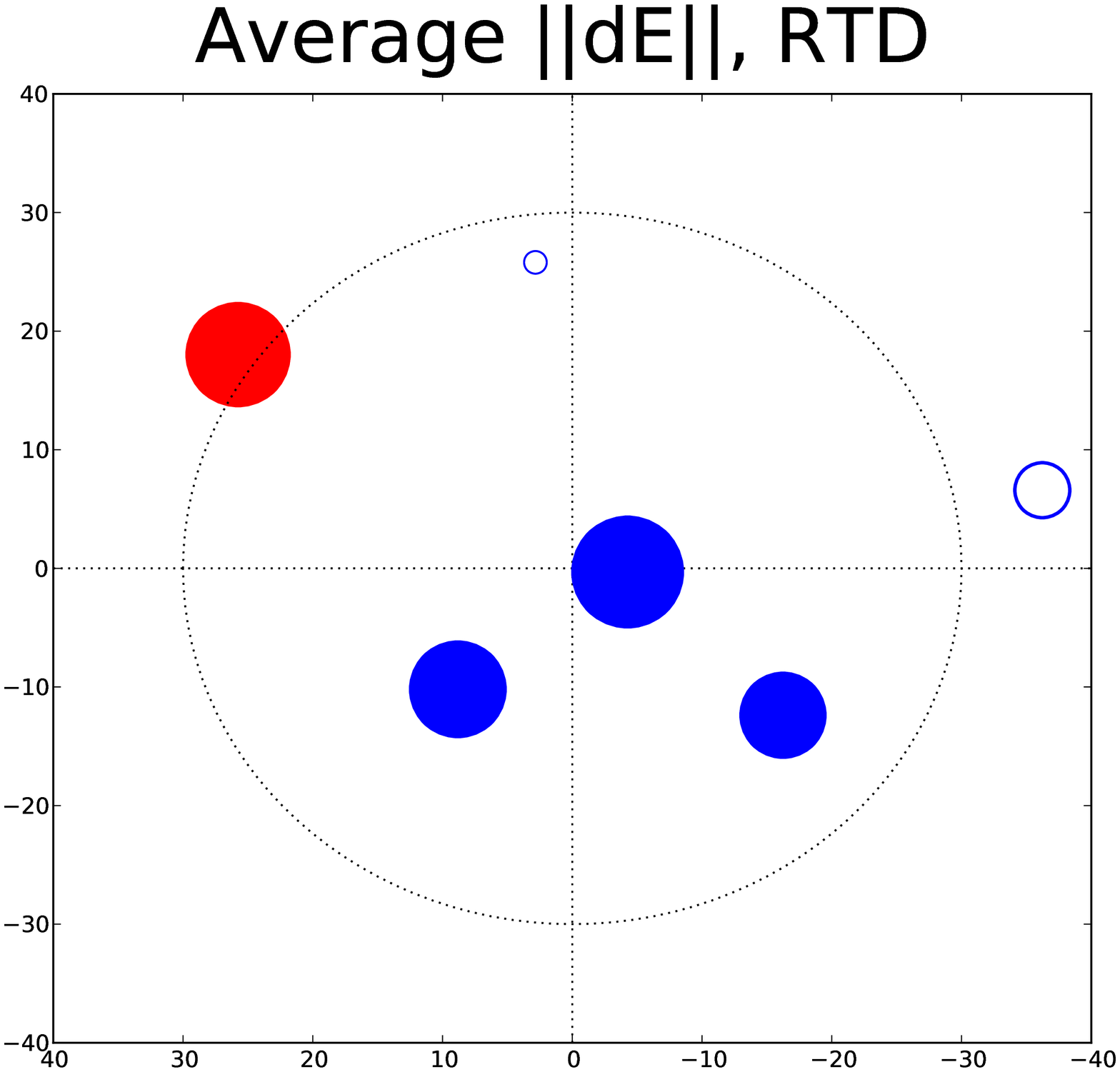} 
\end{tabular}
\caption{\label{fig:rogues-2003}``Rogues gallery'' plot for the 2003 observation. This shows the 12-hour average $||\Delta\jones{E}{}||$ per source, as seen by each antenna. Blue circles correspond to values of $||\Delta\jones{E}{}||>1$, red circles to values of $||\Delta\jones{E}{}||<1$, and areas are proportional to $|\,||\Delta\jones{E}{}||-1\,|$. Line thickness indicates the statistical significance of $|\,||\Delta\jones{E}{}||-1\,|$; filled circles are for detections of over $3\sigma$. The large grid circle is at radius $30\arcmin$.}
\end{figure}

\begin{figure}
\centering
\begin{tabular}{@{}c@{}c@{}c@{}c@{}c@{}}
\includegraphics[width=\roguewidth]{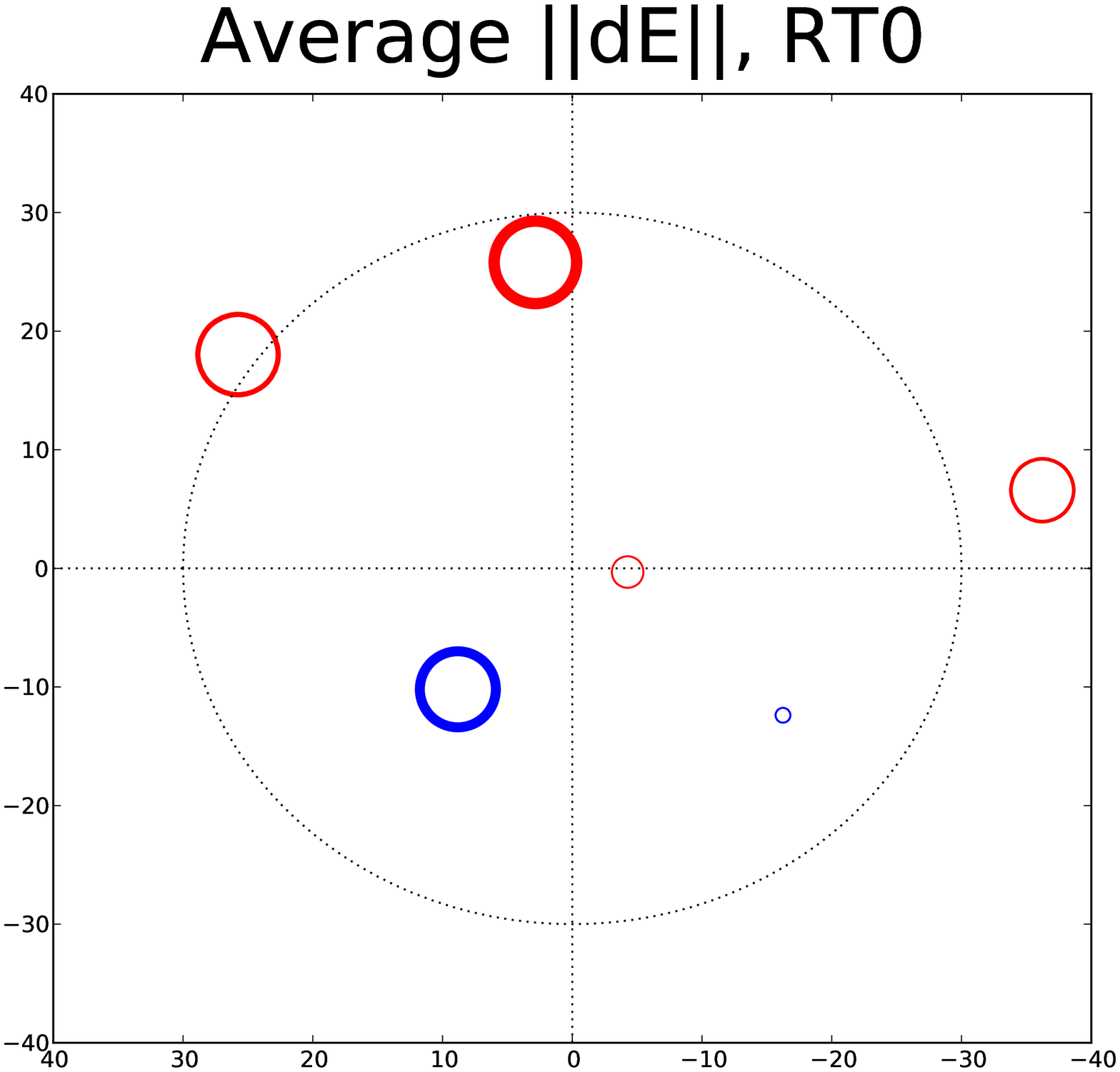} &
\includegraphics[width=\roguewidth]{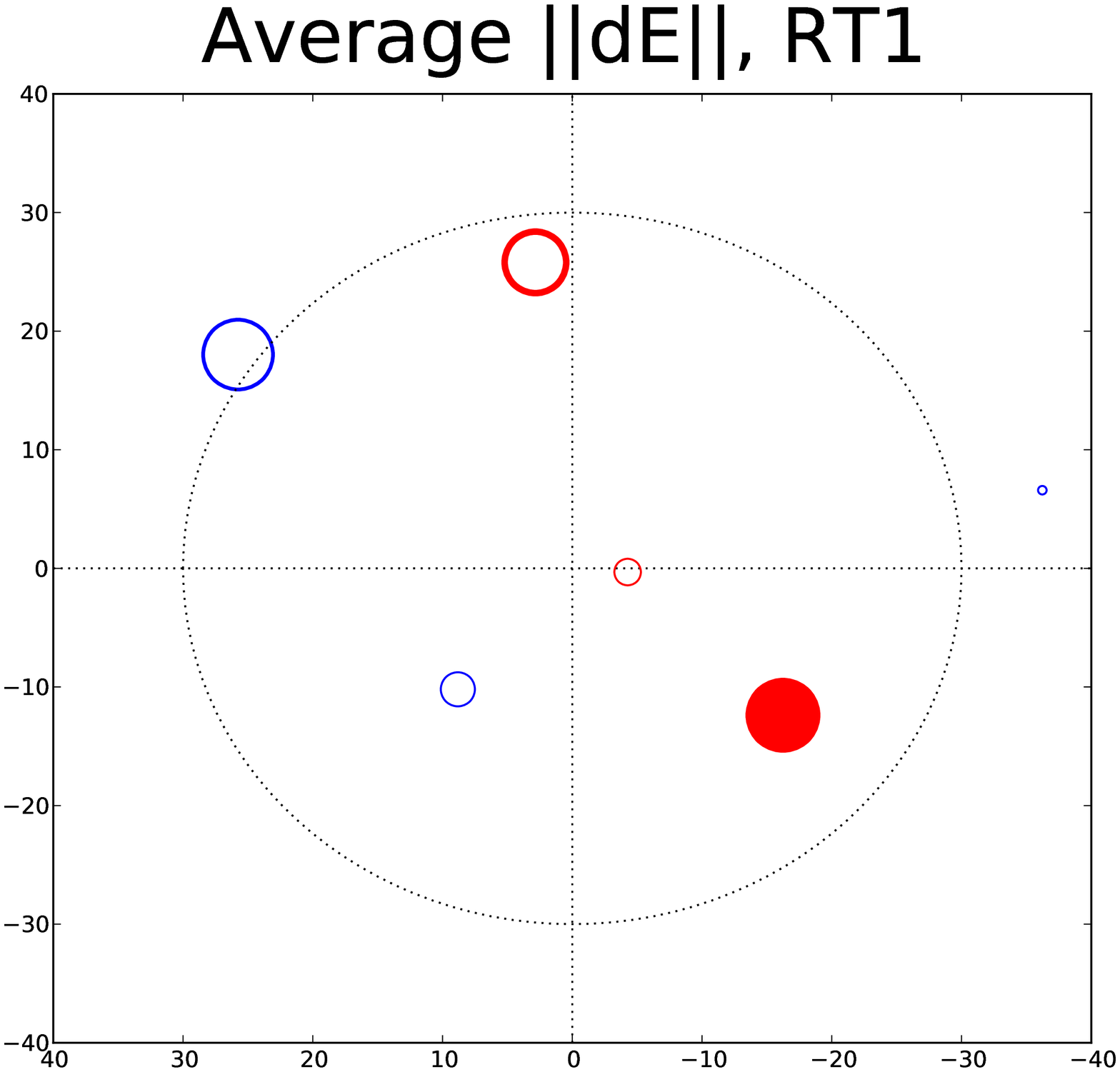} &
\includegraphics[width=\roguewidth]{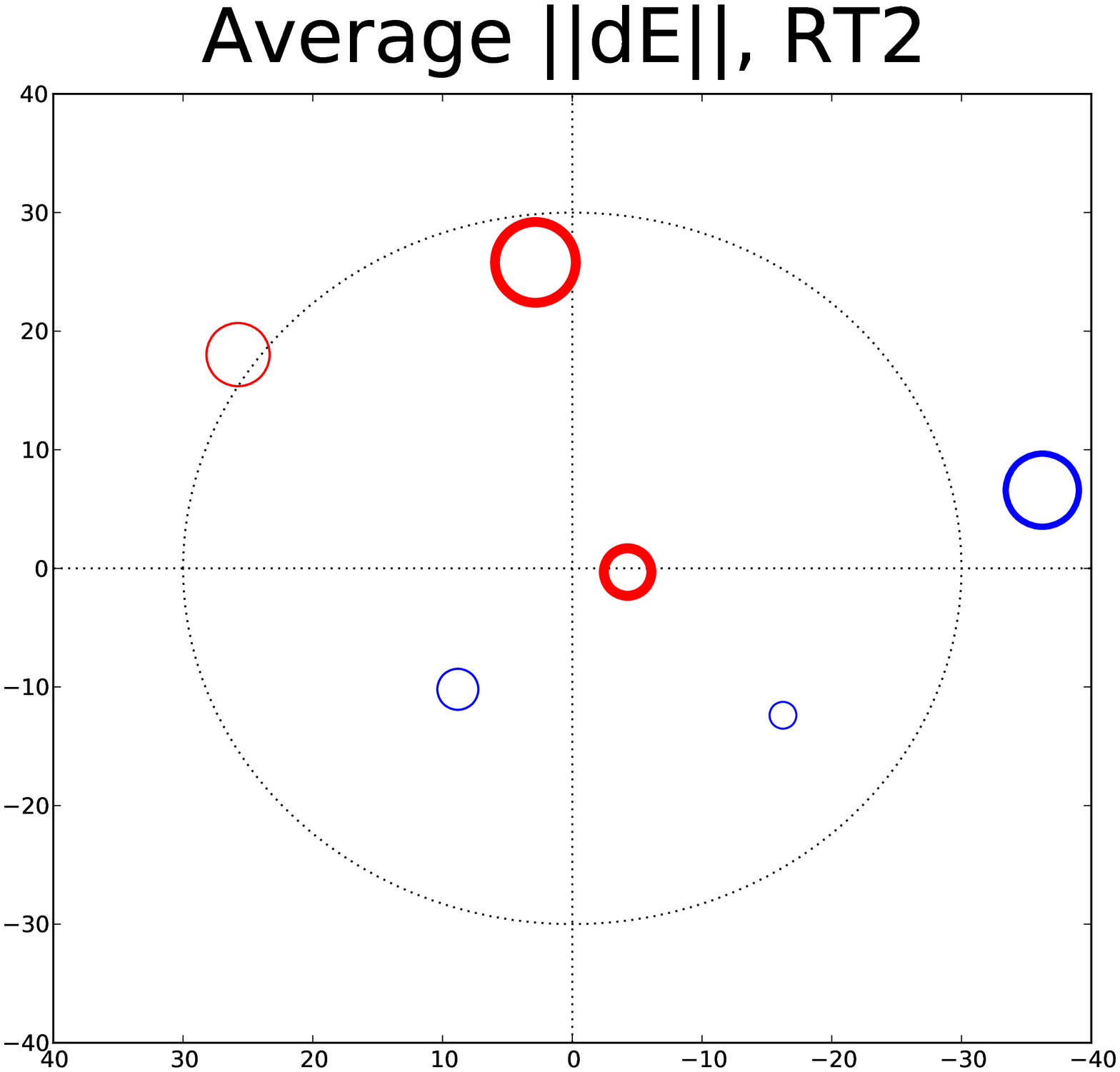} &
\includegraphics[width=\roguewidth]{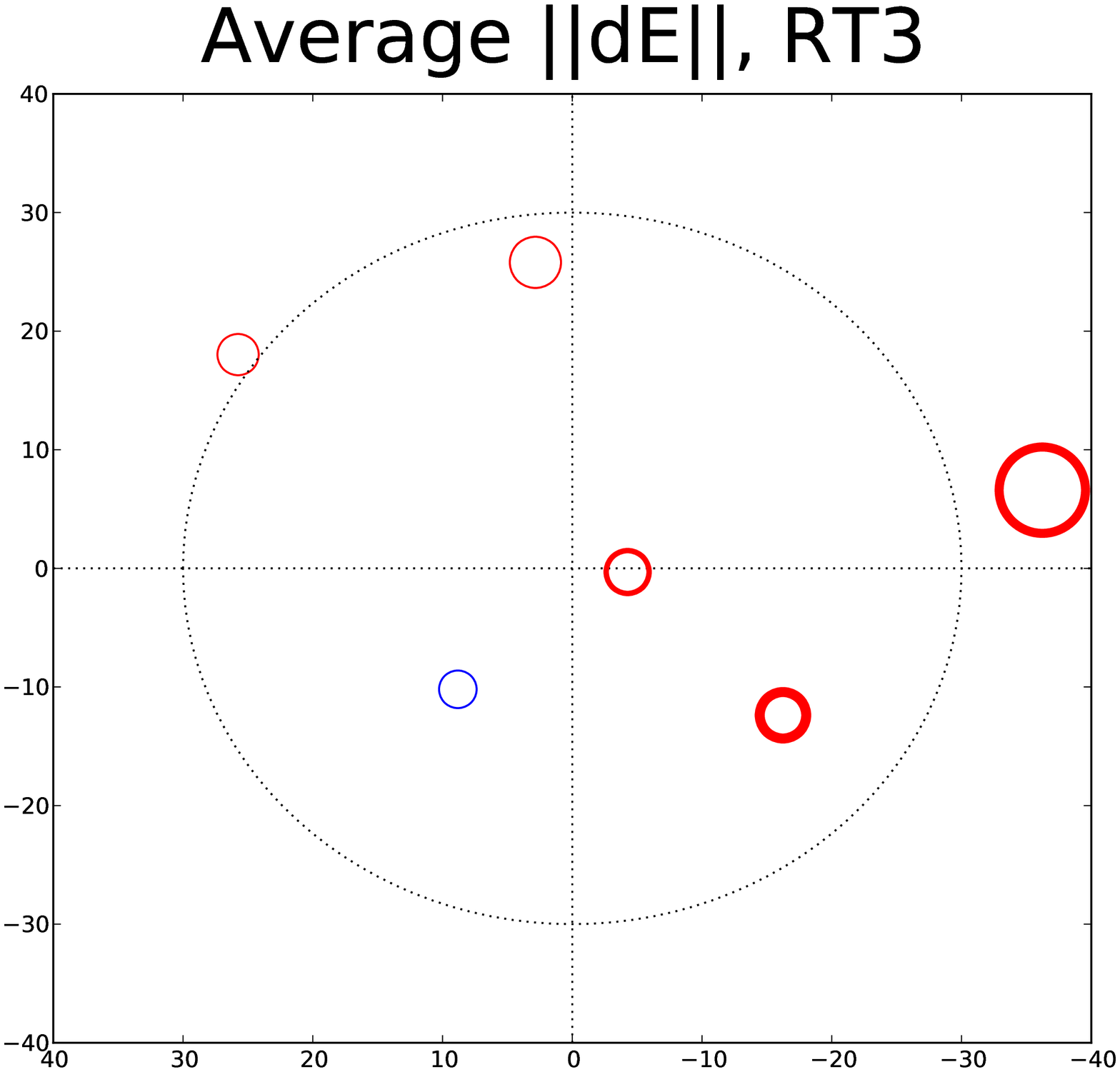} &
\includegraphics[width=\roguewidth]{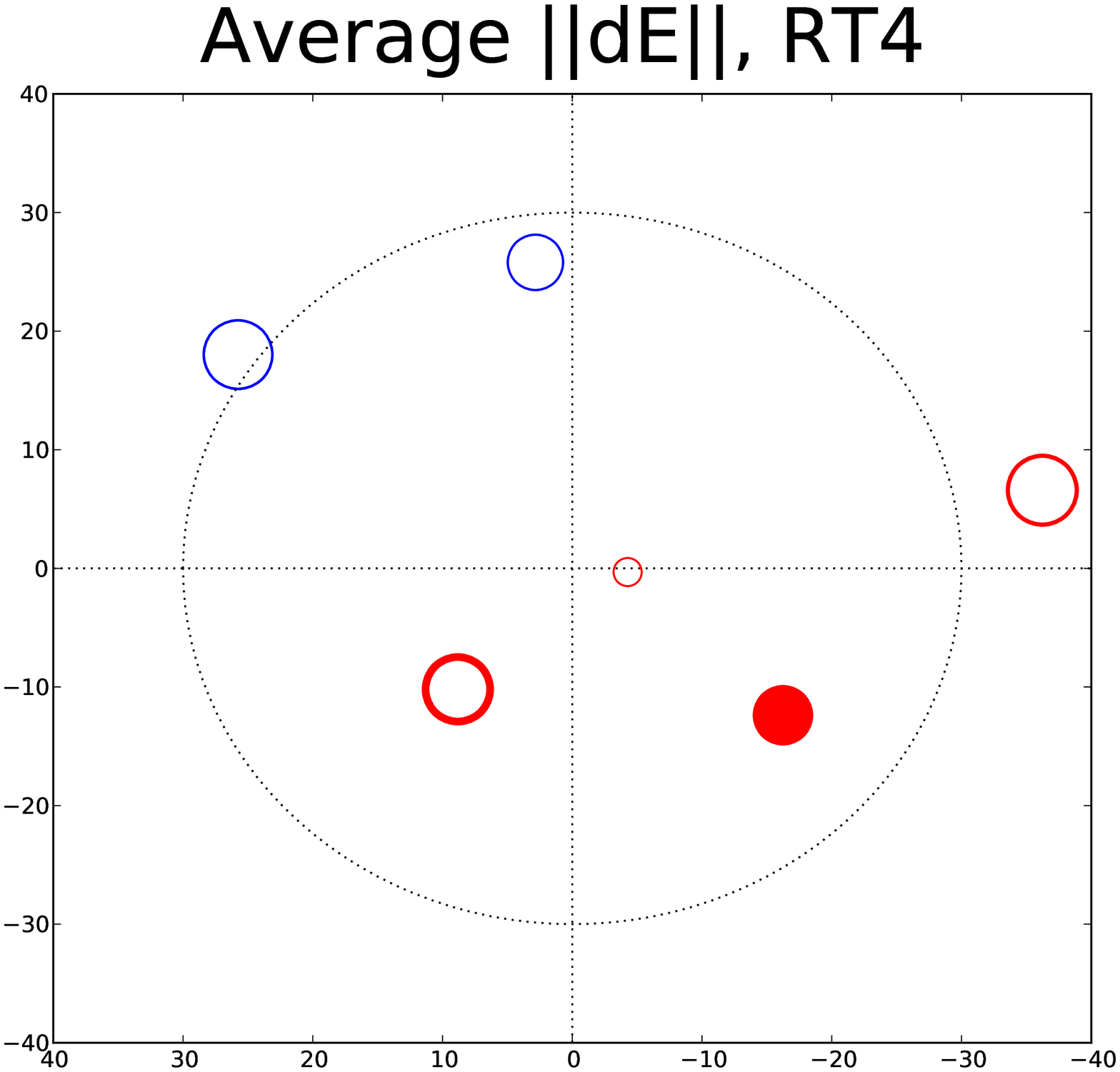} \\
\includegraphics[width=\roguewidth]{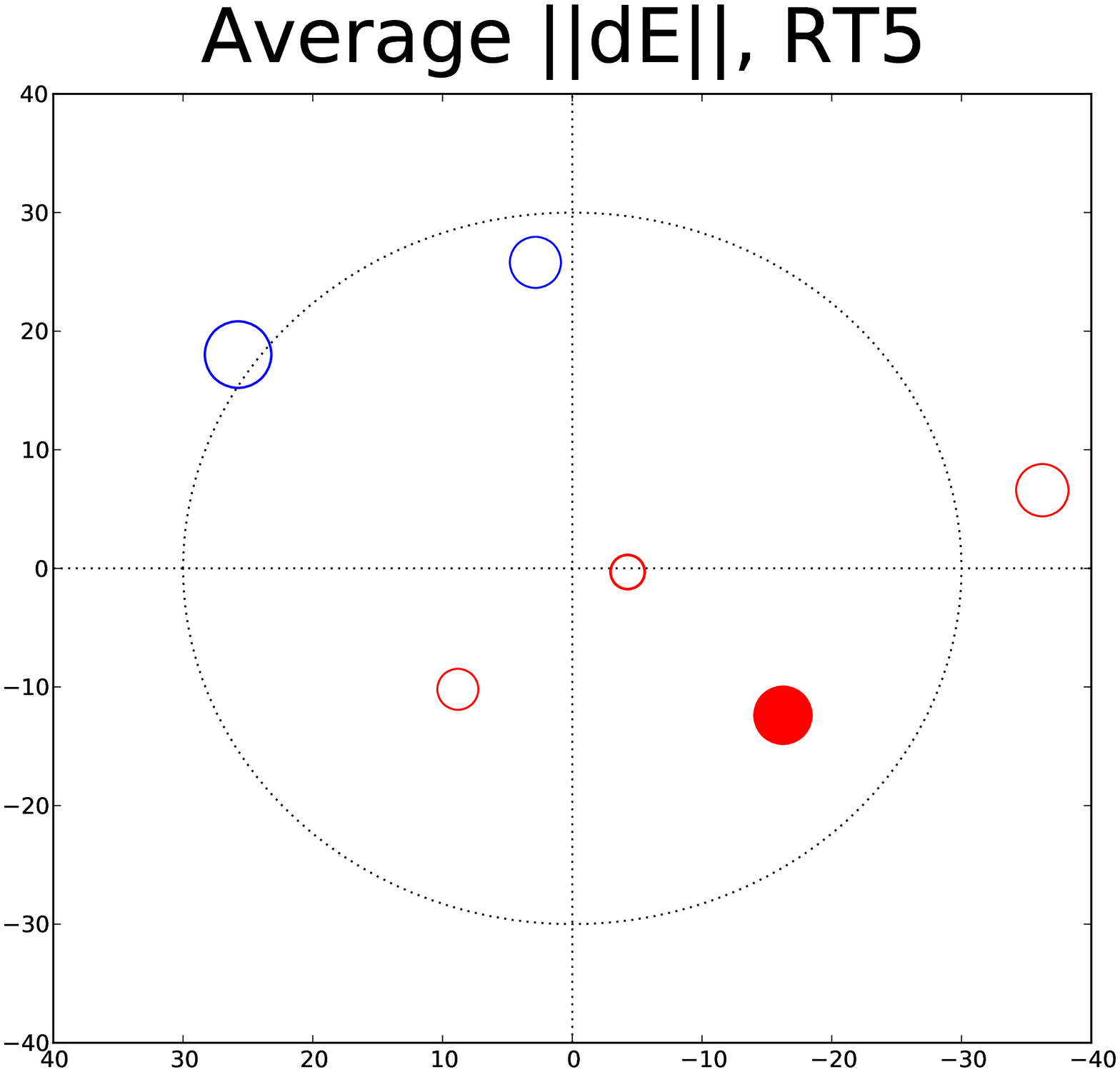} &
\includegraphics[width=\roguewidth]{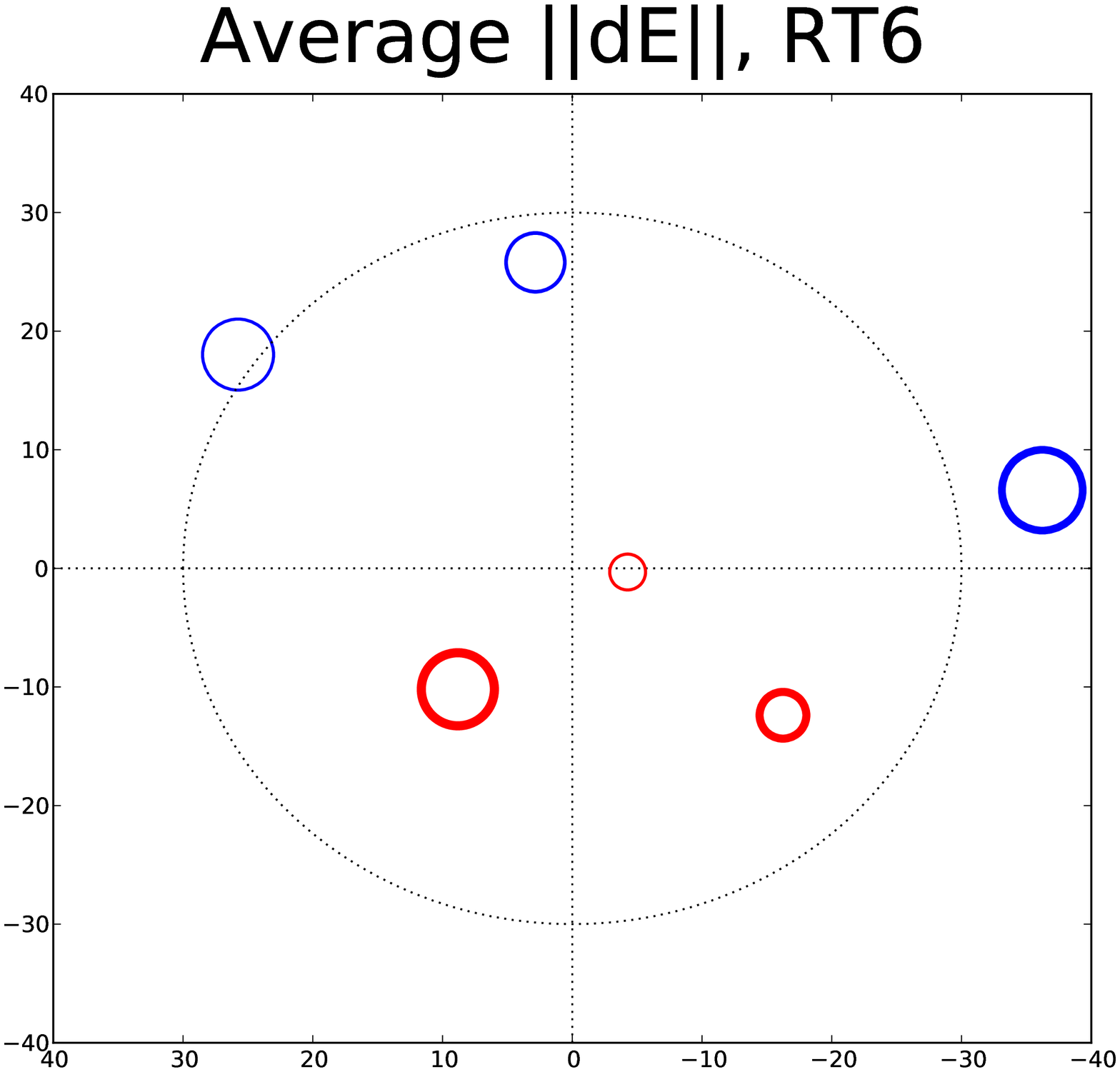} &
\includegraphics[width=\roguewidth]{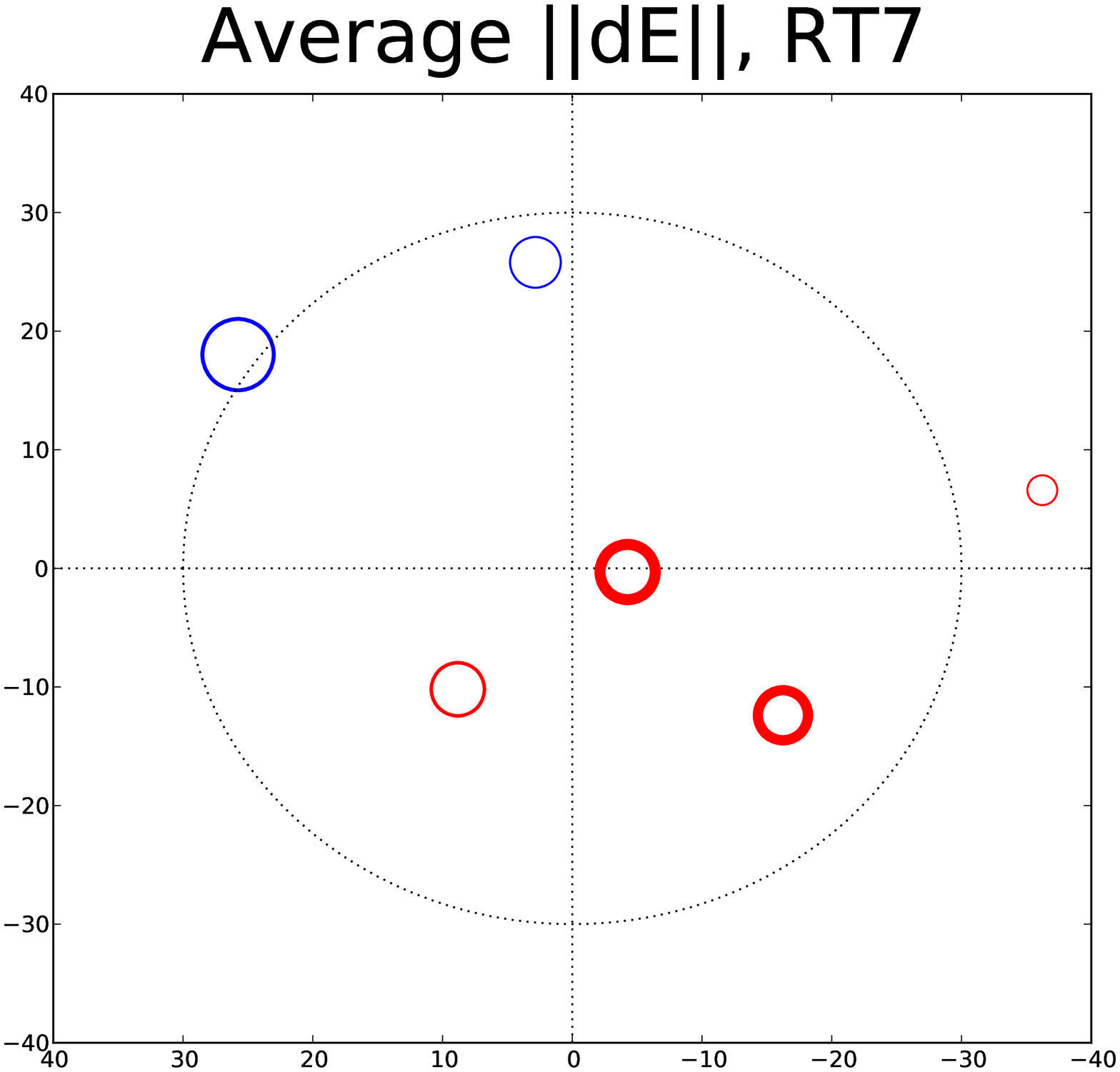} &
\includegraphics[width=\roguewidth]{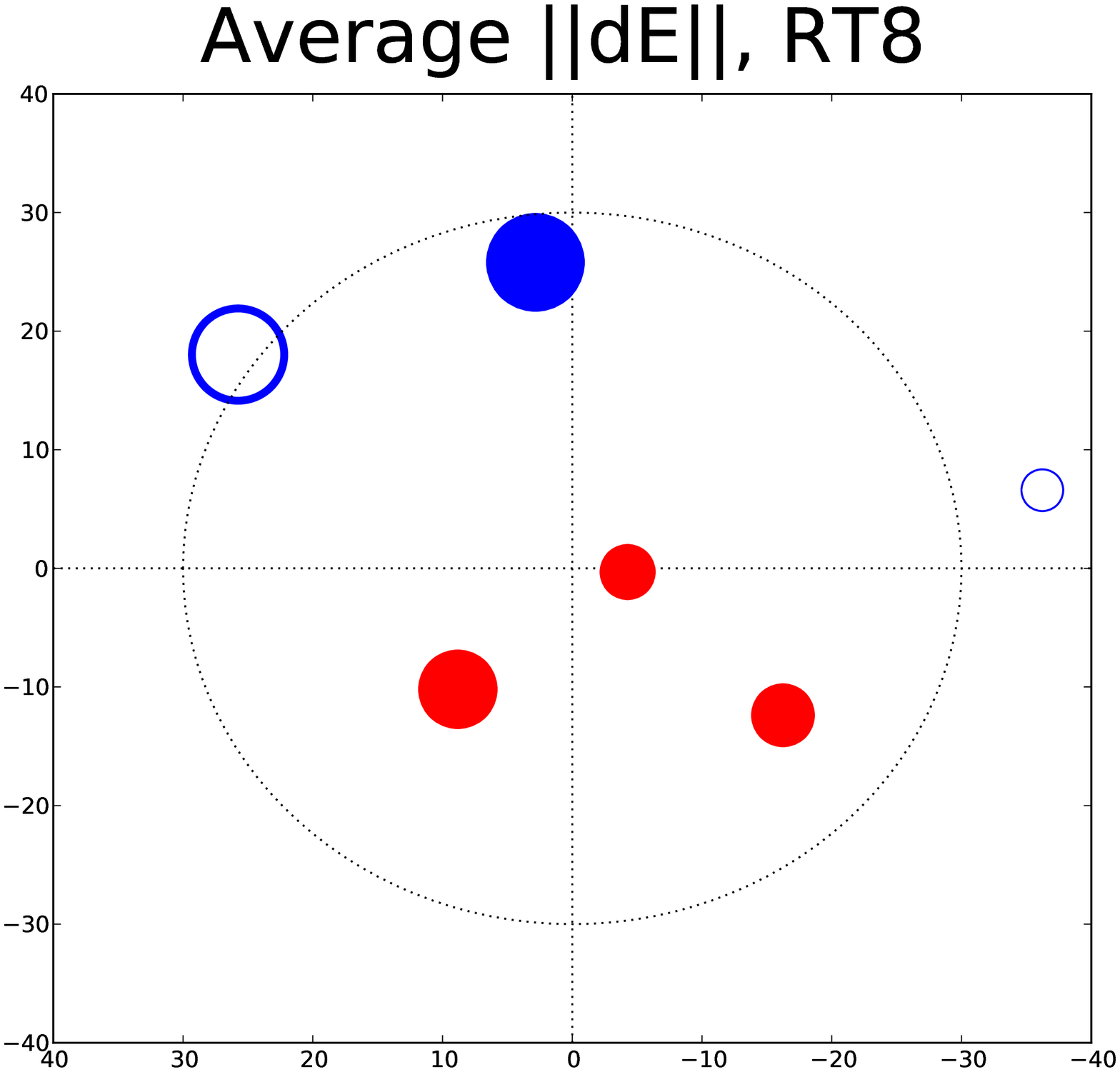} &
\includegraphics[width=\roguewidth]{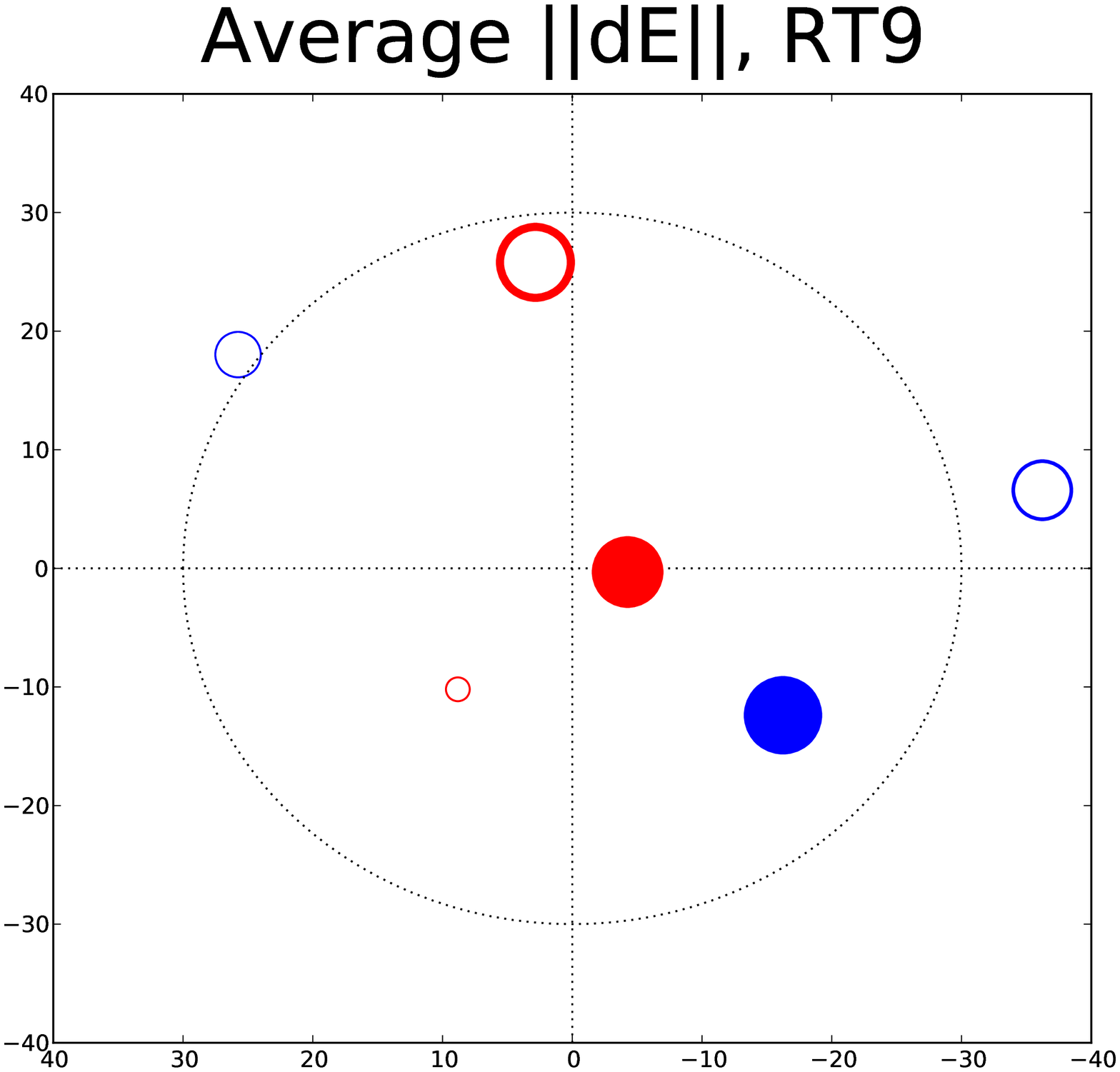} \\
\includegraphics[width=\roguewidth]{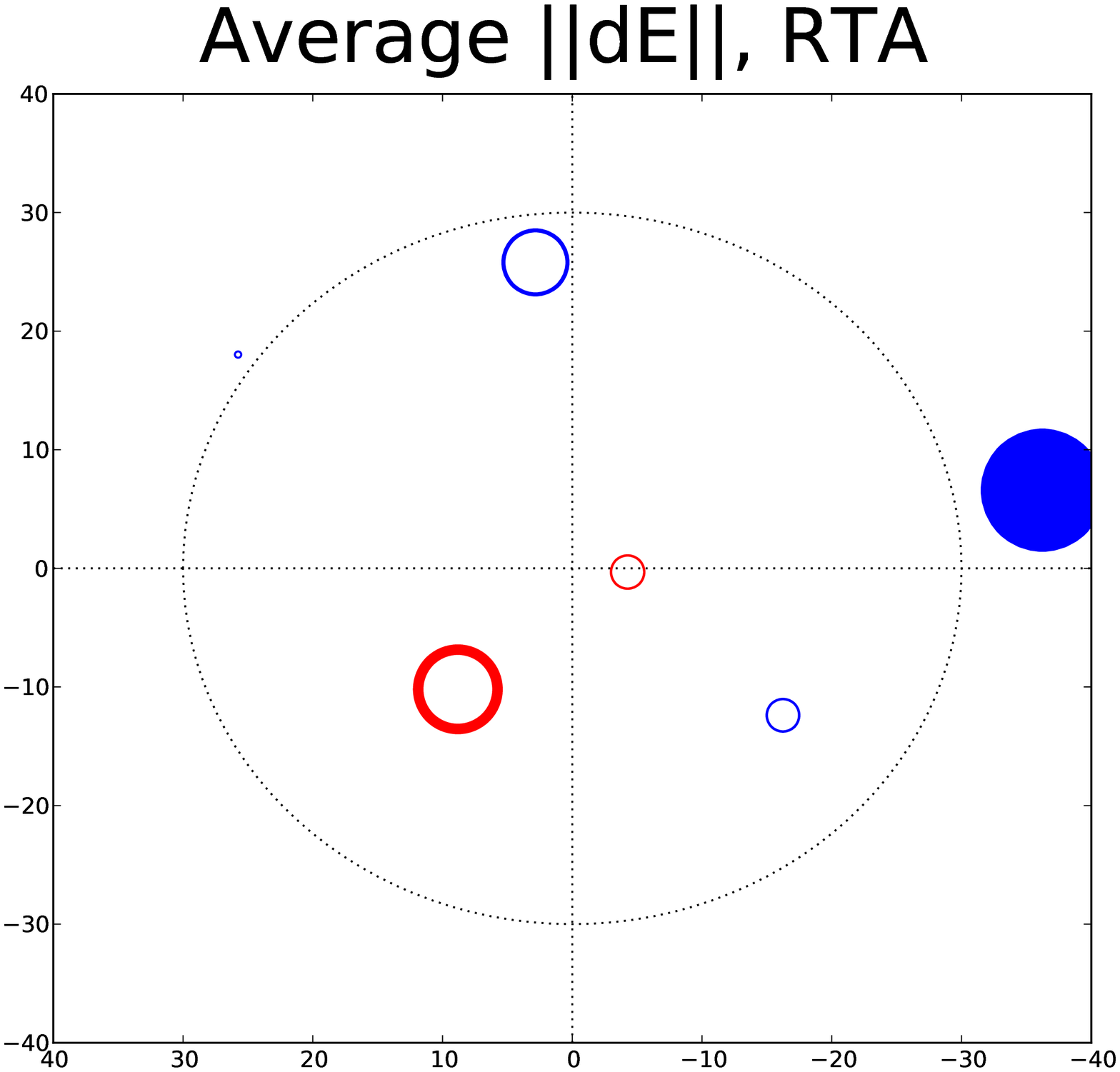} &
\includegraphics[width=\roguewidth]{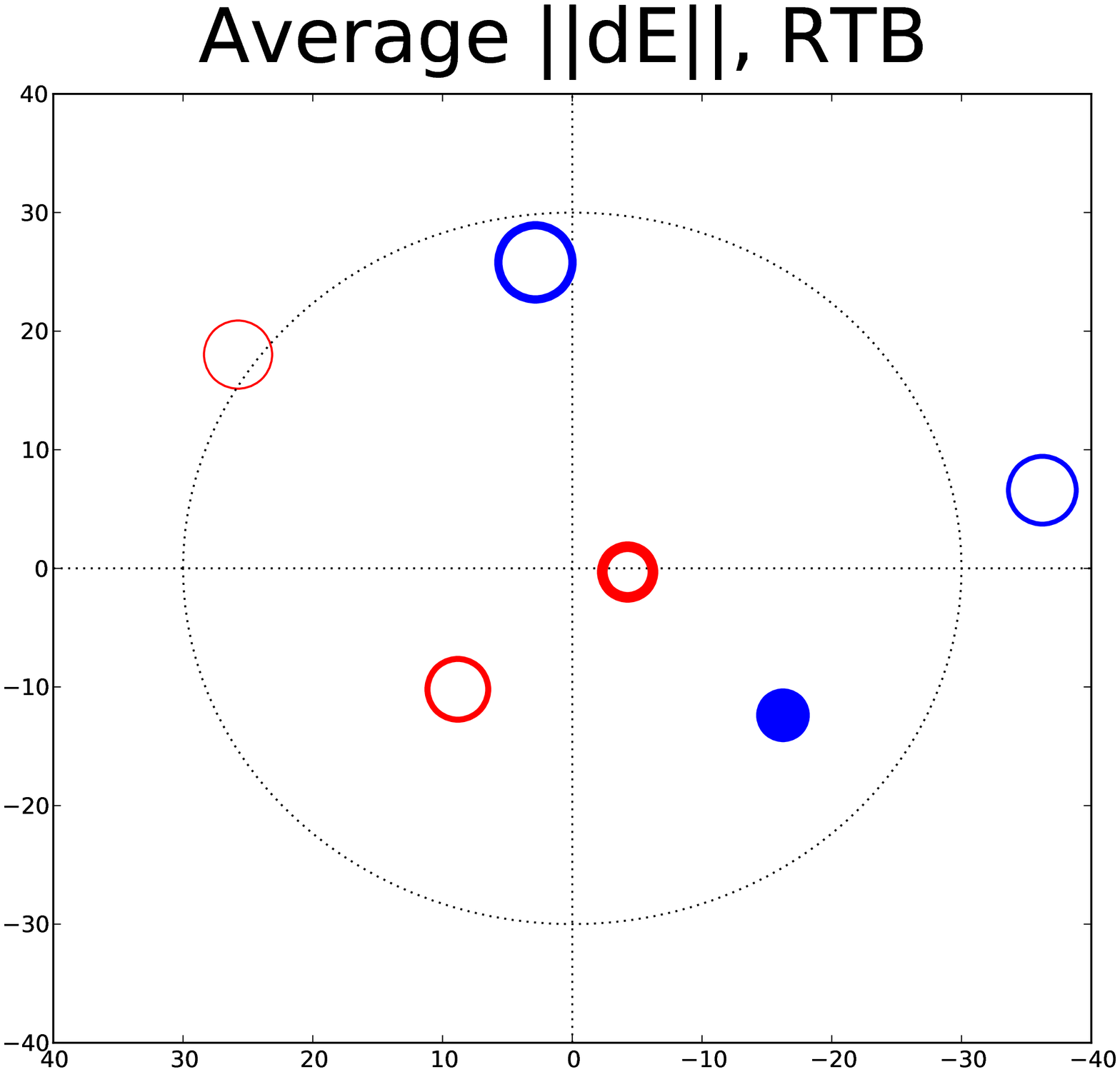} &
\includegraphics[width=\roguewidth]{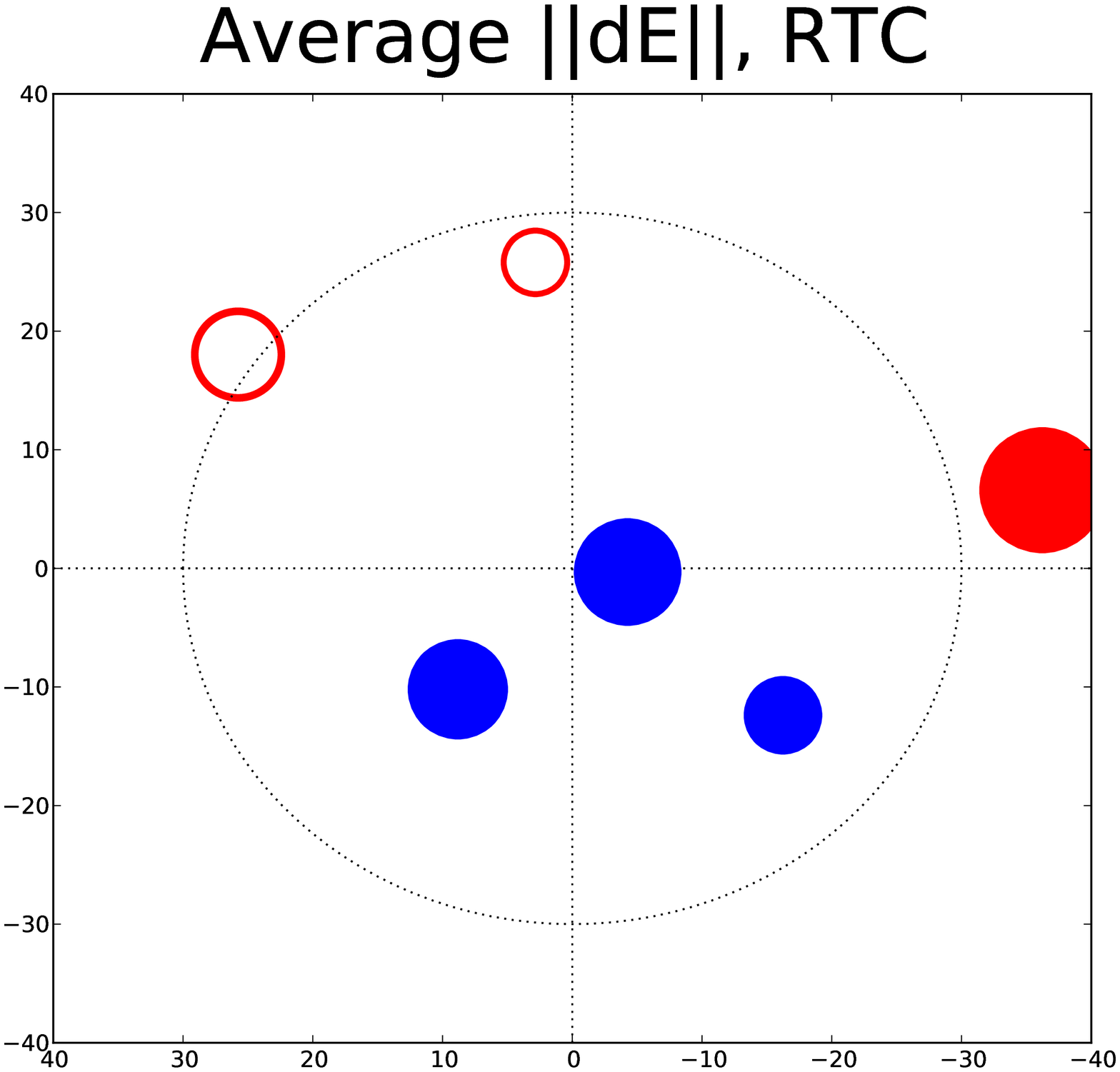} &
\includegraphics[width=\roguewidth]{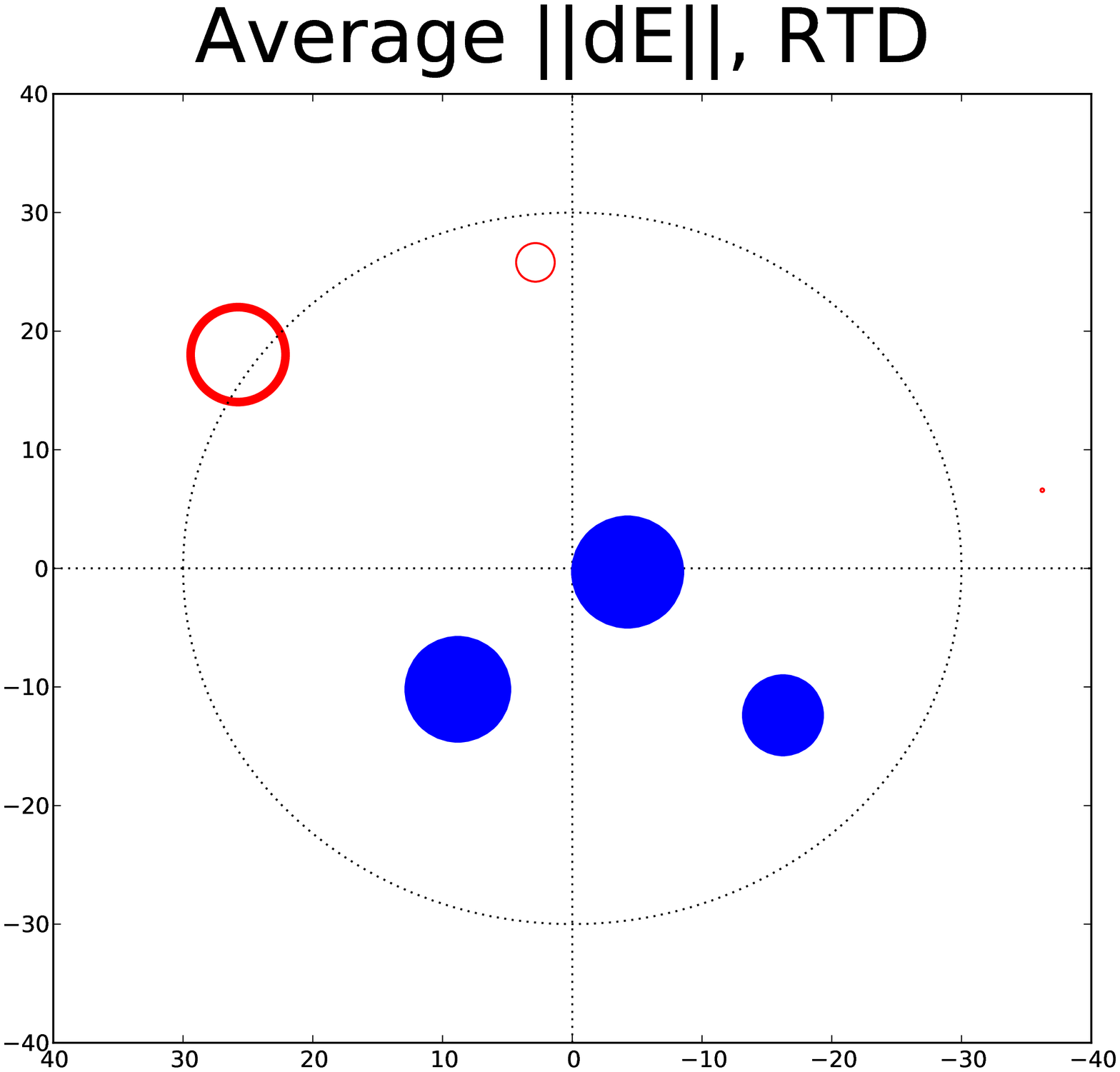} 
\end{tabular}
\caption{\label{fig:rogues-2006}``Rogues gallery'' plot for the 2006 observation, using the same scale as Fig.~\ref{fig:rogues-2003}}
\end{figure}

The galleries show exactly such a pattern for antennas RT5 through RT8 (and perhaps RT4), for both the 2003 and, to a lesser extent, the 2006 observations. It is a little bit strange that 4 (or even 5) adjacent antennas would so consistently mispoint North, and do the same three years later. Perhaps this is another, poorly understood consequence of unmodelled source structure. Some pointers to this are that antennas RT4--8, being in the middle of the array, form up predominantly shorter baselines, and that the long-baseline antennas RTC and RTD exhibit the opposite behaviour. (What hinders such an analysis is the unfortunate fact that the three brightest off-axis sources all exhibit some structure, and all three lie in the bottom half of the field.) Another puzzling feature is the consistently low $||\Delta\jones{E}{}||$ for sources F, H and K on antenna RTC in 2003 (and to a far lesser extent in 2006). If due to source structure, why does it not repeat on RTD? Perhaps RTC is mispointing to the South?

Antennas RT0--2, RT9 and RTA, on the other hand, show completely different patterns, with little to no similarity between 2003 and 2006. Some of these are consistent with a static mispointing. Some antennas (RT8 and RT9, and RTB especially) also show a hint of time variability in $||\Delta\jones{E}{}||$.

In any case, it is clear that the complicated interaction between source structure and differential gain-amplitudes makes the latter extremely difficult to interpret. Note also that my (or rather de Bruyn's) source model was built by NEWSTAR based on regular selfcal, so there's bound to be some contamination from DDE-related artefacts in the source parameters. 
Truly robust methods for disentangling source structure from DDEs have yet to be developed. It is also clear that an approach that parametrizes the DDEs in a ``global'' way, such as pointing selfcal, is the way forward -- what is not yet clear is how much the global solution itself can be affected by unmodelled structure in the brighter sources, and what to do about it.

\subsection{Phase behaviour}


\begin{figure}
\centering
\includegraphics[width=\columnwidth]{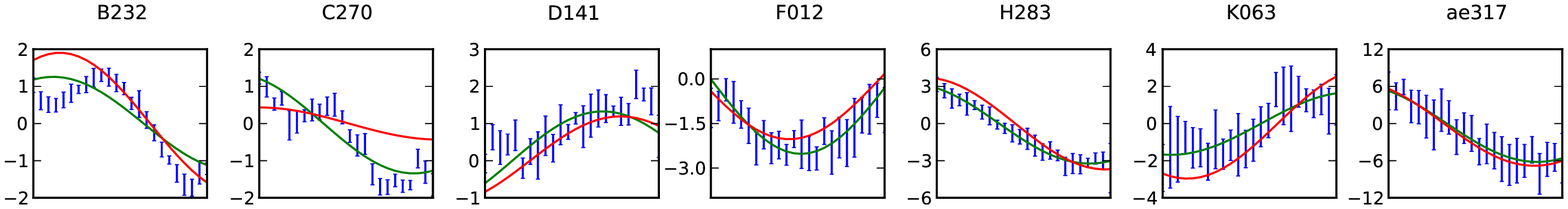}\\
\includegraphics[width=\columnwidth]{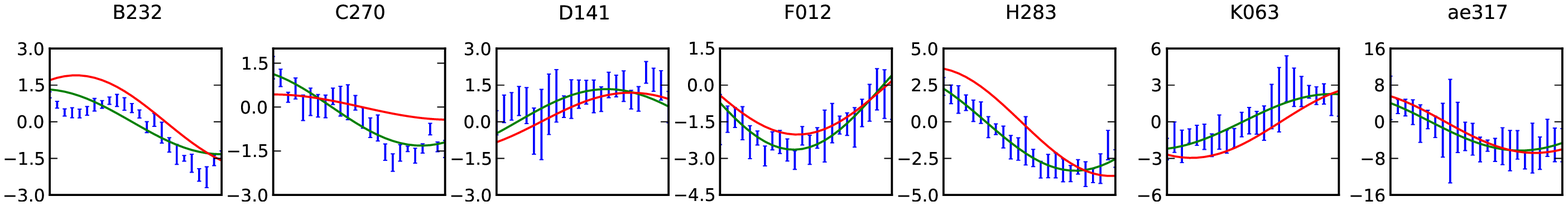}
\caption{\label{fig:dEphase-slope}Phase slopes over the array as a function of time (in deg/km) in the direction of the seven sources for the 2003 (top) and 2006 observations (bottom). The green lines indicate phase slopes corresponding to the fitted position offsets (Fig.~\ref{fig:dEphase-dlm}), the red lines -- phase slopes corresponding to an overall field rotation of $45\arcsec$.}
\end{figure}

\begin{figure}
\centering
\includegraphics[width=.5\columnwidth]{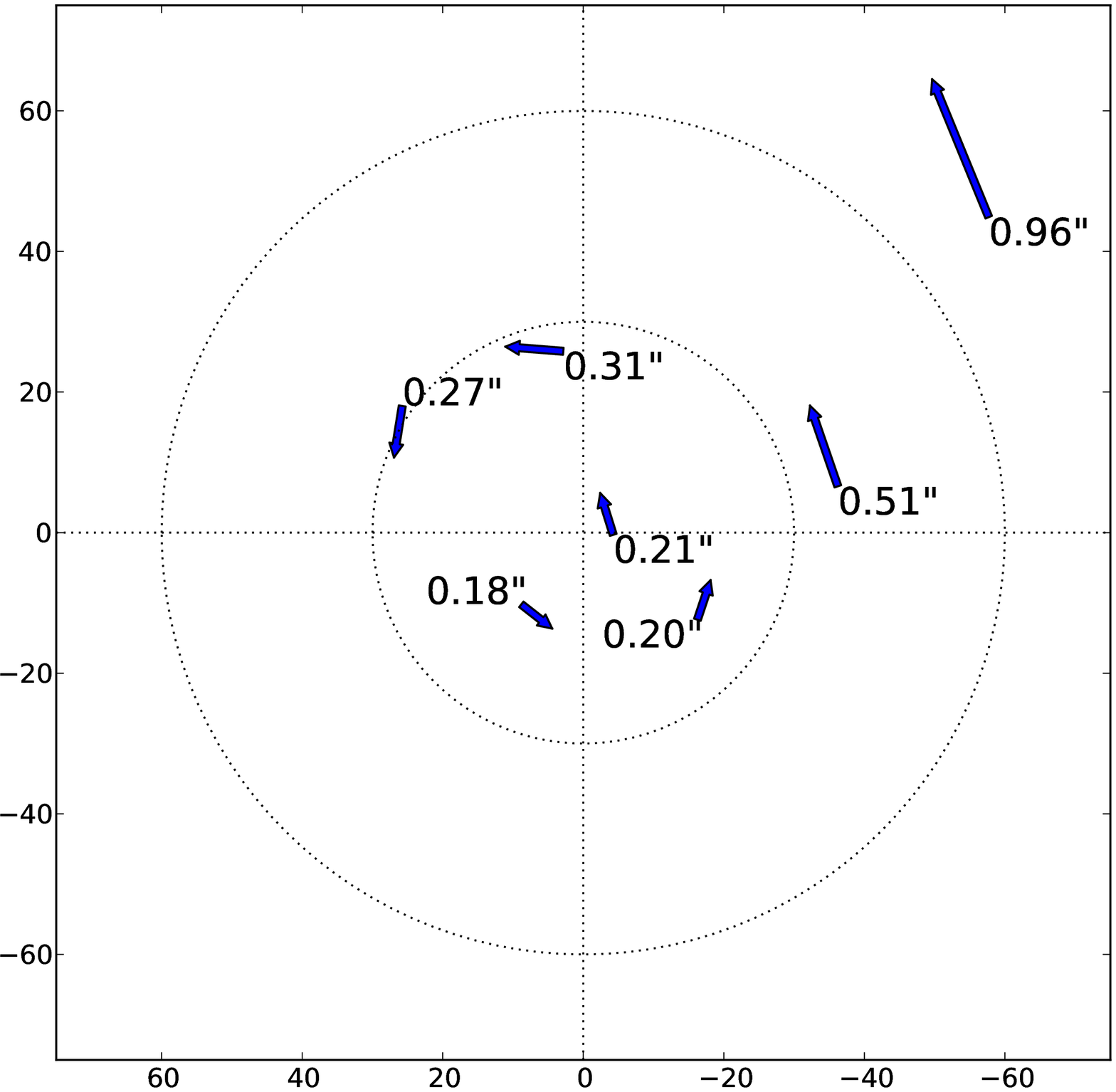}%
\includegraphics[width=.5\columnwidth]{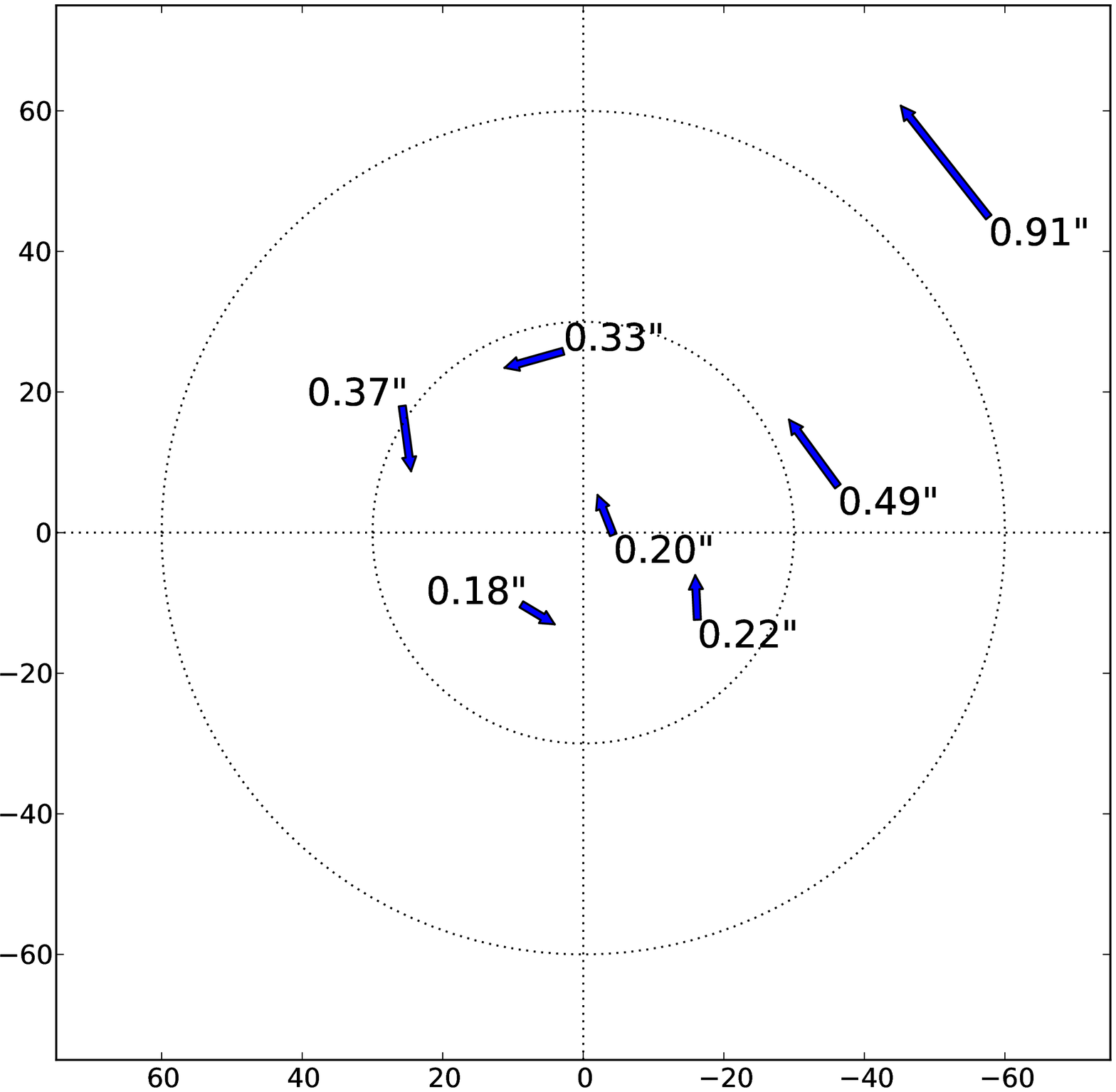}\\
\caption{\label{fig:dEphase-dlm}Fitted position offsets corresponding to the phase slopes of Fig.~\ref{fig:dEphase-slope} (2003 observation on the left, 2006 on the right). The length of the arrows is exaggerated by a factor of 1200: the biggest offset is in fact just under $1\arcsec$.}
\end{figure}

A rather prominent feature of the phase plots in Fig.~\ref{fig:dEphase} is their continuity from antenna to antenna, and the fact that the phases at the two ends of the array exhibit opposite temporal trends. This is suggestive of an evolving phase slope over the array. Fitting such a slope (per source) produces some very striking results (Fig.~\ref{fig:dEphase-slope}). The dominant phase effect is clearly global rather than antenna-based, and is extremely consistent across both observations.

A phase slope over the array can be interpreted as an apparent position offset. It appears that the slope behaviour in Fig.~\ref{fig:dEphase-slope} can be fitted quite well by \emph{constant} position offsets. 
The best-fitting position offsets are indicated in Fig.~\ref{fig:dEphase-dlm}, and the corresponding slope curves are plotted in green on Fig.~\ref{fig:dEphase-slope}. 

Figure~\ref{fig:dEphase-dlm} immediately suggests a field rotation. And indeed, the entire collection of phase slopes (for both the 2003 and 2006 observation), is, to first order, consistent with a rotation of $45\arcsec$ around the phase centre. The corresponding slope curves are plotted in red. While there are some significant differences in the brighter sources, it seems clear that the dominant effect is not an instrumental DDE at all, but a systematic rotation of the sky model. The model positions are derived by NEWSTAR from direct fits to the visibilities, and de Bruyn (priv. comm.) has independently cross-checked the positions of distant sources against the NVSS, which seems to preclude a rotation in the model itself. Note that a $45\arcsec$ rotation can also be introduced by a clock error of about 2.9 s, or a corresponding rotational error in conversion of $uvw$ coordinates from apparent to J2000. Since NEWSTAR and MeqTrees use completely different tool chains and visibility data formats, I cannot exclude a coordinate conversion error somewhere along the line. This needs to be urgently investigated. If indeed the entire sky model is slightly rotated, then perhaps the image of Fig.~\ref{fig:3c147} can even be improved upon!

\subsection{Feeding differential gains back into the sky model\label{sec:model-improvement}}

The results above suggested that I could improve my sky model by feeding back in some information extracted from the $\Delta\jones{E}{}$ solutions. In the previous section, I obtained a correction to the model positions of the seven sources.\footnote{For the moment, I've left aside the issue of whether the positional offsets are ultimately due to a global field rotation. Improving the positions of seven of the brightest off-axis sources should already produce an superior sky model.} Following the discussion of Sect.~\ref{sec:de-analysis-model}, I could also provide corrections for the $I$ and $Q$ fluxes by applying the per-source average $\Delta\jones{E}{}$ amplitudes:

\begin{eqnarray*}
\coh{B}{s}^\mathrm{(corr)} & = & \overline{|\Delta\jones{E}{s}|} \coh{B}{s} \overline{|\Delta\jones{E}{s}|} \\
\overline{|\Delta\jones{E}{s}|} & \equiv & \matrixtt{\overline{|\Delta e_{xs}|}}{0}{0}{\overline{|\Delta e_{ys}|}} = \frac{1}{N_\mathrm{ant}N_t N_\nu} \sum_{p,i,j} |\Delta\jones{E}{sp}(t_i,\nu_j)|,
\end{eqnarray*}

where $t_i$ and $\nu_j$ represent the time and frequency solution intervals of $\Delta\jones{E}{}$. In terms of the $I$ and $Q$ fluxes, the correction becomes:

\begin{eqnarray*}
I^\mathrm{(corr)} = \Sigma \cdot I + \Delta \cdot Q, & \; &  Q^\mathrm{(corr)} = \Delta \cdot I + \Sigma \cdot Q, \\
\Sigma = \frac{1}{2}\left( \overline{|\Delta e_{xs}|}^2 + \overline{|\Delta e_{ys}|}^2 \right), 
& \; & \Delta = \frac{1}{2}\left( \overline{|\Delta e_{xs}|}^2 - \overline{|\Delta e_{ys}|}^2 \right).
\end{eqnarray*}

\begin{figure}
\begin{centering}
\includegraphics[width=.5\columnwidth]{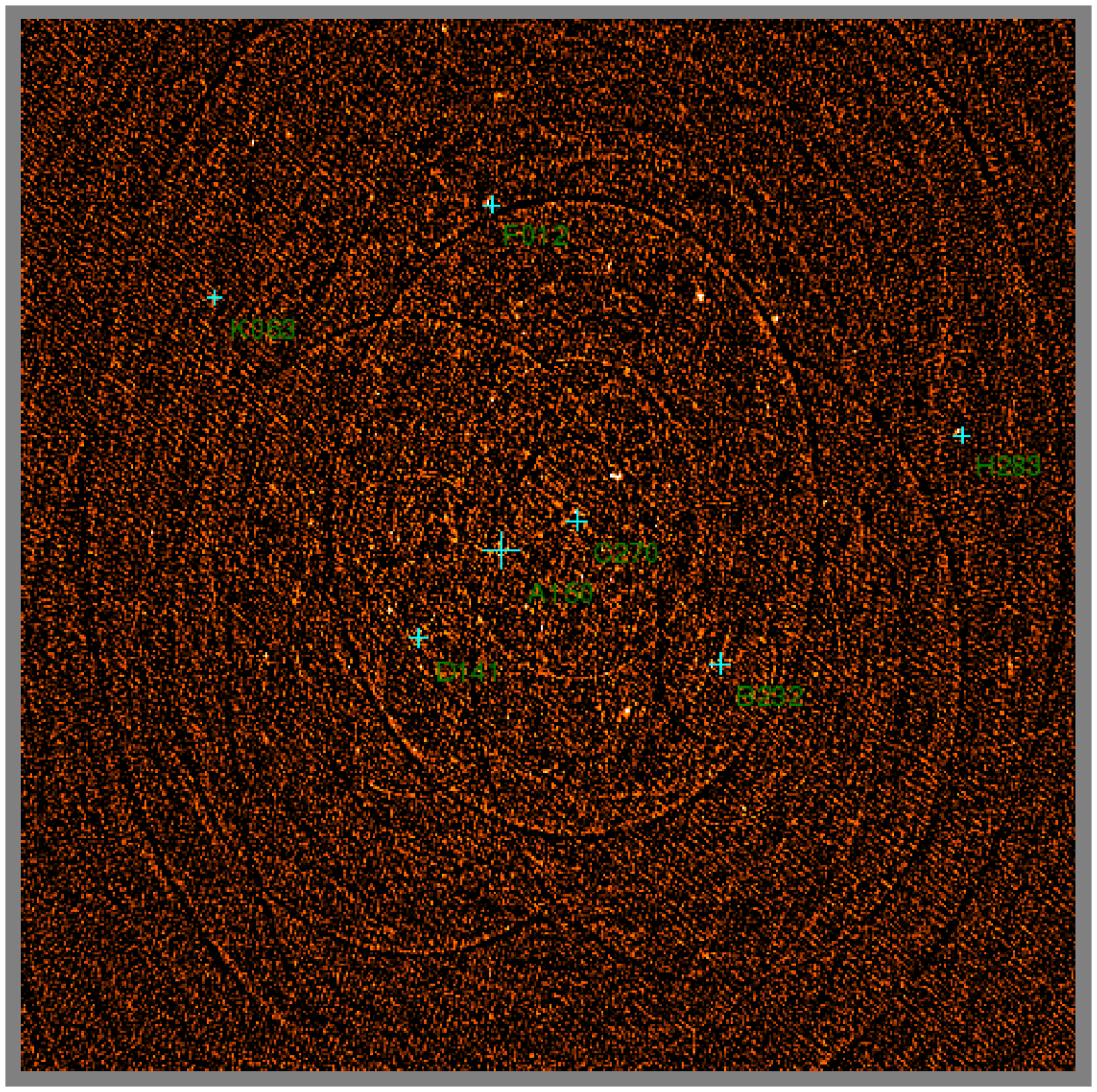}%
\includegraphics[width=.5\columnwidth]{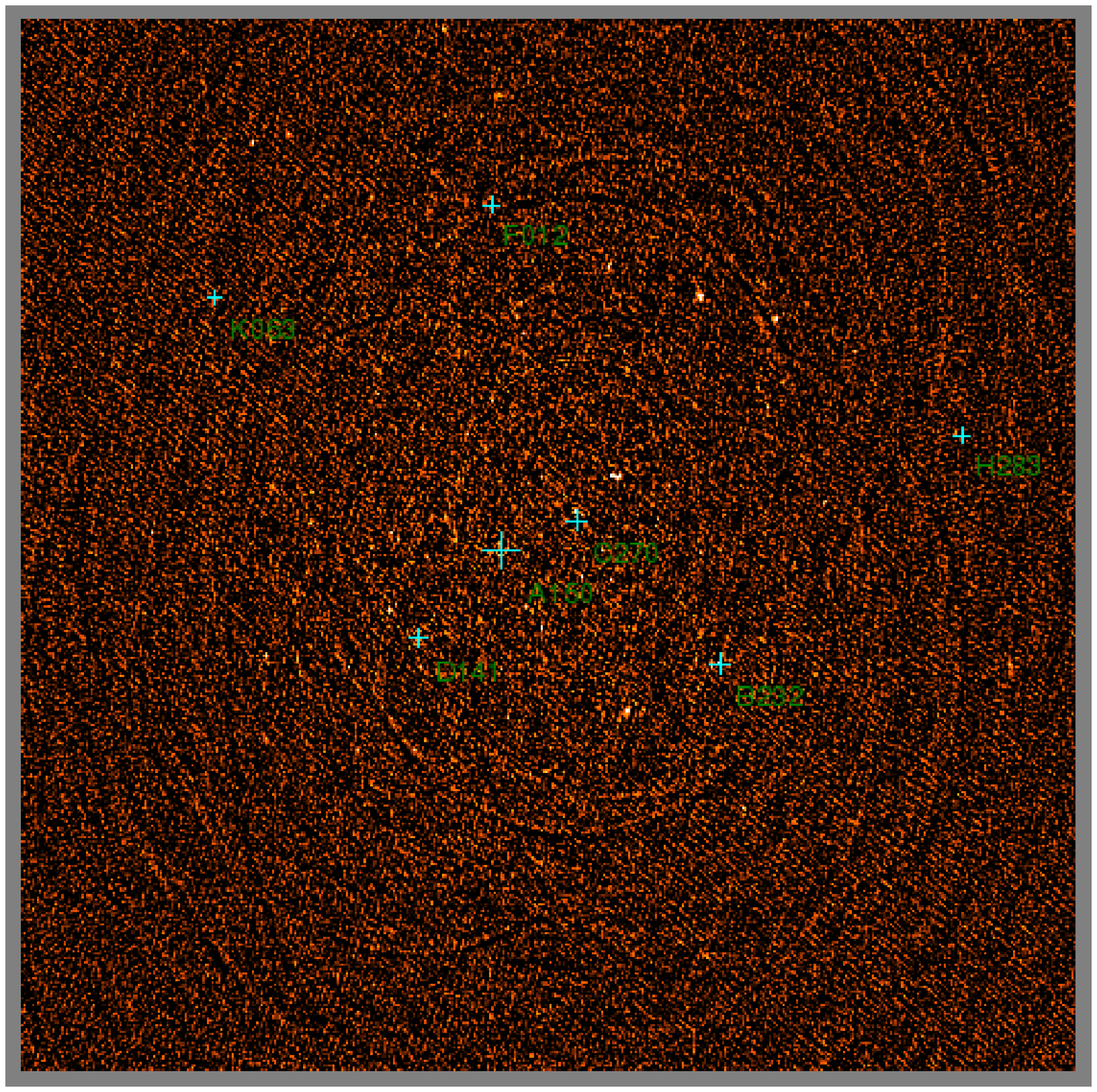}\par
\end{centering}
\caption{\label{fig:residuals-newmodel}Calibration with an improved sky model. This shows single-band residual images after $\jones{G}{p}$ and $\coh{M}{pq}$ solutions. The left image is from the original reduction, the right image uses a sky model improved via my $\Delta\jones{E}{}$ analysis. Crosses indicate the positions of sources for which the model was improved, plus 3C 147 itself (source A).}
\end{figure}

I therefore applied these corrections for $I$, $Q$, and position to my sky models (independently for the 2003 and 2006 observations), and repeated the calibration procedure. An improvement in single-band residuals was immediately apparent (Fig.~\ref{fig:residuals-newmodel}) -- after $\jones{G}{p}$ and $\coh{M}{pq}$ solutions, the seven off-axis sources subtracted noticeably better. Residuals after $\Delta\jones{E}{sp}$ solutions, on the other hand, looked pretty much the same (this is not surprising, since differential gains had already taken care of the visible off-axis errors in the original reduction), with a very slight improvement around 3C 147 itself,
which can be explained by improved $\jones{G}{p}$ solutions due to the more accurate sky model.

\subsection{Phase behaviour II\label{sec:de-analysis-phase2}}

Presumably, the remaining residual structures in Fig.~\ref{fig:residuals-newmodel} are more representative of the instrumental DDEs per se, since inaccuracies in the sky model have been significantly reduced. We should also expect the differential gain solutions to be more indicative of the actual DDEs (apart from the issue of resolved sources affecting $||\Delta\jones{E}{}||$ on antennas RTC and RTD, which the improved sky models do not address at all). Of particular interest is the effect that the improved model has on the differential gain-phases. As for the gain-amplitudes, we would expect them to differ by only an overall per-source scaling factor. Indeed, making the same $||\Delta\jones{E}{}||$ plot as in Fig.~\ref{fig:dEampl} confirms this -- it is, to all intents and purposes, identical (and omitted here to save space), since the plotted amplitudes are renormalized by the per-source average $||\Delta\jones{E}{}||$.

\begin{figure*}
\sidecaption
\parbox[b]{12cm}{
\includegraphics[width=12cm]{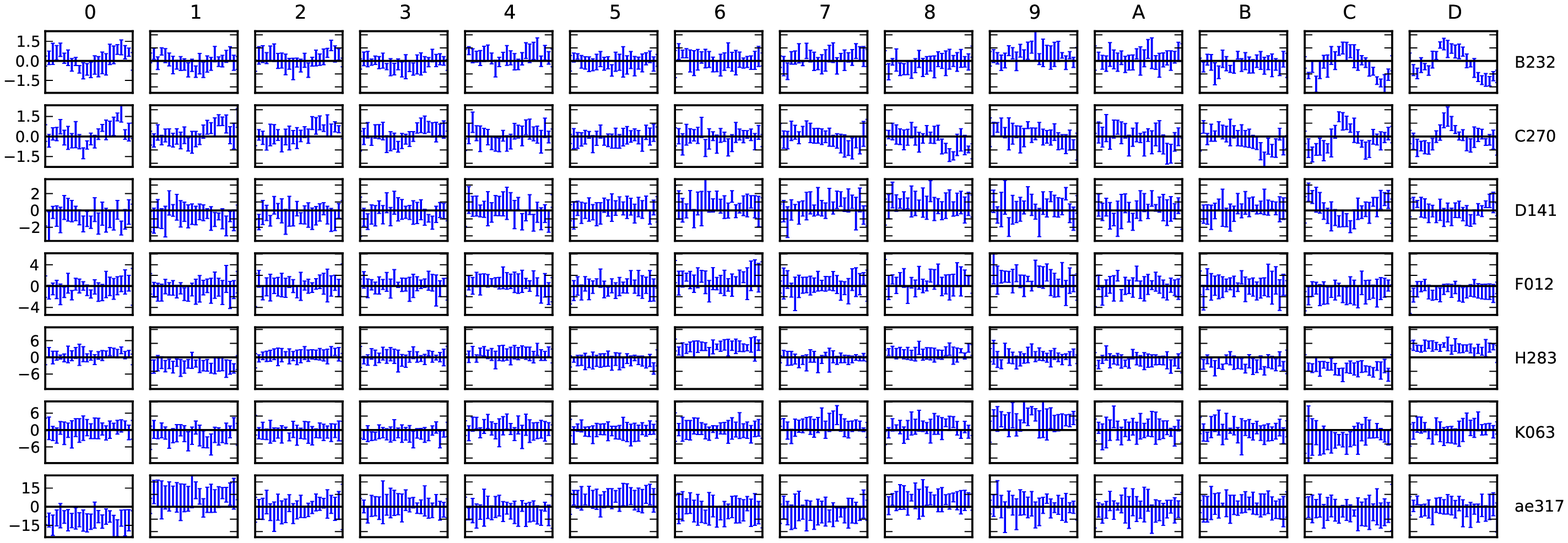}
\includegraphics[width=12cm]{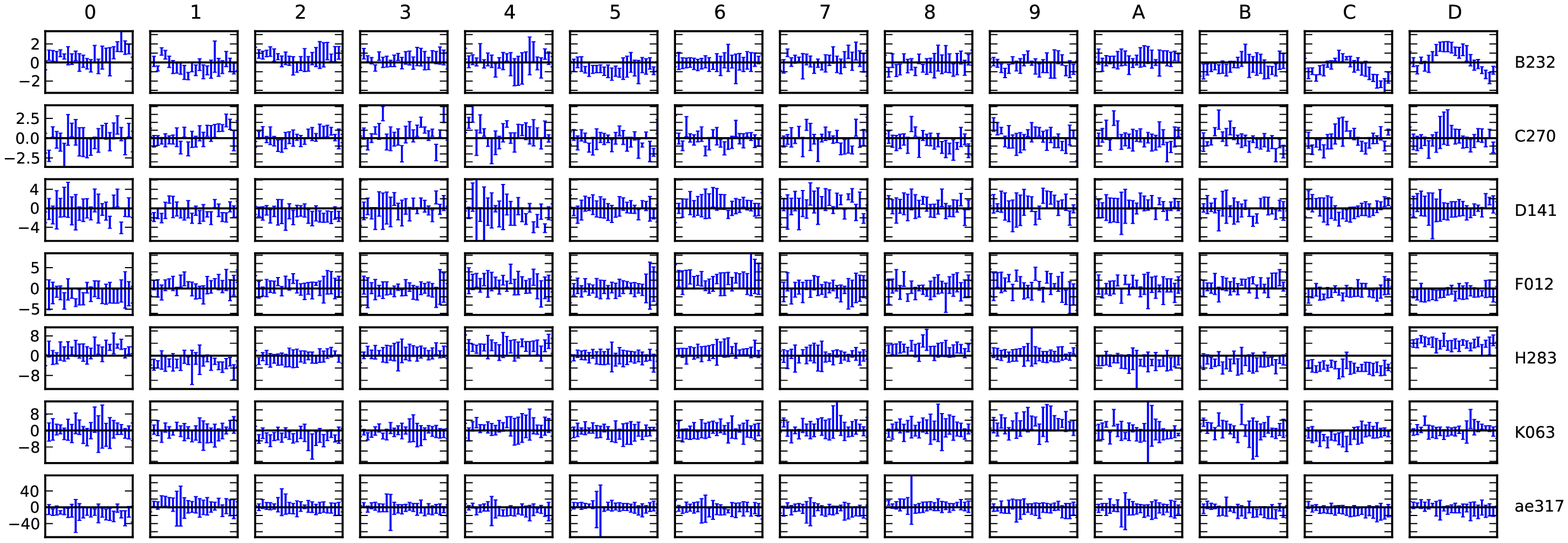}
}
\caption{\label{fig:new-dEphase}Differential gain-phases ($\arg\Delta\jones{E}{}$, in degrees) as a function of time, using improved sky models for the 2003 (top) and 2006 (bottom) observations. Compare to Fig.~\ref{fig:dEphase}.}
\end{figure*}

\begin{figure}
\centering
\includegraphics[width=\columnwidth]{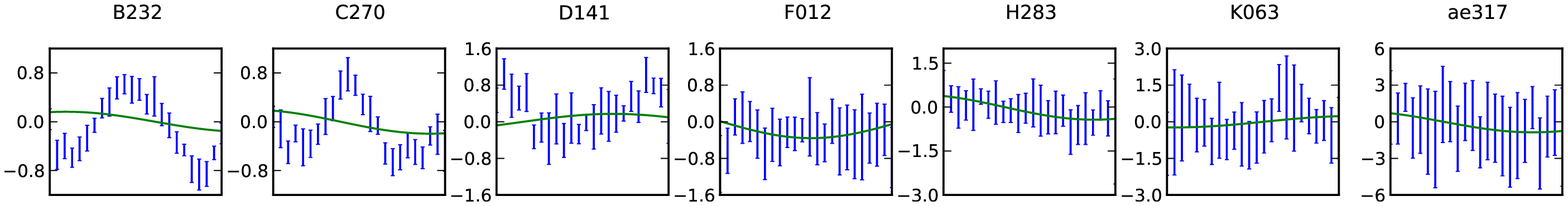}\\
\includegraphics[width=\columnwidth]{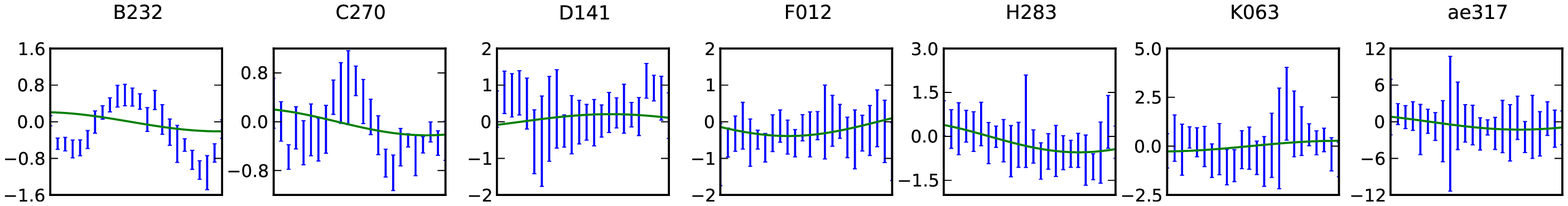}
\caption{\label{fig:new-dEphase-slope}Phase slopes over the array as a function of time (in deg/km), using improved sky models for the 2003 (top) and 2006 observations (bottom). The green lines indicate phase slopes corresponding to the fitted position offsets (Fig.~\ref{fig:new-dEphase-dlm}). Compare to Fig.~\ref{fig:dEphase-slope}.}
\end{figure}

\begin{figure}
\centering
\includegraphics[width=.5\columnwidth]{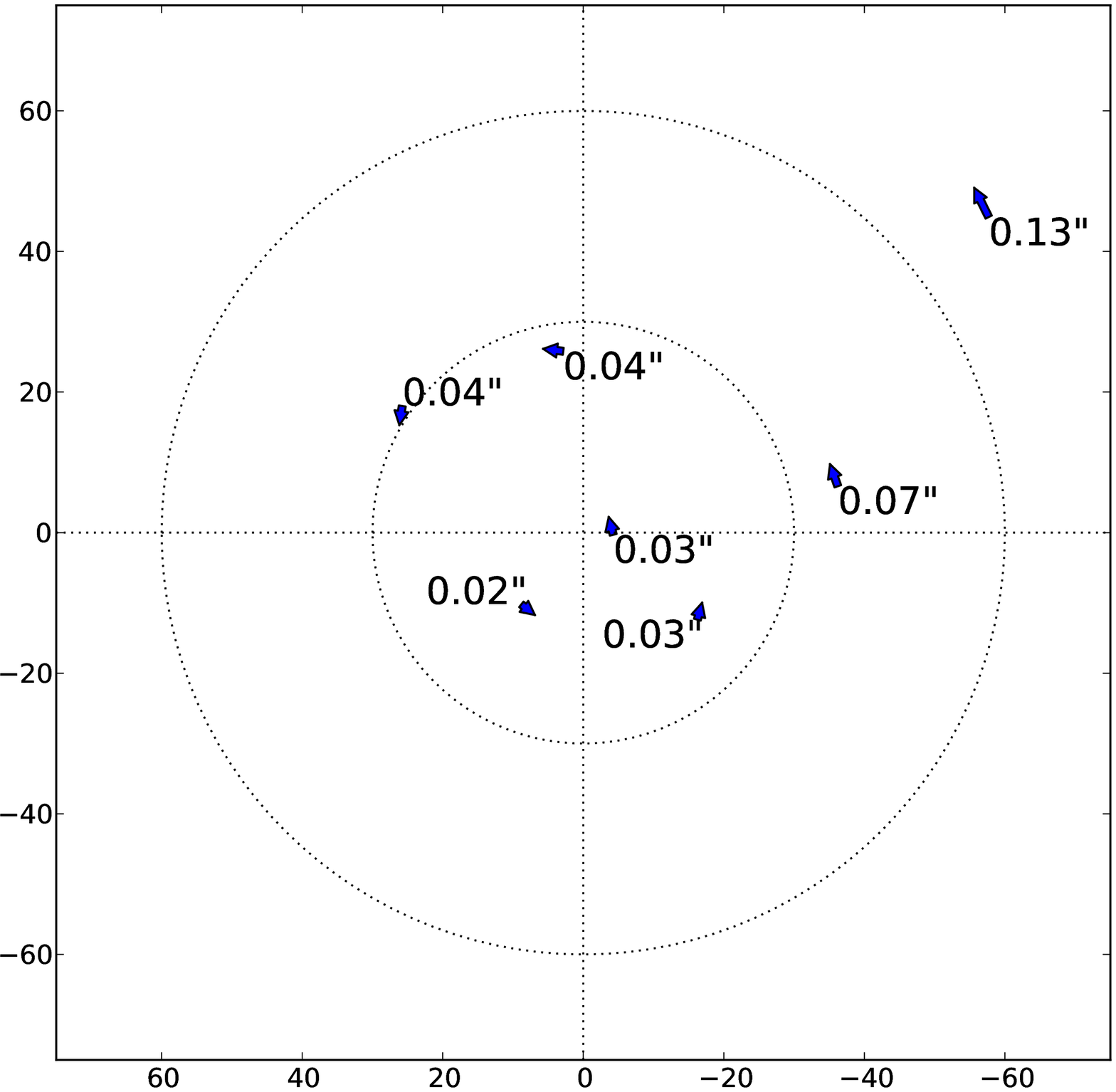}%
\includegraphics[width=.5\columnwidth]{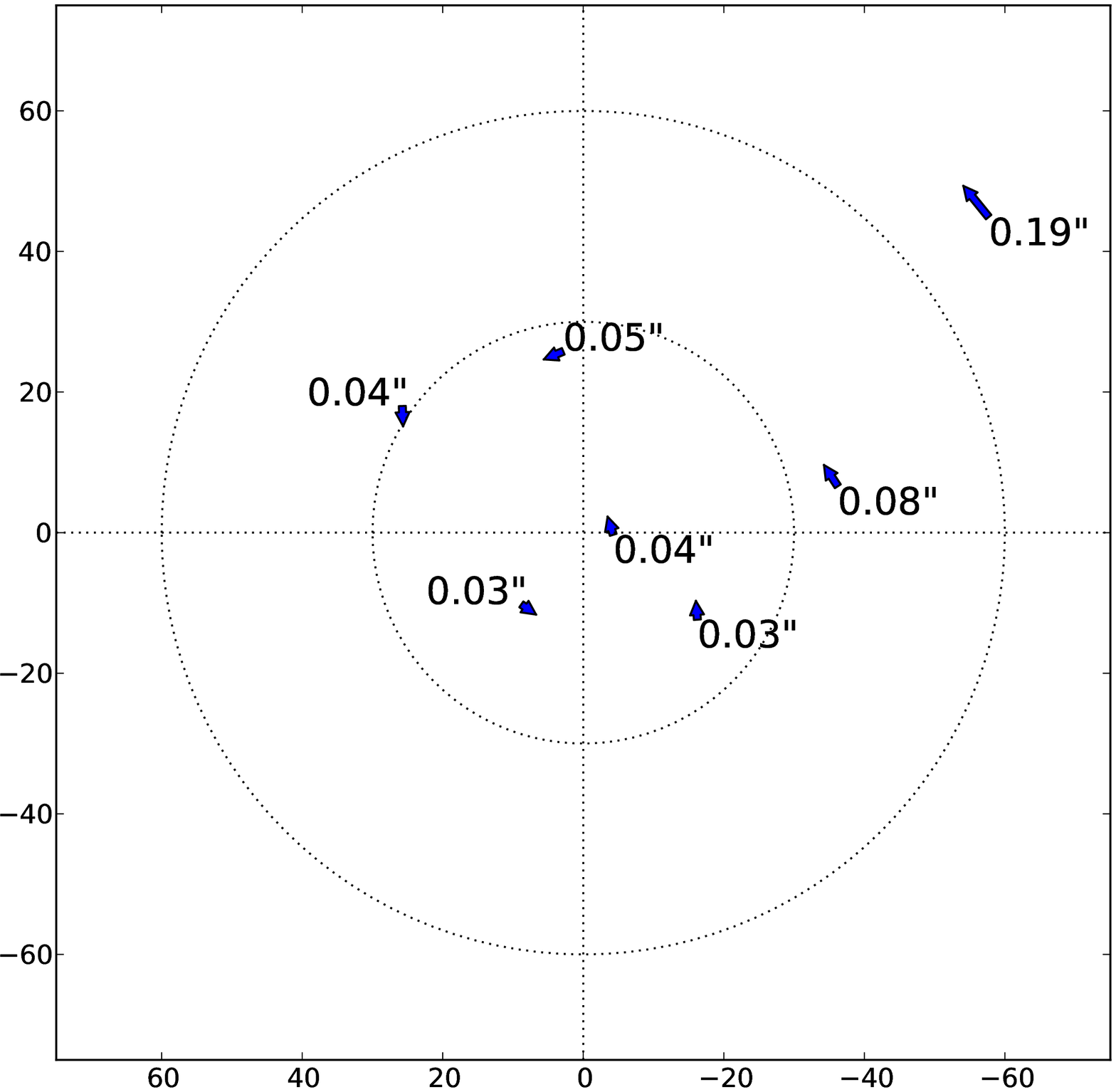}
\caption{\label{fig:new-dEphase-dlm}Fitted position offsets corresponding to the phase slopes of Fig.~\ref{fig:new-dEphase-slope} (2003 observation on the left, 2006 on the right). The length of the arrows is exaggerated by a factor of 1200: the biggest offset is in fact just under $0.2\arcsec$. Compare to Fig.~\ref{fig:dEphase-dlm}.}
\end{figure}

The phases, on the other hand, show a marked difference, since the formerly dominant effect -- that of position offsets -- has been taken out. The $\arg\Delta\jones{E}{}$ solutions themselves are shown in Fig.~\ref{fig:new-dEphase}. Phase slopes are still very much in evidence, as can be seen in Fig.~\ref{fig:new-dEphase-slope}. Somewhat surprisingly, these slopes indicate that some residual position offsets remain, at a level of 15\% to 20\% of the original offsets (Fig.~\ref{fig:new-dEphase-dlm}). This suggests that my procedure of fitting phase slopes to $\arg\Delta\jones{E}{}$ solutions, followed by fitting position offsets to the slopes, systematically \emph{underestimates} the true position offsets. This is possibly an effect of the complex averaging implicit in having one $\Delta\jones{E}{}$ solution per a relatively large solution interval (20 MHz by 30 minutes). If so, this could perhaps be incorporated as a multiplicative correction factor in the model update procedure. Further work is required to fully understand the effect. 


The brighter sources B, C, and (to a lesser extent) D show clear second-order phase effects, both in the phase slopes, and in the phase solutions themselves. The temporal continuity in the phase slopes can be interpreted as a \emph{time-variable} position offset. I can speculatively offer two explanations for such an offset:

\begin{itemize}
\item Unmodelled source structure (again!) For any given hour angle, an E-W array only sees an integrated cross-section through the source in a given direction. If the source is slightly resolved with an asymmetric ``hotspot'', the zero-order moment of each such cross-section will be slightly different. 
\item Differential tropospheric or large-scale ionospheric refraction, including perhaps apparent change of baseline caused by refraction (the Anderson effect).
\end{itemize}

Another puzzling feature of the $\arg\Delta\jones{E}{}$ solutions in Fig.~\ref{fig:new-dEphase}
are the significant and (to first degree) constant phase offsets of some sources (e.g. H, K, ae) on some antennas. The offsets are mostly (though not completely) consistent between the 2003 and 2006 observations. None of the explanations offered above are consistent with a \emph{constant} phase offset! Could this be the phase component of the primary beam? There are too few sources in this reduction to infer any sort of directional dependence, but perhaps the ``QMC Project'' can provide more insights on this effect.

\subsection{The lurking errors\label{sec:deep-errors}}

The two calibrations (with the original and the improved sky models) described above have produced what appear to be identical final maps. This shows that the ``flyswatter'' can accommodate for significant errors in the sky model. On the other hand, the detailed structures in Fig.~\ref{fig:new-dEphase-slope} suggest that (even in the very benign case of 21 cm WSRT observations!) moderately bright off-axis sources still require some form of DDE correction even if the model is perfect. If this is the case, then a legitimate question is: why worry about getting the sky model right, if we need to do $\Delta\jones{E}{}$ solutions anyway, which will absorb any imperfections? (Besides the obvious caveats of the ``flyswatter'' discussed in Sect.~\ref{sec:dE-limitations}, that is.)

\begin{figure}
\begin{centering}
\includegraphics[width=\columnwidth]{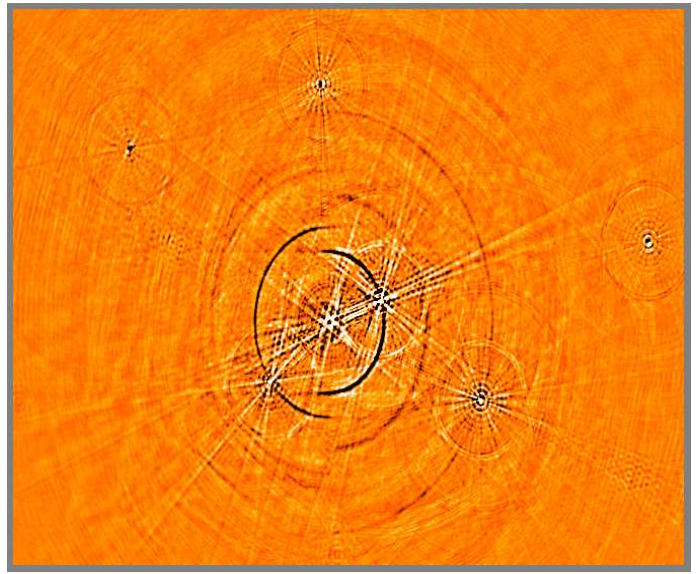}%
\end{centering}
\caption{\label{fig:diff-newmodel}Calibration with an improved sky model. This shows the \emph{difference} between the 8-band residual maps (2003 observation) produced with the original and the improved sky models. Structures around 3C 147 itself and the off-axis sources are mostly due to ``selfcal contamination'' in the original model caused by incorrect off-axis source positions. These are well within the noise: the intensity range of this image is $\pm2 \mu\mathrm{Jy}$, while the 8-band maps have a thermal noise of $13.5 \mu\mathrm{Jy}$.}
\end{figure}

The rather striking image of Fig.~\ref{fig:diff-newmodel} shows that the final maps are not in fact identical, although the difference is buried in the noise. This image was produced by subtracting the original-model 8-band residual map from the improved-model map (2003 observation). 
Since the noise term in both maps is the same, subtraction reveals very faint structures that would normally be hidden in the noise. We're beginning to see more limitations of the ``flyswatter''. In the original reduction, apparent position offsets of the off-axis sources caused phase gradients in $t,\nu$-space in the differential gain-phases. These were approximated by a stepwise $\Delta\jones{E}{}$ solution (since I solved for only one $\Delta\jones{E}{}$ term per 30 minutes, per entire band), which proved to be good enough to drive off-axis errors to a level below the thermal noise. Improving the model positions has effectively ``flattened out'' these gradients, reducing the error made by a stepwise approximation even further. Figure~\ref{fig:diff-newmodel} demonstrates the improvement. The radial spokes correspond to ``jumps'' at the boundaries of the solution intervals, but the other structures (especially the half-circles) are rather more difficult to explain, and will have to be addressed in follow-up work.

The implications of this result is that any errors in the sky model (or uncorrected DDEs) will propagate into the selfcal solutions, and result in faint but highly coherent structures in the residual maps. We may think we are reaching the thermal noise, but in the process, we are producing ``submerged'' calibration artefacts at levels below the noise, where we can't even see that something is still going wrong! This is of particular concern to ongoing work on detection of the Epoch of Reionization (EOR) signature, which relies on statistical analysis of residual images to find sub-noise artefacts of astrophysical origin \citep{EOR-LOFAR,EOR-MWA}. Such analysis will have to reckon with these lurking selfcal artefacts.

\section{Conclusions}

One of the biggest selling points of the RIME formalism is the flexibility it offers for describing observational effects. Unfortunately, to date only three software packages have exploited the power of the RIME (CASA, MeqTrees, and the LOFAR BBS system). Of these, only MeqTrees allows for truly arbitrary forms of the RIME. This paper has explored some practical applications of one such form of the RIME: a form that includes differential gain terms. I have demonstrated that the differential gain approach (the ``flyswatter'') can be a powerful way of dealing with DDEs on a source-by-source basis. This has been used with WSRT data to produce artefact-free maps of 3C 147 at record dynamic ranges of well over a million-to-one. While the differential gain solutions themselves absorb inaccuracies in the sky model as well as the DDEs themselves, I have demonstrated that at least flux and positions corrections can be recovered, so iterative improvements to the sky model are possible. 

The latter may also prove be necessary: I have demonstrated that even a perfect-looking map produced using differential gains contains a large number of selfcal artefacts hidden in the thermal noise, which can be significantly reduced by improving the sky models. These ``invisible'' artefacts have hitherto been ignored, but they should be of particular concern to projects relying on statistical signal extraction, such as the ongoing search for the EOR signature.

The nature of the remaining DDEs (as seen in the differential gain solutions) has not yet been adequately explained. Some of the amplitude effects are consistent with pointing error. There phase behaviour is even more difficult to understand, but may be due to unmodelled source structure. Further work is required on the subject. 

I have shown that differential gain-phase solutions can be used to detect position shifts to within small fractions of the synthesized beam size. Offsets of less than $0.05\arcsec$ (well under 0.01 of the PSF size!) have been reliably detected. There is a very clear indication of a systematic rotational offset of $\sim45\arcsec$ in the sky model generated by NEWSTAR, when interpreted using MeqTrees. This is may be due to a coordinate conversion error somewhere in the visibility data processing tool chain, and needs to be investigated further.

Finally, I should consider some wider implications of my results. All currently mooted schemes of DDE calibration for LOFAR \citep{JEN:LOFAR3}, the MWA \citep{Mitchell:MWA-cal} and the ionosphere in general \citep{Intema:SPAM,Cotton:FBC} revolve around the use of ``beacon sources'' to probe the ionosphere and/or the primary beam. It is rather difficult to envisage a closed-loop scheme without beacons (how else would one sample a DDE?), so future telescopes such as the SKA will most likely need to use something very similar. Any such scheme predicates on there being a sufficient number of sufficiently bright in-beam beacons for any direction on the sky. This is not a problem at the LOFAR and MWA end of the spectrum, since the low-frequency sky is so much brighter, but it has been a bit of a worry for the higher frequencies, where FoVs are narrower and sources are fainter.

My 3C 147 results suggest that calibration beacons can be a lot fainter than previously thought. What has been established is that for this particular configuration of the WSRT, sources as faint as 2~mJy can provide meaningful DDE solutions. This result can be scaled to future telescope designs by comparing their expected sensitivity with that of the 3C 147 observation.

\begin{acknowledgements}

I would like to thank a succession of managers for putting up with me all these years, and Jan Noordam for 
making this process considerably easier (especially for the managers), and for many other things besides. Ger de Bruyn has been more than generous with data, models and wisdom. Johan Hamaker started it all, and Wim Brouw has provided an avalanche of insights. Last but certainly not least, the rest of the MeqTrees team has been instrumental in making everything work.

\end{acknowledgements}

\bibliographystyle{aa}

\bibliography{16082}

\end{document}